%% file: main.tex
\documentclass[11pt, a4paper]{article}
\usepackage{jcappub}
\usepackage{makecell}
\usepackage{xcolor}
\definecolor{dgreen}{RGB}{26,148,49}
\definecolor{xlinkcolor}{cmyk}{1,1,0,0}
\hypersetup{linkcolor=xlinkcolor,citecolor=xlinkcolor,urlcolor=xlinkcolor}
\usepackage{hyperref}
\usepackage{orcidlink}
\usepackage{multirow}
\usepackage{pdflscape} 

\usepackage{amsmath,amsthm,amssymb,amsfonts}
\usepackage{graphics}
\usepackage{listings}
\usepackage{enumitem}
\usepackage{url}
\usepackage{appendix}
\usepackage{booktabs}
\usepackage{rotating}

\usepackage{mathtools}
\usepackage[super]{nth}

\DeclareMathOperator{\erf}{erf}

\title{CIBER $\times$ galaxy cross-correlations reveal a bright, low-redshift NIR background}
\author[a, b]{Richard M. Feder\,\orcidlink{0000-0002-9330-8738}}
\author[c]{Grigory Heaton\,\orcidlink{0009-0003-5681-2956}}
\author[c]{James J. Bock\,\orcidlink{0000-0002-5710-5212}}
\author[c]{Yun-Ting Cheng\,\orcidlink{0000-0002-5437-0504}}
\author[d]{Yi-Kuan~Chiang\,\orcidlink{0000-0001-6320-261X}}
\author[c]{Phillip M. Korngut}
\author[e]{Shuji Matsuura\,\orcidlink{0000-0002-5698-9634}}
\author[c]{Jordan Mirocha\,\orcidlink{0000-0002-8802-5581}}
\author[g]{Kohji Tsumura\,\orcidlink{0000-0001-7143-6520}}
\author[h,f]{Michael Zemcov\,\orcidlink{0000-0001-8253-1451}}

\affiliation[a]{Berkeley Center for Cosmological Physics, University of California, Berkeley, CA 94720,
USA}
\affiliation[b]{Lawrence Berkeley National Laboratory, Berkeley, California 94720, USA}
\affiliation[c]{California Institute of Technology, 1200 E California Blvd, 91125, USA}
\affiliation[d]{Academia Sinica Institute of Astronomy and Astrophysics (ASIAA), No. 1, Section 4, Roosevelt Road, Taipei 10617, Taiwan}
\affiliation[e]{School of Science and Technology, Kwansei Gakuin University, Sanda, Hyogo 669-1337, Japan}
\affiliation[f]{Jet Propulsion Laboratory, California Institute of Technology, 4800 Oak Grove Drive, Pasadena, CA 91109, USA}
\affiliation[g]{Frontier Research Institute for Interdisciplinary Science, Tohoku University, 980-8578 Miyagi, Japan}
\affiliation[h]{School of Physics and Astronomy, Rochester Institute of Technology, 1 Lomb Memorial Drive, Rochester, New York 14623, USA}
\emailAdd{rmfeder@berkeley.edu}
\abstract{We perform the first tomographic analysis of near-IR extragalactic background light (EBL) anisotropies, cross-correlating \textit{CIBER} 1.1 and 1.8 $\mu$m imager data with photometric galaxy catalogs from DESI Legacy Survey DR8 and Hyper-Suprime-Cam Ultra-Deep Survey. We measure significantly higher cross-power than expectations from an integrated galaxy light (IGL) model on scales $\ell < 2000$, concentrated at low redshift ($z\lesssim 0.6$). Cluster member galaxies and associated structure account for 15-20\% of the large-angle cross-power, indicating that group- and galaxy-scale halos contribute the bulk of the signal. Through a parametric halo model decomposition, we detect two-halo and one-halo clustering in cross-power at high significance, with amplitudes that decline smoothly across $z=0{-}1$. The inferred one-halo cross-power is of similar amplitude between DESI-LS and the deeper HSC catalog, implying a scenario in which low-redshift EBL fluctuations are amplified by contributions from lower-mass halos with satellites and/or diffuse intra-halo light (IHL). Converting our two-halo fits into estimates of $b_I \times dI/dz$, we find that standard IGL predictions underestimate our measurements, even when assuming an intensity bias as high as 3, similar to that of large SZ clusters, suggesting that a higher $dI/dz$ is required to reconcile observed discrepancies. Lastly, we find that correlated large-scale structure (LSS) at $z<1$ accounts for a substantial fraction of the \emph{CIBER} auto-power reported in earlier work. These results identify low-redshift LSS as a significant and previously unappreciated contributor to near-IR EBL fluctuation measurements, setting the stage for cross-correlation science with \emph{CIBER}-2, \emph{SPHEREx} and a variety of LSS tracers.}

\begin{document}
\maketitle
\input{introduction}

\input{cl_calc}
\input{ciber}
\input{phot_samples}

\input{mock_tests}
\input{cross_cl_results}
\input{modeling}
\input{discussion}
\acknowledgments

This material is based upon work supported by the National Aeronautics and Space Administration under APRA research grants NNX10AE12G,
NNX16AJ69G, 80NSSC20K0595, 80NSSC22K0355, and
80NSSC22K1512. R.M.F. acknowledges support from the Berkeley Center for Cosmological Physics. The authors would like to thank Simone Ferraro, Tzu-Ching Chang and Martin White for useful discussions. The authors acknowledge the \textit{CIBER}-1 instrument team for collecting the imager data used in this work.

This publication makes use of data products from the Two Micron All Sky Survey, which is a joint project of the University of Massachusetts and the Infrared Processing and Analysis Center/California Institute of Technology, funded by the National Aeronautics and Space Administration and the National Science Foundation. 

This work used Bridges-2 \citep{Bridges2} at Pittsburgh Supercomputing Center through allocation PHY240026 from the Advanced Cyberinfrastructure Coordination Ecosystem: Services \& Support (ACCESS) program, which is supported by National Science Foundation grants \#2138259, \#2138286, \#2138307, \#2137603, and \#2138296.

The DESI Legacy Imaging Surveys consist of three individual and complementary projects: the Dark Energy Camera Legacy Survey (DECaLS), the Beijing-Arizona Sky Survey (BASS), and the Mayall z-band Legacy Survey (MzLS). DECaLS, BASS and MzLS together include data obtained, respectively, at the Blanco telescope, Cerro Tololo Inter-American Observatory, NSF NOIRLab; the Bok telescope, Steward Observatory, University of Arizona; and the Mayall telescope, Kitt Peak National Observatory, NOIRLab. NOIRLab is operated
by the Association of Universities for Research in Astronomy (AURA) under a cooperative agreement with the National Science Foundation. Pipeline processing and analyses of the data were supported by NOIRLab and the Lawrence Berkeley National Laboratory. Legacy Surveys also uses data products from the Near-Earth Object Wide-field Infrared Survey Explorer (NEOWISE), a project of the Jet Propulsion Laboratory/California Institute of Technology, funded by the National Aeronautics and Space Administration. Legacy Surveys was supported by: the Director, Office of Science, Office of High Energy Physics of the U.S. Department of Energy; the National Energy Research Scientific Computing Center, a DOE Office of Science User Facility; the U.S. National Science Foundation, Division of Astronomical Sciences; the National Astronomical Observatories of China, the Chinese Academy of Sciences and the Chinese National Natural Science Foundation. LBNL is managed by the Regents of the University of California under contract to the U.S. Department of Energy.

\emph{Software:} \texttt{matplotlib}, \texttt{numpy} \citep{numpy}, \texttt{emcee} \citep{2013PASP..125..306F}, \texttt{corner} \citep{corner}, \texttt{pyfftw}.

\appendix
\input{app_phot_dIdz}
\input{app_catalogs}

\newpage
\bibliography{references}{}
\bibliographystyle{aasjournal}

\end{document}

%% file: introduction.tex
\section{Introduction}

The extragalactic background light (EBL) comprises the integrated light from all sources outside of the Milky Way across cosmic history \citep{scott_ebl}. Through its spectral and spatial characteristics, the EBL can be used to understand the astrophysical processes that drive cosmic light production \citep{hauser_dwek} and to probe star formation across a broad range of redshifts \citep{raue_ebl_sfrd, cooray16}.

While measurements of the EBL have reached agreement at far-infrared wavelengths using several methods and datasets \citep{viero13, viero15, varnish_pd_analysis, akari_fir, ykc25_farir}, the landscape of measurements at optical and near-infrared (NIR) wavelengths is more complicated. Recent absolute intensity measurements with \emph{LORRI} at optical wavelengths \citep{lauer_2021, lauer_2022, postman_2024} and indirect measurements using gamma-ray absorption \citep{banerjee26, alfaro26} have come into closer agreement with estimates from deep galaxy catalogs \citep{carter25}, (i.e., the claimed ``optical EBL convergence" \citep{ebl_biteau}); however, the measurements are in direct tension with absolute NIR measurements from the Cosmic Infrared Background ExpeRiment \citep[\emph{CIBER};][]{matsuura_ciber} and inferred CIB monopole intensities from DIRBE \citep{cosmoglobe_cibmono}, both of which imply a factor of $\sim 3\times$ higher intensity for observed wavelengths $\lambda = 1-2$ $\mu$m. Diffuse light measurements with the Hubble Space Telescope (HST) through the SKYSURF project also reveal a higher than expected sky intensity after removing known contributions \citep{carleton, tompkins_skysurf}, which may either be cosmological or explained by an additional isotropic Zodiacal light (ZL) component \citep{obrien_skysurf}.

Fluctuation-based measurements of the EBL bypass the conventional challenge of absolute photometric measurements, namely degeneracy with ZL, which has been shown to be spatially smooth on large angular scales \citep{arendt16, pyo_akari}. Unlike line-emission intensity mapping (LIM), the signal from broadband, near-IR intensity mapping is dominated by the stellar continua of all galaxies. \cite{zemcov14} pursued fluctuation measurements at 1.1 and 1.6 $\mu$m using imaging data from the second and third flights of \emph{CIBER}. This measurement revealed fluctuations on angular scales $\theta > 5^{\prime}$ that exceeded predictions with integrated galactic light (IGL), integrated stellar light (ISL) and diffuse Galactic light (DGL). In \cite{feder25a, feder25b}, we performed a fluctuation analysis of fourth flight \emph{CIBER}-1 data, finding similar results to \cite{zemcov14} but with higher sensitivity and improved methodology. Notably, on scales $5^{\prime} < \theta < 30^{\prime}$ ($300< \ell<2000$), the 1.1 and 1.8 $\mu$m auto-power spectrum measurements were an order of magnitude higher than existing models that include IGL, ISL and DGL at $\sim 10\sigma$ significance. While that work found no evidence for local and Galactic foregrounds as the primary source of the observed \emph{CIBER} fluctuation power, the contributions associated with linear and non-linear galaxy clustering, in particular at low redshift, are less well understood.

The physical origin of the excess fluctuation power has direct bearing on the broader EBL tension. If the signal originates from two-halo-dominated clustering of unresolved galaxies, it implies an enhanced bias-weighted intensity kernel $W_I(z) = b_I(z)\times dI/dz$, pointing toward a large, unresolved galaxy contribution to the NIR background. If instead, the signal is dominated by or one-halo processes -- such as intra-halo light (IHL) produced by tidal stripping \citep{cooray12, zemcov14, mitchellwynne}, or enhanced emission from satellite galaxies in massive halos -- it reflects brighter and more biased emission from halos rather than a missing population of sources. Distinguishing these scenarios from auto-spectra alone is challenging and requires wide surveys to clearly separate their spatial signatures.  

A powerful tool for deciphering EBL fluctuations is redshift tomography \citep{cheng22_cnib_tomography}. Building on the ``clustering redshift" technique from \cite{menard_clusred} that characterizes the bias-weighted redshift distribution of photometric galaxy samples through cross-correlations with spectroscopic reference samples, EBL redshift tomography can be used to constrain $W_I(z)$ \citep{ykc_uv, ykc25_farir}. Cross-correlating the intensity field with galaxy tracers enables a tomographic decomposition as a function of both angular scale and redshift, allowing further interpretation of the full signal. In this work, we build on \cite{zemcov14, chengihl, feder25a, feder25b} by cross-correlating the \emph{CIBER} \nth{4} flight imaging dataset against a number of photometric galaxy samples detected at optical and infrared wavelengths. We use catalogs derived from the DESI Legacy Survey \citep[DESI-LS][]{legacy_survey} and Hyper-Suprime-Cam \cite[HSC;][]{hsc} imaging datasets, utilizing available photometric redshift estimates to bin tracers across $0<z<1$. 

The paper is organized as follows. We describe the halo model underpinning our cross-correlation investigation in \S \ref{sec:cib_modl}. We then introduce \emph{CIBER} and present initial sensitivity forecasts for the \nth{4}-flight dataset in \S \ref{sec:ciber}. In \S \ref{sec:phot_samples} we introduce our photometric tracer samples and the associated processing for fluctuation measurements. After detailing the baseline CIB mocks and predictions in \S \ref{sec:cib_predictions}, we present and validate our pseudo-$C_{\ell}$ pipeline for the unbiased recovery of auto- and cross-power spectra (\S \ref{sec:mock_tests}). Our primary cross-correlation measurements are presented in \S \ref{sec:results}, after which in \S \ref{sec:model_interp} we pursue parametric cross-spectrum fits to interpret the one- and two-halo contributions of correlated LSS at $z<1$. Lastly, in \S \ref{sec:conclusion} we discuss the implications of the measurements and limitations of our analysis, along with future prospects for wider-area intensity mapping.

%% file: cl_calc.tex
\section{CIB Modeling}
\label{sec:cib_modl}
\subsection{Halo model for cross-correlations}

We employ a halo model framework to decompose galaxy and intensity auto- and cross-power spectra into one-halo and two-halo contributions, following \cite{cheng22_cnib_tomography}. The total angular power spectrum for any pair of tracers $\lbrace \rm X, Y \rbrace$ can be expressed as

\begin{equation}
    C_{\ell}^{\rm XY} = C_{\ell}^{\rm XY, 1h} + C_{\ell}^{\rm XY, 2h} + C_{\ell}^{\rm XY, P},
\end{equation}
where the one-halo term captures correlations between objects within the same dark matter halo, the two-halo term captures correlations between objects in separate halos tracing the large-scale matter distribution, and $C_{\ell}^{\rm XY, P}$ is the Poisson contribution from the discrete nature of the tracers. On large angular scales ($\ell \lesssim 1000$, $\theta \gtrsim 10^{\prime}$), two-halo clustering typically dominates the power spectrum, while on smaller scales the one-halo term becomes more significant. The relative contribution of each component depends on the observed angular scale and tracer redshift distribution, among other details regarding the tracer galaxies' halo occupation distribution.

We integrate three-dimensional power spectra to angular power spectra using the Limber approximation,
\begin{equation}
    C_\ell^{\rm XY} = \int dz\, \frac{H(z)}{c}\, 
    \frac{W_{\rm X}(z)\, W_{\rm Y}(z)}{\chi^2(z)}\, 
    P_{\rm XY}\!\left(k = \frac{\ell + 1/2}{\chi(z)}, z\right),
    \label{eq:limber}
\end{equation}
where $\chi(z)$ is the comoving distance, $H(z)$ is the Hubble parameter, and $W_X(z)$ is the geometric radial kernel for tracer $X$. We focus on scales $\ell\geq 300$, for which the Limber approximation is accurate. We define the two-halo galaxy kernel
\begin{equation}
    W_g^{\rm 2h}(z) \equiv b_g(z)\frac{dN_g}{dz},
    \label{eq:galaxy_kernel}
\end{equation}
where $dN_g/dz$ denotes the galaxy redshift distribution per steradian and $b_g(z)$ is the linear galaxy bias, which depends on the halo bias $b_h(M,z)$, weighted by the halo mass distribution of the sample:
\begin{equation}
    b_g(z) = \frac{1}{\overline{n}_g} \int dM \frac{dn}{dM} b_h(M,z) \langle N_g \rangle.
    \label{eq:galaxy_bias}
\end{equation}
We define a similar kernel for the NIR intensity field,
\begin{equation}
    W_I^{\rm 2h}(z, \lambda) \equiv b_I(z, \lambda)\frac{d(\lambda I_{\lambda})}{dz}.
    \label{eq:intensity_kernel}
\end{equation}
Here, $b_I(z, \lambda)$ is the \emph{intensity bias}, defined as the luminosity-weighted bias of all sources contributing to the intensity field at redshift $z$ and observed wavelength $\lambda$,
\begin{equation}
b_I(z,\lambda) = \frac{1}{\langle I\rangle} \int dM \, \frac{dn}{dM} \, b_h(M,z) \, L_I(M,z),
\label{eq:bI_def}
\end{equation}
while $d(\lambda I_{\lambda})/dz$ is the specific intensity redshift kernel ($dI/dz$ hereon). With access to intensity monopole information, one can constrain $b_I$ as done in \cite{ykc25_farir}, while conversely, one can place constraints on $dI/dz$ if assuming prior knowledge of $b_I$ (e.g., from simulations).

\subsection{Two-halo clustering}
\label{sec:2h_terms}
On large scales, where both the galaxy and intensity fields trace the same underlying matter power spectrum $P_{\rm m}(k,z)$, we assume $W_g \approx W_g^{\rm 2h}$ and $W_I\approx W_I^{\rm 2h}$ and approximate the two-halo power spectra as
\begin{align*}
C^{\rm gg, 2h}_\ell &= \int dz \, \frac{H(z)}{c\chi^2(z)} \, W_g^2(z) \, P_{\rm m}(k_\ell,z); \\
C^{\rm II, 2h}_\ell &= \int dz \, \frac{H(z)}{c\chi^2(z)} \, W_I^2(z) \, P_{\rm m}(k_\ell,z); \\
C^{\rm Ig, 2h}_\ell &= \int dz \, \frac{H(z)}{c\chi^2(z)} \, W_g(z) \, W_I(z) \, r_\ell^{\rm Ig}(z) \, P_{\rm m}(k_\ell,z),
\end{align*}
with $k_{\ell} \equiv (\ell + 1/2)/\chi(z)$. The parameter $r_{\ell}^{\rm Ig}(z)$ describes any intrinsic de-correlation between intensity and matter fluctuations. Because all three spectra share the same $P_{\rm m}$ factor, their amplitudes are related by the ratio between $W_I$ and $W_g$. In particular,
\begin{equation}
    \frac{C_\ell^{\rm Ig, 2h}}{C_\ell^{\rm gg, 2h}} 
    \approx \frac{W_I(z)}{W_g(z)} = 
    \frac{b_I \times dI/dz}{b_g \times dN/dz},
    \label{eq:ratio_estimator}
\end{equation}
which forms the basis for tomographic reconstruction of $W_I(z)$ from cross-correlations \citep{menard_clusred, ykc_tsz, ykc25_farir, ykc_26}. We note that this relation only holds exactly for narrow galaxy redshift slices (i.e., $\Delta z_g \ll 1$). In the limit where the linear bias is large, non-linear bias corrections become important \citep{fernandez_komatsu}.

\subsection{One-halo contributions and Poisson fluctuations}

On smaller scales, the one-halo term describes power from pairs of objects or intensity residing within the same dark matter halo. The intensity-galaxy cross-spectrum is proportional to

\begin{equation}
    C_\ell^{\rm Ig, 1h} \propto \int dM\, \biggl[\frac{dn}{dM}\, 
    \langle L_I(M,z) \rangle \, 
    \langle N_g \rangle
    \times\, u_I(k_\ell|M,z)\, u_g(k_\ell|M,z) \biggr],
    \label{eq:1h_cross}
\end{equation}
where $dn/dM$ is the halo mass function, $L_I(M,z)$ is the halo luminosity, $\langle N_g \rangle$ is the mean galaxy occupation, and $u_I(k|M)$, $u_g(k|M)$ are the normalized radial profiles of the intensity and galaxy distributions within halos, respectively. For simplicity, we assume both follow a Navarro-Frenk-White (NFW) profile \citep{nfw_paper}, i.e.,
\begin{equation}
    u_{\rm NFW}(k|M) = \frac{1}{M}4\pi \int dr r^2 \rho(r|M)j_0(kr),
\end{equation} 
where
\begin{equation}
\quad \rho(r|M) = \frac{\rho_0}{\frac{r}{R_s}\left(1 + \frac{r}{R_s}\right)^2}.
\end{equation}
For galaxy auto spectra, the strength of the one-halo term depends on the halo mass function and galaxy occupation distributions, while for $C_{\ell}^{\rm Ig}$ it depends on the total luminosity in halos associated with each tracer. 

In the limit of unresolved sources (i.e., $u(k|M,z) \to 1$), both the one-halo clustering and discreteness of the galaxy sample lead to a scale-independent Poisson level. The galaxy auto-spectrum shot noise is given by $C_{\ell}^{\rm gg, P} = \bar{n}_g^{-1}$, while for the cross-spectrum,
\begin{equation}
    C_{\ell}^{\rm Ig, P} = \left(\frac{dN_g}{d\Omega}\right)^{-1} \Delta z_g \, \frac{d(\lambda I_{\lambda})}{dz}\bigg|_{z_g},
    \label{eq:poisson_cross}
\end{equation}
i.e., it is proportional to the mean intensity of the tracer galaxies. While the one-halo term contributes a scale-independent component from correlated pairs residing within the same halo, the Poisson contribution in \eqref{eq:poisson_cross} arises from random, unclustered sampling of discrete sources.  

%% file: ciber.tex
\section{Cosmic infrared background experiment (CIBER)}
\label{sec:ciber}

\subsection{Fourth flight dataset}
\emph{CIBER}\footnote{\url{https://ciberrocket.github.io/}} is a rocket-borne instrument designed to characterize the NIR EBL through imaging and spectral measurements \citep{zemcov13, ciber_lrs}. We use imager data \citep{bock13} from the fourth and final (non-recovered) flight, launched at 3:05 UTC 2013 June 6 from Wallops Flight Facility in Virginia on a four-stage Black Brant XII rocket. The imager data are comprised of five science fields covering 20 deg$^2$ and were used for the fluctuation measurements in \cite{feder25b}. Details on the data reduction and processing can be found in that work and in \cite{feder25a}. 

For this study (and unless otherwise specified), we adopt the fiducial \emph{CIBER} science masks from \cite{feder25b}, in which 
sources brighter than $J_{\rm AB}<16.9$ and $H_{\rm AB}<17.3$ are masked for 1.1 $\mu$m and 1.8 $\mu$m, respectively. This masks the majority of stars while removing only the brightest galaxies, retaining the cross-correlation signal from fainter CIB sources. For our lowest redshift cross-correlations in \S \ref{sec:results},  $0.0<z<0.2$, we relax the masking threshold to $J_{\rm AB}<15.0$, $H_{\rm AB}<15.3$. We combine these masks with the coverage and bright star masks produced by the external tracer catalogs (described in \S \ref{sec:phot_samples}) when applicable. 


\subsection{Sensitivity forecasts}
\label{sec:sensitivity_forecast}
Let $C_{\ell}^{\rm II}$ and $C_{\ell}^{\rm gg}$ denote the intensity and galaxy auto power spectrum, respectively. Assuming a diagonal Gaussian covariance, the cross-power spectrum variance can be expressed as
\begin{equation}
    (\delta C_{\ell}^{\rm Ig})^2 = \frac{1}{n_{\ell}}\left[(C_{\ell}^{\rm Ig})^2 + (C_{\ell}^{\rm II} + N_{\ell}^{\rm II})(C_{\ell}^{\rm gg} +\overline{n}^{-1})\right],
    \label{eq:cross_ps_uncertainty}
\end{equation}
where $n_{\ell}=2\ell+1$ is the number of modes per bandpower, $N_{\ell}^{\rm II}$ is the noise bias from instrument read noise and photon noise, and $\overline{n}^{-1}$ denotes the galaxy shot noise level. We model the variance on the galaxy auto spectrum as a combination of sample variance and shot noise,
\begin{equation}
    (\delta C_{\ell}^{\rm gg})^2 = \frac{2}{n_{\ell}}(C_{\ell}^{g} + \overline{n}^{-1})^2.
    \label{eq:auto_ps_uncertainty}
\end{equation}

\begin{figure*}
    \centering
    \includegraphics[width=0.95\linewidth]{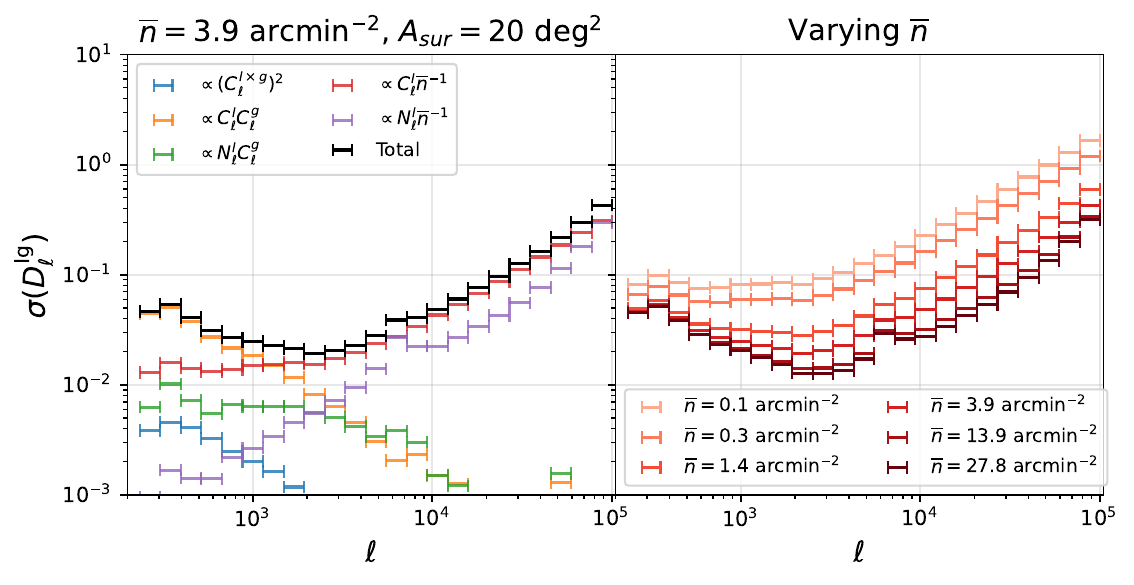}
    \caption{\emph{CIBER} $\times$ galaxy 1$\sigma$ cross-power spectrum sensitivity in bandpowers for 1.1 $\mu$m. We show $D_{\ell}^{\rm Ig} = \ell(\ell+1)C_{\ell}^{\rm Ig}/2\pi$ (with unit nW m$^{-2}$ sr$^{-1}$). In the left panel, we show the different components from Eq. \eqref{eq:cross_ps_uncertainty} for a fiducial tracer density matching that of the DESI-LS $z<1$ catalog with $\overline{n}=1.4\times 10^4$ deg$^{-2}$ (3.9 arcmin$^{-2}$). The right panel shows the total sensitivity for different tracer densities. We assume a typical masking fraction $f_{\rm mask}=0.3$ when computing the effective area. These forecasts highlight the necessity for dense photometric samples in order to achieve sample-variance-dominated measurements on scales $\ell<5000$.}
    \label{fig:clx_forecast}
\end{figure*}

In Figure \ref{fig:clx_forecast} we present nominal bandpower sensitivity estimates for 1.1 $\mu$m, assuming an average mask fraction of 30\% and with varying galaxy shot noise levels. These forecasts are informed by measurements of the \emph{CIBER} auto-power spectra, their noise biases, and DESI-LS galaxy auto- and cross-spectrum predictions (presented later in \S \ref{sec:mock_validation}). On large scales ($\ell<3000$), our measurements are dominated by sample variance of the \emph{CIBER} auto-spectrum, highlighted by the orange curve ($\propto C_{\ell}^{\rm II} C_{\ell}^{\rm gg}$). On smaller scales, our measurement sensitivity is limited by galaxy shot noise, with further degradation from \emph{CIBER} read noise near $\ell=6000$. In the right-hand panel of Fig. \ref{fig:clx_forecast} we show the dependence of our sensitivity for different tracer densities, which are chosen to bracket the tracer densities of catalogs used in this work. We note that these forecasts assume a flat noise power spectrum for the galaxy field, which is broken in the case of systematic variations that contribute to tracer non-uniformity.

%% file: phot_samples.tex
\section{Photometric galaxy samples}
\label{sec:phot_samples}

We use photometric galaxy samples observed at optical and infrared wavelengths, with properties that are summarized in Table \ref{tab:photometric_catalogs}. While spectroscopic catalogs provide superior redshift precision, our \emph{CIBER} forecasts suggest that such cross-correlations are strongly shot noise-limited for the scales of interest ($\ell > 300$), motivating the use of denser photometric tracers. In this section, we describe the properties of these samples and detail the processing applied before performing cross-correlations. 

\begin{table}[ht]
    \centering
    \begin{tabular}{c|c|c|c|c}
        Survey Dataset & Fiducial selection & Coverage & $\bar{n}_g$ [deg$^{-2}$] & $\bar{n}_g$ [arcmin$^{-2}$] \\
        \hline
        DESI Legacy Survey (DR8) & $z_{\rm AB}<22$, $z_{\rm phot}<1$ & Full & $1.4\times 10^4$ & 3.9\\
        DESI-LS CMGs (DR9) & see \cite{wen_cmg_24} & Full & $5.0\times 10^2$ & 0.14 \\
        HSC PDR3 (UltraDeep) & $i_{\rm AB}<25$, $z_{\rm phot}<1$ & 1/5 fields & $1.0\times 10^5$ & 27.8 \\
        
    \end{tabular}
    \caption{Photometric samples used in this work and their properties. All magnitudes are quoted in monochromatic (AB) units. The HSC-UDS catalog is available in the SWIRE ELAIS-N1 field but not the other four \emph{CIBER} fields.}
    \label{tab:photometric_catalogs}
\end{table}

\subsection{Galaxy catalogs}
\subsubsection{DESI Legacy Survey}
The DESI Legacy Survey (DESI-LS hereon) is comprised of optical $griz$ photometry obtained through a combination of DECaLS and MzLS+BASS imaging \citep{legacy_survey}, along with WISE photometry at 3.4 $\mu$m and 4.6 $\mu$m. We use the photo-z catalog of \cite{duncan22}, which is derived from the DESI-LS DR8 dataset and includes a $z<22$, extinction-corrected magnitude cut for the purpose of sample uniformity. The \emph{CIBER} fields fall in both the DECaLS and MzLS/BASS footprints. For samples binned by photometric redshift, we use \texttt{The Tomographer}\footnote{Link to code: \url{https://github.com/yuvoonng/tomographer}, other examples on \url{tomographer.org}.} \citep{tomographer_ykc} in order to obtain $b_g \times dN/dz$, the bias-weighted redshift distribution. While our cross-correlation measurements are limited to the \emph{CIBER} $2^{\circ} \times 2^{\circ}$ field of view, we compute $C_{\ell}^{\rm gg}$ from larger $6^{\circ} \times 6^{\circ}$ regions surrounding each \emph{CIBER} field in order to reduce sample variance in the galaxy auto spectrum results presented in \S \ref{sec:results} and \S \ref{sec:model_interp}.

In addition to the DR8 sample, we also use a sample of cluster member galaxy (CMG) candidates from the DESI-LS DR9 imaging \citep{wen_cmg_24}. The clusters are identified by searching for local overdensities in narrow redshift slices around massive galaxies ($\sim 50$ deg$^{-2}$), after which groups of galaxies with $N_{\rm gal}\geq 6$ within $r_{500}(z, M_{\star})$ and a richness $\lambda_{500}\geq 10$ are used to select clusters at high confidence. The cluster redshift distribution has a mean of $z=0.6$ and primarily covers $z<1$. 


\subsubsection{Hyper-Suprime-Cam}
\label{sec:hsc}
The Hyper-Suprime-Cam Subaru Strategic Program (HSC-SSP) is a wide-field imaging survey conducted with the 8.2 meter \emph{Subaru} telescope, covering 670 deg$^2$ of the sky in multiple optical and near-infrared bands \citep{hsc_ssp}. The \emph{CIBER} SWIRE field is within HSC's Deep/Ultra-Deep footprint in the ELAIS-N1 field, which covers $\sim 36$ deg$^2$ and reaches point source depths of $m_{\rm AB}\gtrsim 25-26$ in all five $grizy$ bands. We use $i$-band selected catalogs from the third HSC Public Data Release \citep{hsc_pdr3} for cross-correlation. The catalogs also contain photo-z estimates from a variety of independent methods \citep{tanaka_hsc_photz}, with typical uncertainties $\hat{\sigma_z} = 0.03-0.07$. We use the photo-z estimates from \texttt{DNNz}\footnote{\url{https://hsc-release.mtk.nao.ac.jp/doc/index.php/photometric-redshifts__pdr3/}}, a neural network-based code.

We supplement the \emph{CIBER} science mask with a bright source mask unique to the HSC imaging data. In particular, we mask regions with obvious diffraction spikes near the brightest stars in the FOV, since these regions tend to have higher numbers of spurious sources in the HSC catalog. This increases the masking fraction for SWIRE ELAIS-N1 from 39\% to 42\%.

\begin{figure}[ht]
    \centering
    \includegraphics[width=0.6\linewidth]{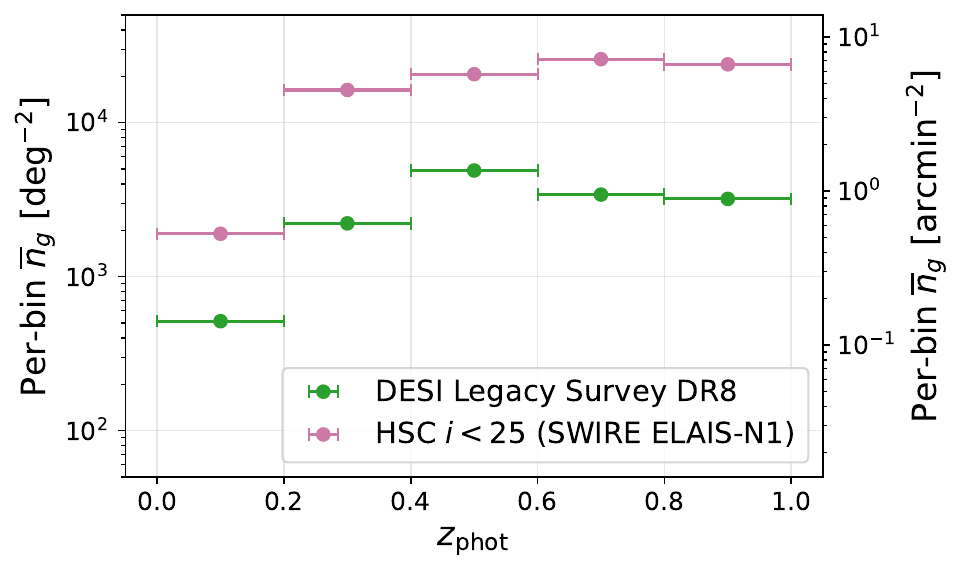}
    \caption{Photometric redshift distribution for DESI-LS (green) and HSC (pink) in redshift bins of $\Delta z=0.2$ spanning $0.0<z<1.0$. These correspond to the same samples defined in Table \ref{tab:photometric_catalogs}.}
    \label{fig:ls_hsc_ng}
\end{figure}

While the HSC catalog only overlaps with one of our science fields, it extends several magnitudes deeper than the DESI-LS catalogs. We highlight this in Fig. \ref{fig:ls_hsc_ng} by plotting the DESI-LS and HSC tracer densities in coarse redshift bins spanning $0.0<z_{\rm phot}<1.0$. Our cross-correlation sensitivity forecasts (in \S \ref{sec:sensitivity_forecast}) suggest that relatively dense tracers are needed in order to achieve sample variance-limited measurements with \textit{CIBER}. However, in order to utilize the full depth of the HSC catalog, it is necessary to quantify the impact of selection non-uniformity across the field. 

\subsection{Galaxy overdensity and randoms correction}
\label{sec:randoms}
To prepare the observed photometric galaxy catalogs for cross-correlation, we derive the galaxy overdensity field $\delta_g(\pmb{x})$,
\begin{equation}
    \delta_g(\pmb{x}) = \frac{n_g(\pmb{x}) - \alpha n_r(\pmb{x})}{\overline{n}_g}.
\end{equation}
where $n_g(\pmb{x})$ is the observed galaxy counts field, $n_r(\pmb{x})$ is the field estimated from random catalogs capturing survey geometry and other observational effects, and $\alpha$ rescales the total random counts to the data, i.e., $\alpha = N_g/N_r$, where $N_r \gg N_g$ to minimize noise on the correction. We describe the random catalog construction for both DESI-LS and HSC tracers in Appendix \ref{sec:randoms}. 

In Appendix \ref{sec:randcorr_app}, we inspect the impact of the randoms corrections for DESI-LS and HSC auto- and cross-power spectra. For DESI-LS, the randoms correction modifies the galaxy auto spectrum by $\lesssim 20\%$, while the cross-spectra have corrections that are slightly larger ($\pm 60\%$ for bandpowers with $\ell<2000$). The impact of corrections for the HSC sample is more significant, however they depend strongly on the choice of HSC magnitude cut. These results motivate our fiducial HSC selection of $i<25$, for which corrections at $\ell<2000$ are $<10\%$ on the auto and $\sim$tens of percent on the crosses.

%% file: mock_tests.tex
\section{CIB mocks and predictions}
\label{sec:cib_predictions}
\subsection{Semi-empirical IGL model}
\label{sec:igl_modl}
We construct mock intensity maps and power spectrum predictions using the \texttt{Ares} semi-empirical galaxy formation model \citep{mirocha26a}. While \texttt{Ares} is designed to accommodate a range of astrophysical scenarios and extensions, we adopt a standard IGL model configuration for comparison, for which a flexible parameterization of galaxy properties is calibrated as a function of redshift and halo mass against measurements of the stellar mass function \citep{moustakas13, tomczak14}, the star-forming main sequence \citep{salim07, whitaker14, tomczak16}, and UV luminosity functions \citep{arnouts05, page25, williams18}. The model generates self-consistent populations of central galaxies with dust-attenuated SEDs spanning UV to infrared wavelengths, with associated halo masses.

For the power spectrum recovery tests described in \S \ref{sec:mock_tests}, we construct IGL mocks by painting the \texttt{Ares} source populations onto lognormal density field realizations drawn from the linear matter power spectrum, assigning galaxy positions according to Poisson realizations of the galaxy density field. We do not implement galaxy biasing in the mocks used to test pipeline recovery, however we do incorporate galaxy bias in our IGL predictions (see \S \ref{sec:twohalo_pred}). The mocks cover a combined area of 100 deg$^2$. For both the mocks and predictions, we apply masking- and tracer-based magnitude selections consistent with each \emph{CIBER} $\times$ galaxy combination. 

\subsection{Power spectrum predictions}
\label{sec:ps_prediction}
\subsubsection{Two-halo clustering and galaxy bias}
\label{sec:twohalo_pred}
To predict $C_{\ell}^{\rm 2h}$, we combine estimates of $b_g(z)$ for our tracers with the projected linear matter power spectrum following \S\ref{sec:2h_terms}.
\begin{itemize}
    \item For DESI-LS, we derive $b_g(z)$ directly from cross-correlation measurements of the \cite{duncan22} DR8 sample using \texttt{The Tomographer}, which returns the bias-weighted distribution $b_g(z)\times dN/dz$. Using the direct $dN/dz$ estimates from the catalog, we derive $b_g^{\rm DESI-LS}(z)$ per bin. We then fit a smooth form to $b_g$ estimates in five $\Delta z = 0.2$ bins between $0 < z < 1$, obtaining
\begin{equation}
    b_g^{\rm DESI-LS}(z) = 0.94 - 0.21z + 1.22z^2.
\end{equation}
\item For HSC, we adopt $b_g(z) = 1.0 + 0.84z$, following measurements from \cite{jean_hsc} for an HSC $i < 24.5$ sample. Although our fiducial selection is 0.5 mag deeper, the difference in effective bias is small compared to our cross-spectrum uncertainties. 
\end{itemize}
We apply these as effective biases describing the combined central+satellite population for each sample. While we assume $b_I = 1$ in our fiducial predictions, we consider alternate models for $b_I(z)$ in \S \ref{sec:model_interp} when interpreting our halo model fits. 

\subsubsection{One-halo prescription}
\label{sec:onehalo_pred}
To incorporate one-halo contributions into our IGL predictions, we add satellites using a sub-halo model prescription tied to the \texttt{Ares} central galaxies and their halos. The one-halo cross-spectrum can be expressed as an integral over the product of galaxy kernel $J_{\rm g}(k|M)$ and intensity kernel $J_{\rm I}(k|M)$ over mass, weighted by $dn/dM$:
\begin{equation}
    P_{\rm 1h}^{\rm Ig}(k) = \frac{1}{\overline{n}_{\rm g}} \int dM \frac{dn}{dM}J_{\rm I}(k|M)J_{\rm g}(k|M).
\end{equation}
Here, $\overline{n}_g$ denotes the comoving 3D galaxy density. We decompose the galaxy and intensity kernels into centrals and satellites,
\begin{equation}
    J_{\rm g} = J_{\rm g}^{\rm cen} + J_{\rm g}^{\rm sat}; \quad J_{\rm I} = J_{\rm I}^{\rm cen} + J_{\rm I}^{\rm sat},
\end{equation}
where
\begin{equation}
    J_{\rm g, cen} = \langle N_{\rm cen} \rangle; \quad J_{\rm g, sat}= \langle N_{\rm sat} \rangle u_g(k|M)
\end{equation}
and
\begin{equation}
    J_{\rm I}^{\rm cen}(M) =\langle L_{\rm cen}(M) \rangle \langle N_{\rm cen}\rangle; \quad J_{\rm I}^{\rm sat}(M) = \langle L_{\rm sat}(M)\rangle \langle N_{\rm sat}(M) \rangle u_I(k|M).
    \label{eq:cen_sat_I}
\end{equation}
As the mass dependence of satellite occupancy and luminosity can vary significantly across HOD scenarios while producing similar $J_{\rm I}^{\rm sat}$, we define the effective quantity $f_{\rm sat}^{\rm L}(M)$, the ratio of integrated satellite luminosity to total (central + satellite) luminosity:
\begin{equation}
    f_{\rm sat}^{\rm L}(M) \equiv \frac{L_{\rm sat, tot}(M)}{L_{\rm cen}(M) + L_{\rm sat, tot}(M)},
    \label{eq:fsat_definition}
\end{equation}
where $\langle L_{\rm sat, tot}\rangle = \langle L_{\rm cen}\rangle \frac{f_{\rm sat}^{\rm L}}{1-f_{\rm sat}^{\rm L}}$. With this parameterization, the intensity kernel can be rewritten as
\begin{equation}
    J_{\rm I}(k|M)=\langle L_{\rm cen} \rangle \langle N_{\rm cen} \rangle \left[ 1 + \frac{f_{\rm sat}^{\rm L}}{1-f_{\rm sat}^{\rm L}} u_I(k|M) \right].
\end{equation}
We adopt a universal form for the sub-halo mass function from \citet{tinker_wetzel}, which we use to estimate the number of detected satellites $\langle N_{\rm sat}(M)\rangle$. We assume a $L_{\rm sat, tot}(M_h)$ dependence that follows \cite{purcell07}, assigning high fractional contribution from satellites in high-mass systems. For all predictions, we integrate over the halo mass range $10 \leq \log_{10} M_h \leq 15$ and project to $C_{\ell}^{\rm Ig}$ using \eqref{eq:limber}. We use the same one-halo formalism to generate predictions for $C_{\rm \ell, 1h}^{\rm gg}$ and $C_{\rm \ell, 1h}^{\rm II}$.

\begin{figure*}
\centering
\includegraphics[width=\linewidth]{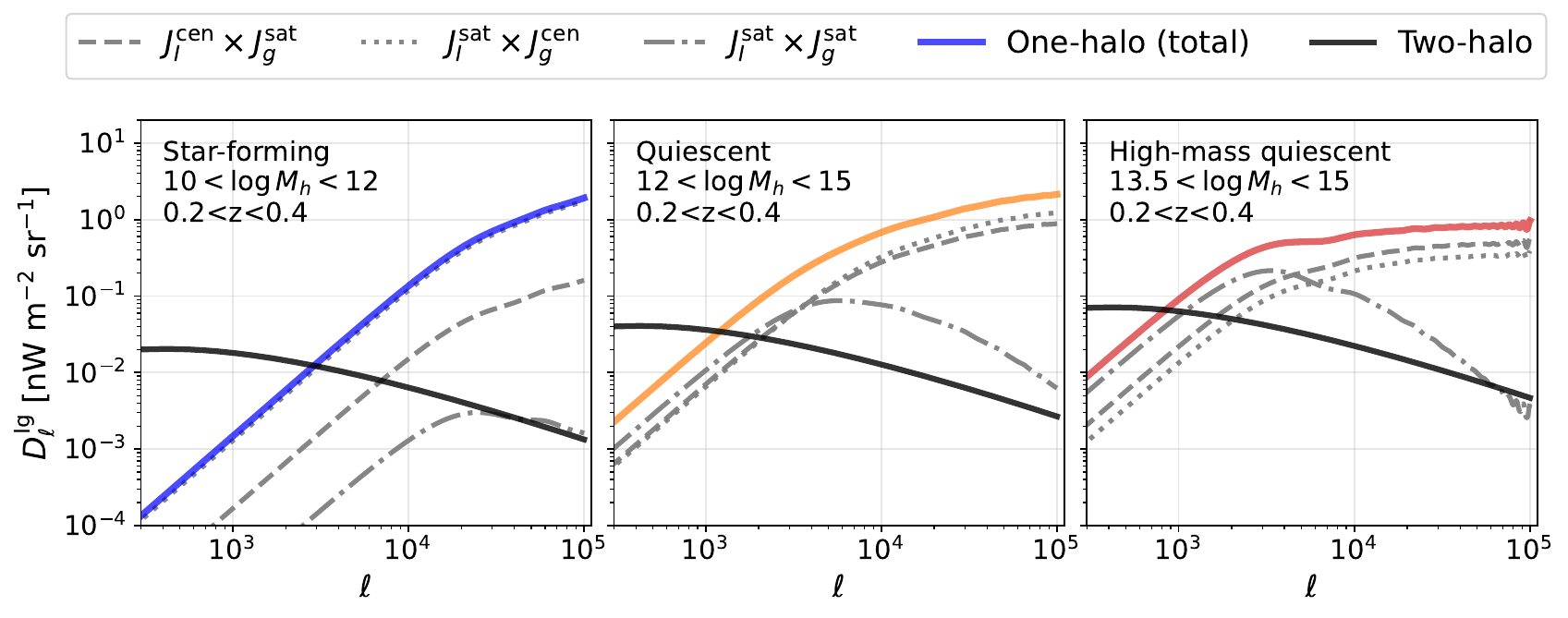}
\includegraphics[width=\linewidth]{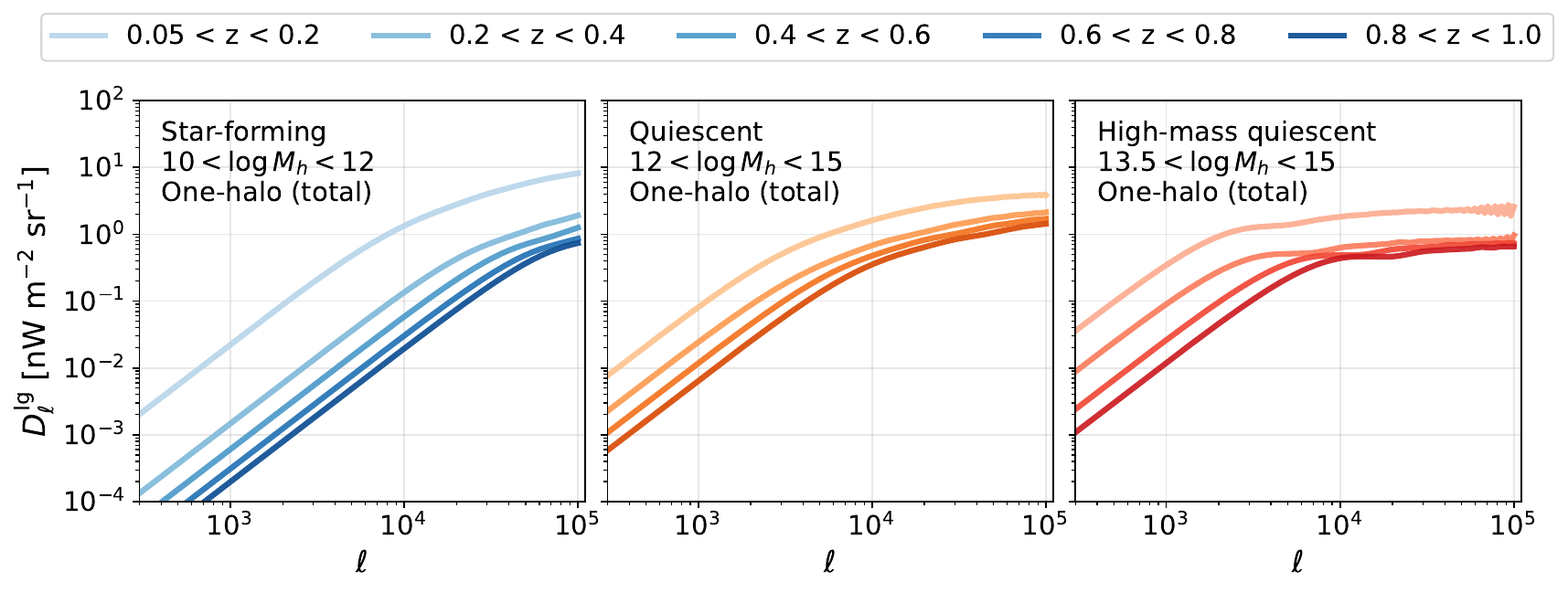}
\caption{\textit{CIBER} 1.1 $\mu$m one-halo predictions derived from the \texttt{Ares} model. In the top row, we show the three scale-dependent cross-terms of $J_I$ and $J_g$ in a single example redshift bin ($0.2<z<0.4$) for star-forming galaxies (blue, left), quiescent galaxies (orange, middle) and high-mass quiescent systems (red, right), alongside the corresponding two-halo predictions (black). In the bottom row, we show the combined one-halo predictions across our five redshift bins. At low redshift, our one-halo predictions are highly sensitive to the most massive halos, and so we integrate from $0.05<z<0.2$ from our simulations.}
\label{fig:pop_terms_censat}
\end{figure*}

In Figure \ref{fig:pop_terms_censat} we highlight the different central and satellite contributions comprising our one-halo predictions, for star-forming galaxies, quiescent galaxies, and high-mass quiescents (roughly matching the halo masses of the DESI-LS-based cluster catalog). For the star-forming population, the one-halo term is dominated by the $J_{\rm I}^{\rm sat} \times J_{\rm g}^{\rm cen}$ term on all angular scales. However, for higher-mass systems, both the $J_{\rm I}^{\rm cen} \times J_{\rm g}^{\rm sat}$ and $J_{\rm I}^{\rm sat}\times J_{\rm g}^{\rm sat}$ terms are more prominent, in particular for scales $\ell < 10^4$. In \S \ref{sec:model_interp}, we employ a star-forming/quiescent decomposition in our phenomenological one-halo model to capture potential deviations with respect to our fiducial predictions that are driven by a different mass dependence.

\subsubsection{Poisson level}
While we use the galaxy fluxes from \texttt{Ares} mocks to determine the ``centrals-only" Poisson level, we adopt satellite fractions informed by previous studies to estimate the corrected Poisson level of our tracer samples, which contain both centrals and satellites. 
\begin{itemize}
    \item For DESI-LS, the halo occupation distribution (HOD) fits to DESI-like luminous red galaxies suggest a satellite fraction $f_{\rm sat} \sim 10-15\%$ for a sample with photometric redshifts in the range $0.4 < z < 0.9$ \citep{zhou2021_desilike_lrgs}.  
    \item Similar HOD studies using the HSC-SSP shape catalog \citep{ishikawa_satellite} constrain $f_{\rm sat} \sim 15-20\%$ across the redshift range $0.3 < z < 1.0$, with higher/lower $M_{\star}$-based selections yielding lower/higher $f_{\rm sat}$.
\end{itemize}
In the absence of a more refined tracer selection and central+satellite model, we assume $f_{\rm sat}(z) \approx f_{\rm sat} = \lbrace 0.12, 0.17\rbrace$ for the DESI-LS and HSC samples, respectively. The impact on the cross-shot noise level depends on the relative emissivity of satellites and centrals, $\langle j_{\nu}^{\rm sat}\rangle / \langle j_{\nu}^{\rm cen}\rangle$, i.e., 
\begin{equation}
    C_{\ell}^{\rm Ig, P} = \frac{1}{\overline{n}_g}\left(\overline{n}_{\rm cen}\langle j_{\nu}^{\rm cen}\rangle + \overline{n}_{\rm sat}\langle j_{\nu}^{\rm sat}\rangle \right),
\end{equation}
where $\overline{n}_g = \overline{n}_{\rm cen} + \overline{n}_{\rm sat}$ and
\begin{equation}
    \langle j_{\nu}^{\rm cen} \rangle \equiv \frac{\Delta z^{-1}d(\lambda I_{\lambda})/dz|_{\rm cen}}{dN_g/d\Omega|_{\rm cen}}; \quad \langle j_{\nu}^{\rm sat} \rangle \equiv \frac{\Delta z^{-1}d(\lambda I_{\lambda})/dz|_{\rm sat}}{dN_g/d\Omega|_{\rm sat}}.
\end{equation}
For simplicity, we assume $\langle j_{\nu}^{\rm sat}\rangle / \langle j_{\nu}^{\rm cen}\rangle = 0.5$. In practice this is a small correction because satellites constitute a minority of each tracer sample.

\section{Power spectrum estimation}
\label{sec:mock_tests}

\subsection{Pseudo-power spectrum pipeline}

We use a similar pseudo-$C_{\ell}$ formalism to that in \cite{feder25a} for our analysis. The observed intensity-galaxy cross-power spectrum, $\hat{C}_{\ell}^{\rm Ig}$, can be written as 
\begin{equation}
    \hat{C}_{\ell}^{\rm Ig} = M_{\ell\ell^{\prime}}B_{\ell'}W^p_{\ell'} C_{\ell'}^{\rm Ig},
\end{equation}
in which $C_{\ell'}^{\rm Ig}$ is the true cross-spectrum, $M_{\ell\ell'}$ is the mode coupling matrix (from masking and map filtering), $B_{\ell}$ is the \emph{CIBER} beam transfer function and $W_{\ell}^p$ is the pixel window function. 

For the galaxy auto power spectrum, we first compute the galaxy overdensity field following the procedure described in \S \ref{sec:randoms}. After computing the pseudo-power spectrum of the field, we correct for mask and filtering effects using the same mode-mixing correction as for the crosses.


\subsubsection{Noise covariance estimation}
While instrumental noise and flat field errors in the \textit{CIBER} maps do not lead to biases in galaxy cross-correlations, their effects on the cross-spectrum covariance must be modeled. We follow a Monte Carlo approach, drawing 500 \textit{CIBER} noise realizations and computing the dispersion of cross-spectra against the observed galaxy density maps. The realizations include read noise and photon noise, along with contributions from our noisy flat field stacking estimator detailed in \cite{feder25a}.

\subsubsection{Mode-mixing correction}

The mode-mixing matrix $M_{\ell\ell^{\prime}}$ accounts for the combined impact of the instrument and astronomical mask, along with filtering applied to the maps before computing auto-/cross-spectra. Unlike the intensity auto-power spectrum measurements in \cite{feder25b}, the cross-spectra require less aggressive image filtering to obtain unbiased measurements and are immune to the multiplicative flat field bias described in \cite{feder25a}. As such, we apply gradient filtering to the galaxy overdensity and \emph{CIBER} intensity maps. To estimate $M_{\ell\ell^{\prime}}$, we generate 500 Gaussian realizations per bandpower and calculate the response in the pseudo-spectrum after applying the same filtering and masking as the data. 


\subsubsection{Beam and pixel transfer function}

Our beam correction $B_{\ell}$ is derived using the best-fit point spread function (PSF) fits in \cite{chengihl} and evaluated as the average power suppression for a point-like source, averaged over sub-pixel position \citep{feder25b}.

We correct for additional power suppression from the pixel window, which averages over sky fluctuations smaller than the pixel size, captured by $W_{\ell}^{p}$. In particular, we simulate 100 realizations of uncorrelated point sources convolved with the \textit{CIBER} beam and mapped onto detector pixels, compute the cross-power spectrum and take the ratio with respect to the analytic Poisson level estimate. We find that $W_{\ell}^{p}\gtrsim 0.9$ for multipoles $\ell<10^5$. 

\subsection{Mock power spectrum recovery}

In Appendix \ref{sec:mock_validation}, we describe the mocks used to validate our pseudo-$C_{\ell}$ pipeline on an ensemble of 500 \emph{CIBER} mocks. These follow those used in \cite{feder25a}. From these tests, we demonstrate unbiased recovery of both the galaxy auto-spectrum $C_{\ell}^{\rm gg}$ and the \emph{CIBER} $\times$ galaxy cross-spectrum $C_{\ell}^{\rm Ig}$ across the full multipole range of interest. We find that residual source alignment errors (from external catalog position errors and \emph{CIBER} astrometric registration) have a measurable impact on the recovered cross-correlations, suppressing $C_{\ell}^{\rm Ig}$ on scales $\ell > 20000$, which motivates a determination of the alignment error using cross-correlation with stars in App. \ref{sec:ciber_gaia}.

%% file: cross_cl_results.tex
\section{Results}
\label{sec:results}

\subsection{Cross-power spectrum measurements}
\begin{figure*}
    \centering
    \includegraphics[width=\linewidth]{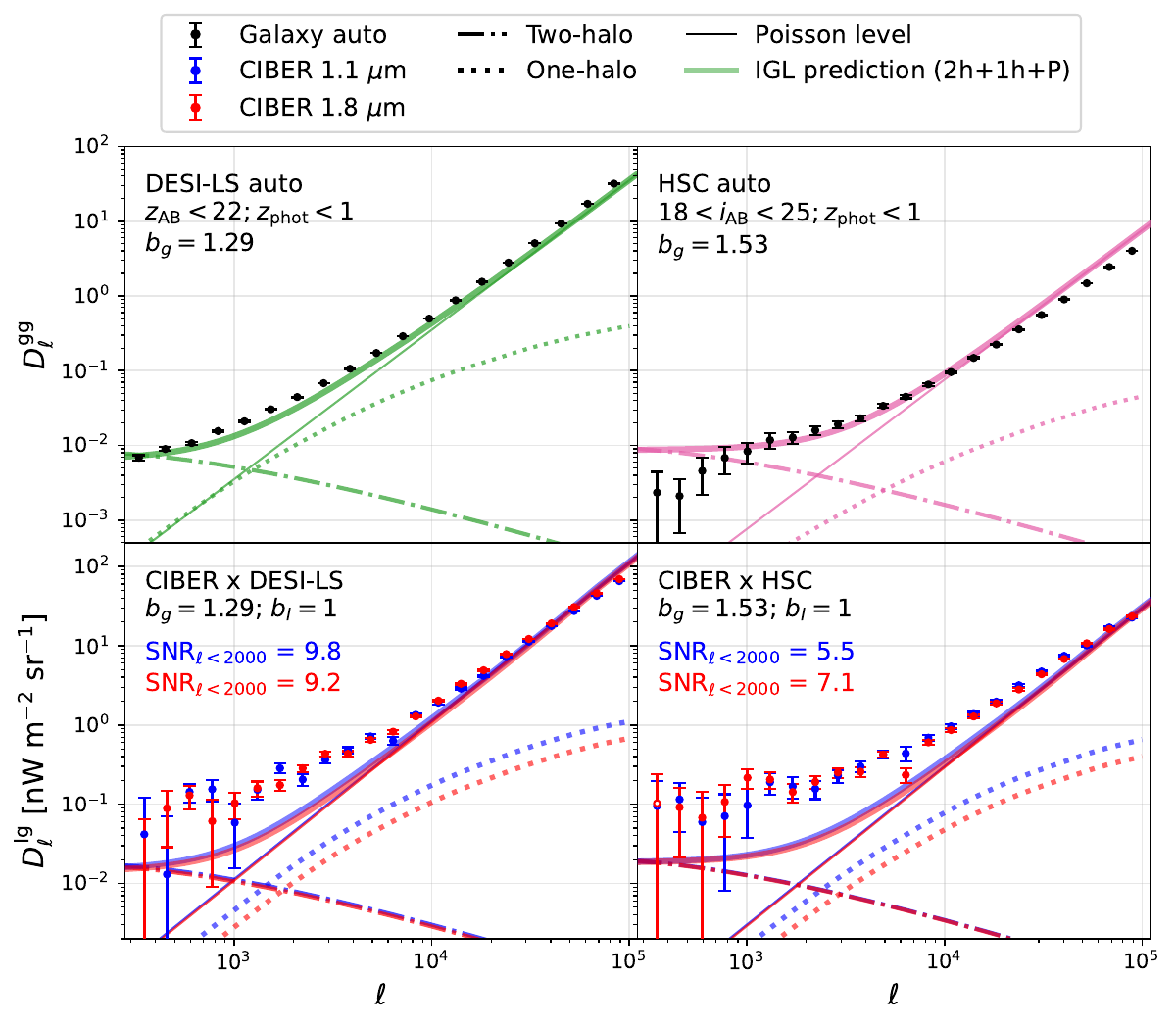}
    \caption{Auto- and cross-power spectra between \emph{CIBER} 1.1/1.8 $\mu$m maps (blue/red) and DESI-LS and HSC $z<1$ galaxies, compared against our standard IGL model. We define $D_{\ell}^{\alpha \beta} = \ell(\ell+1)C_{\ell}^{\alpha \beta}/2\pi$. For this and subsequent figures in \S \ref{sec:results} and \S \ref{sec:model_interp}, open circles indicate negatively valued bandpowers. The DESI-LS galaxy auto spectrum measurements come from larger $6^{\circ}\times6^{\circ}$ fields with coarser bandpowers. We compare our measurements with the IGL predictions described in \S \ref{sec:cib_predictions} (thick solid curves), which are comprised of two-halo (dash-dot), one-halo (dotted) and Poisson (thin solid line) contributions.}
    \label{fig:omnibus_cross}
\end{figure*}

In Figure \ref{fig:omnibus_cross} we present our primary, field-averaged auto- and cross-spectrum measurements between \emph{CIBER} and the DESI-LS and HSC samples. We show bandpower measurements spanning multipoles $304 < \ell < 10^5$, matching the range used in \cite{feder25b}. For each band+tracer configuration, we generate baseline IGL model predictions with two-halo, one-halo and Poisson components following \S \ref{sec:cib_modl}, with consistent magnitude- and redshift-based selections. We compare these predictions with parametric halo model fits to the auto- and cross-power spectra in \S \ref{sec:model_interp}.

These measurements represent the first detections of large-angle (several arcminute to sub-degree) NIR fluctuations explicitly correlated with galaxies. For $304<\ell<2000$, we detect cross-power at $9.8\sigma/9.2\sigma$ significance from \textit{CIBER} $\times$ DESI-LS and $5.5\sigma/7.1\sigma$ with \textit{CIBER} $\times$ HSC. In Appendix \ref{sec:field_consistency} we detail tests of the internal consistency of our DESI-LS cross-spectrum measurements across the five \textit{CIBER} fields, using the individual field measurements and their covariance. We find that the per-field deviations of the galaxy auto- and cross-spectra are consistent with statistical uncertainties, however growing toward the smallest angular scales where the statistical precision is extremely high. 

We assess the internal consistency of the galaxy auto- and cross-power spectra across the five \textit{CIBER} fields for the DESI-LS sample. We defer a full description of this analysis and its results to Appendix \ref{sec:field_consistency}, where we compare per-field bandpower measurements against field-averaged spectra and compute probability-to-exceed (PTE) values under the assumption of Gaussian-distributed uncertainties. We find that the galaxy auto-spectra are consistent across fields, with uniformly distributed PTE values.

On small angular scales ($\ell > 10,\!000$, $\theta \lesssim 1^{\prime}$), the auto- and cross-spectra agree closely with our IGL predictions, reflecting a model that is well-calibrated in the Poisson-dominated regime for both tracers. We find that the small-scale cross-spectra have a sub-Poissonian slope relative to the expectation for Poisson fluctuations, $D_{\ell}\propto \ell^2$. We measure $D_{\ell}^{\rm Ig} \propto \ell^{1.6}$ for \emph{CIBER} $\times$ DESI-LS and $\propto \ell^{1.5}$ for \emph{CIBER} $\times$ HSC. The corresponding cross-spectrum predictions follow $D_{\ell}\propto \ell^{1.8}$ over the same range, due to one-halo contributions. These findings are broadly consistent with fluctuation measurements from \emph{HST} deep imaging \citep{thompson07}. 

Towards larger angular scales ($\ell<10,\!000$, $\theta \gtrsim 1^{\prime}$), our measured cross-spectra exhibit increasing deviations from our IGL predictions. This is most apparent near $\ell=1000$, where the measurements are roughly a factor of three higher than the corresponding models in both \emph{CIBER} bands. For $\ell < 2000$, the measurements exceed our predictions at $7.4\sigma/6.2\sigma$ significance for \emph{CIBER} $\times$ DESI-LS 1.1/1.8 $\mu$m and $4.3\sigma/5.3\sigma$ for \emph{CIBER} $\times$ HSC. Extending to $\ell_{\rm max}=4000$, the disagreement rises to $10.3\sigma/13.2\sigma$ and $7.2\sigma/9.4\sigma$ significance for the same combinations. Over the same range of angular scales, our halo model reproduces $C_{\ell}^{\rm gg}$ with much higher accuracy than $C_{\ell}^{\rm Ig}$, suggesting that the discrepancy is not driven by inaccurate modeling of the galaxy density field.  

We confirm that the measured large-angle cross-power is unique to our galaxy samples, which we demonstrate in App.~\ref{sec:ciber_gaia} by cross-correlating the \emph{CIBER} maps against \emph{Gaia} DR3 stars \citep{gaia_dr3}. While previous work has found that the distribution of Milky Way stars is spatially uniform up to degree scales \citep{zemcov14}, we confirm this for the \emph{Gaia} sample, recovering auto- and cross-power spectra that are dominated by Poisson fluctuations. As a result, we interpret the large-scale discrepancies as originating from the CIB with minimal contamination from local foregrounds.

\subsection{Redshift dependence of DESI-LS cross-spectra}
\label{sec:zdep_clx}
\begin{figure*}
    \centering 
\includegraphics[width=\linewidth]{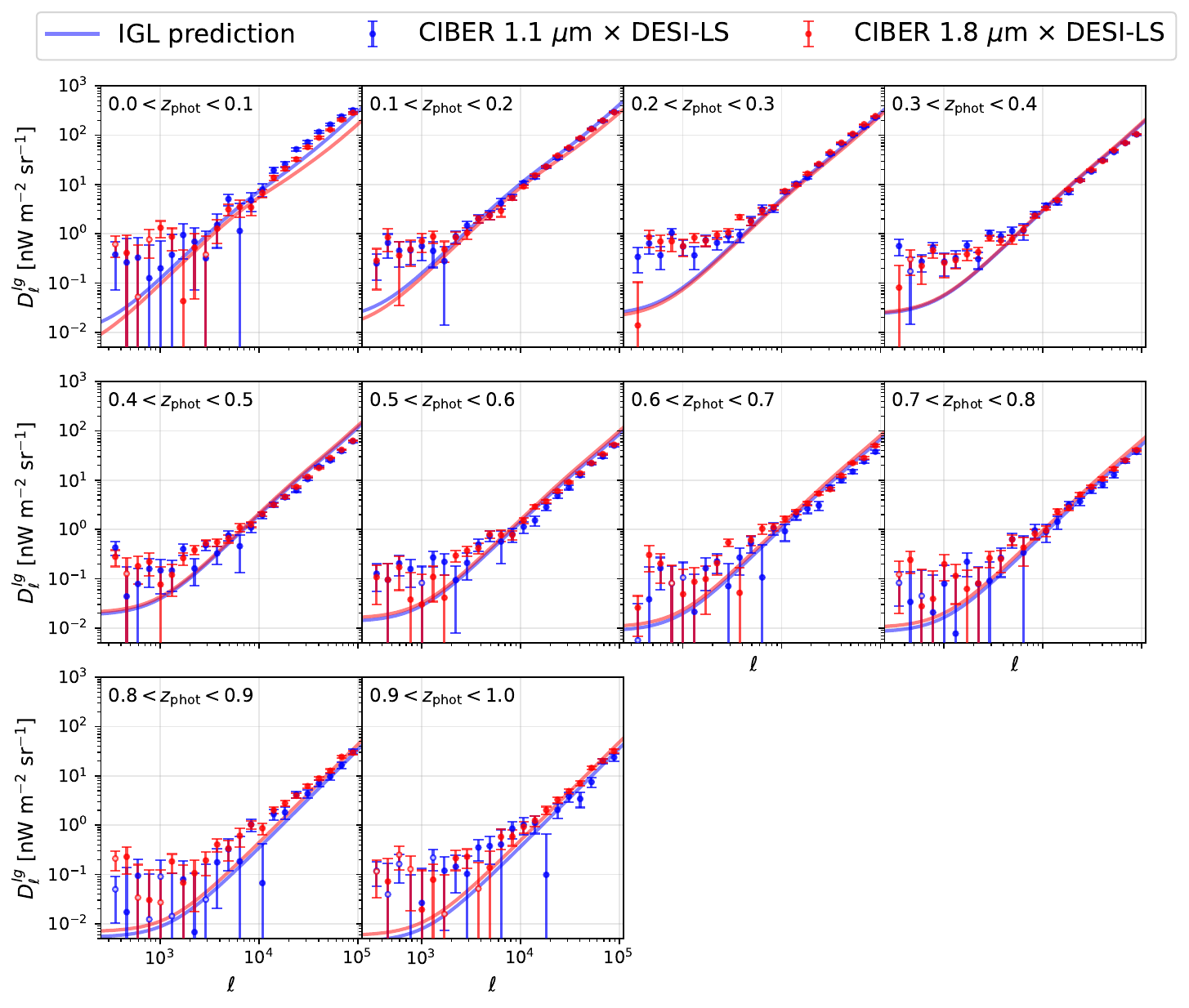}
    \caption{\emph{CIBER} $\times$ DESI-LS cross-power spectra for 1.1 $\mu$m (blue) and 1.8 $\mu$m (red), in $\Delta z=0.1$ redshift bins spanning $0.0<z<1.0$. The curves in each panel show our baseline IGL model predictions, which are described in \S \ref{sec:cib_predictions}.}
    \label{fig:ciber_ls_byz}
\end{figure*}

In this subsection, we split the DESI-LS sample into relatively fine tomographic redshift bins ($\Delta z=0.1$) and compute cross-correlations with \emph{CIBER}. We show the resulting cross-spectra in Fig. \ref{fig:ciber_ls_byz}. We detect Poisson fluctuations for all combinations and, like the $z<1$ results, find close agreement with our baseline IGL predictions on small angular scales. One exception is in the lowest redshift bin ($0.0<z<0.1$), where our IGL predictions are highly sensitive to the bright-end galaxy distribution and the upper tail of the halo mass function. Furthermore, the cross-correlations with our lowest-redshift samples probe scales comparable to the half-light radii of resolved galaxies, which may also impact the small-scale measurements.  

On larger angular scales, the strongest cross-correlation signals arises at low redshift $(0.1<z<0.5)$. In these tomographic bins, we make firm detections of cross-power on several arcminute to degree scales, exceeding our baseline IGL predictions by factors as large as $5-10$ in our $z\in[0.2,0.3)$ and $z\in[0.3,0.4)$ bins. Our higher-redshift crosses are limited by statistical uncertainties, and so we can neither confirm nor exclude the possibility of similar deviations beyond $z=0.6$. 

\begin{figure*}
    \centering 
\includegraphics[width=0.9\linewidth]{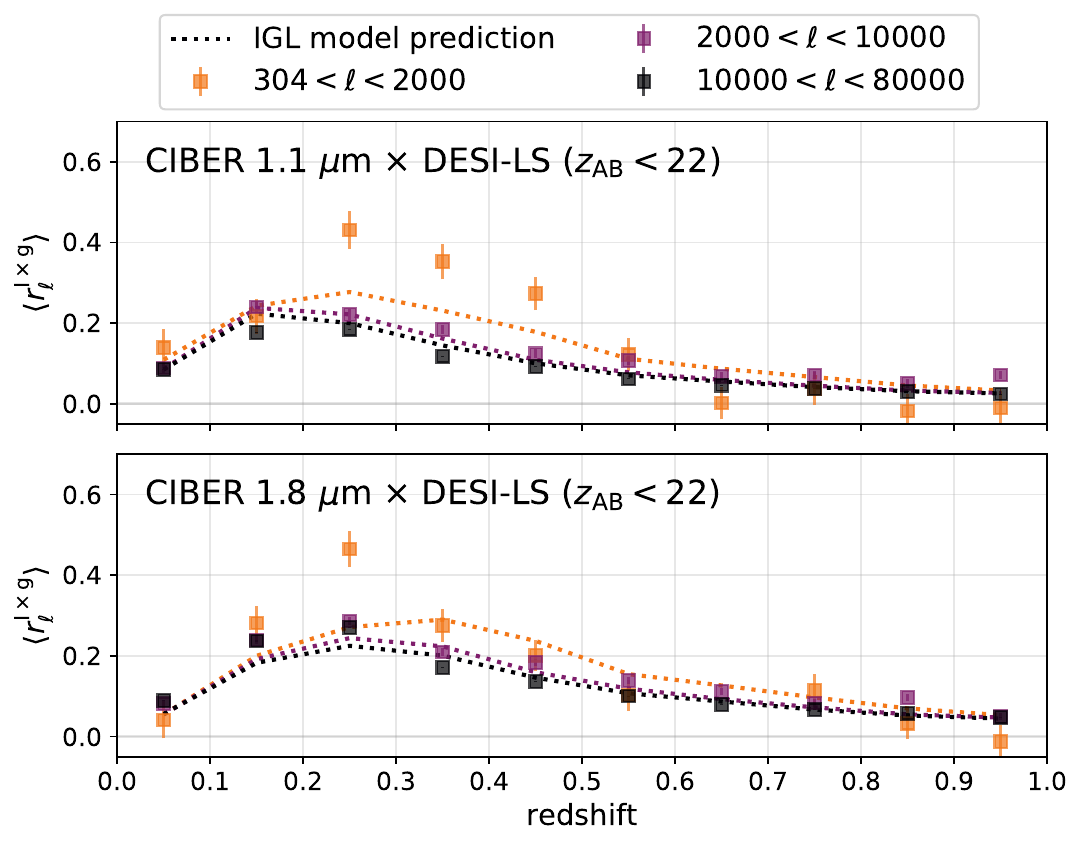}
    \caption{Redshift dependence of \emph{CIBER} $\times$ DESI-LS cross-correlation coefficient $r_{\ell}^{\rm Ig}\equiv C_{\ell}^{\rm Ig}/\sqrt{C_{\ell}^{\rm II}C_{\ell}^{\rm gg}}$, averaged over three broad bandpowers, alongside our fiducial IGL predictions (dotted curves). Note that $\hat{r}_{\ell}(0.6<z<0.7)$ is negative for bandpower $\ell \in [304, 2,\!000)$ due to noise fluctuations.}
    \label{fig:ciber_ls_rl_byz}
\end{figure*}

Using the corresponding galaxy and \emph{CIBER} auto-spectra, we compute the galaxy $\times$ intensity cross-correlation coefficient, $r_{\ell}^{\rm Ig} \equiv C_{\ell}^{\rm Ig}/\sqrt{C_{\ell}^{\rm II}C_{\ell}^{\rm gg}}$, in three broad bandpowers ($304 <\ell<2,\!000$, $2,\!000<\ell<10,\!000$ and $10,\!000<\ell<80,\!000$) and compare with our IGL model in Fig. \ref{fig:ciber_ls_rl_byz}. All cases show a similar trend, in which $r_{\ell}^{\rm Ig}$ peaks at $z\sim 0.2-0.4$ and then declines toward $z=1$. We find that $r_{\ell}^{\rm Ig}$ is highest at low-$\ell$ ($\ell<2000$) and exceeds $r_{\ell, \rm pred}^{\rm Ig}$ in several redshift bins. In redshift bin $z\in[0.2,0.3)$, $r_{\ell}^{\rm Ig}$ peaks in both bands at $0.40-0.45$, compared to predicted $r_{\ell}^{\rm Ig} \sim 0.25-0.30$. This difference is robust to photo-z errors, which we confirm by recomputing $r_{\ell}^{\rm Ig}$ with broader $\Delta z = 0.2$ bins. We find that the 1.1 $\mu$m low-$\ell$ measurements exceed the predicted $r_{\ell}^{Ig}$ primarily between $0.2<z<0.5$, while for 1.8 $\mu$m the deviations arise at $0.1<z<0.3$. As $r_{\ell}^{\rm Ig}$ is defined relative to the total intensity field, the higher correlation coefficients may reflect stronger relative contributions from those redshifts to the total EBL intensity.   

Since the galaxy overdensity field $\delta_g$ may be a less effective tracer of intensity on scales dominated by one-halo and Poisson fluctuations, we use a flux-weighted, coherence-based estimator in \S \ref{sec:recon_auto} to reconstruct the intensity auto-power associated with low-redshift galaxy clustering.

\subsection{Contribution from \texorpdfstring{$z<1$}{} clusters}
\label{sec:cmg_cross}
\begin{figure}
    \centering 
\includegraphics[width=\linewidth]{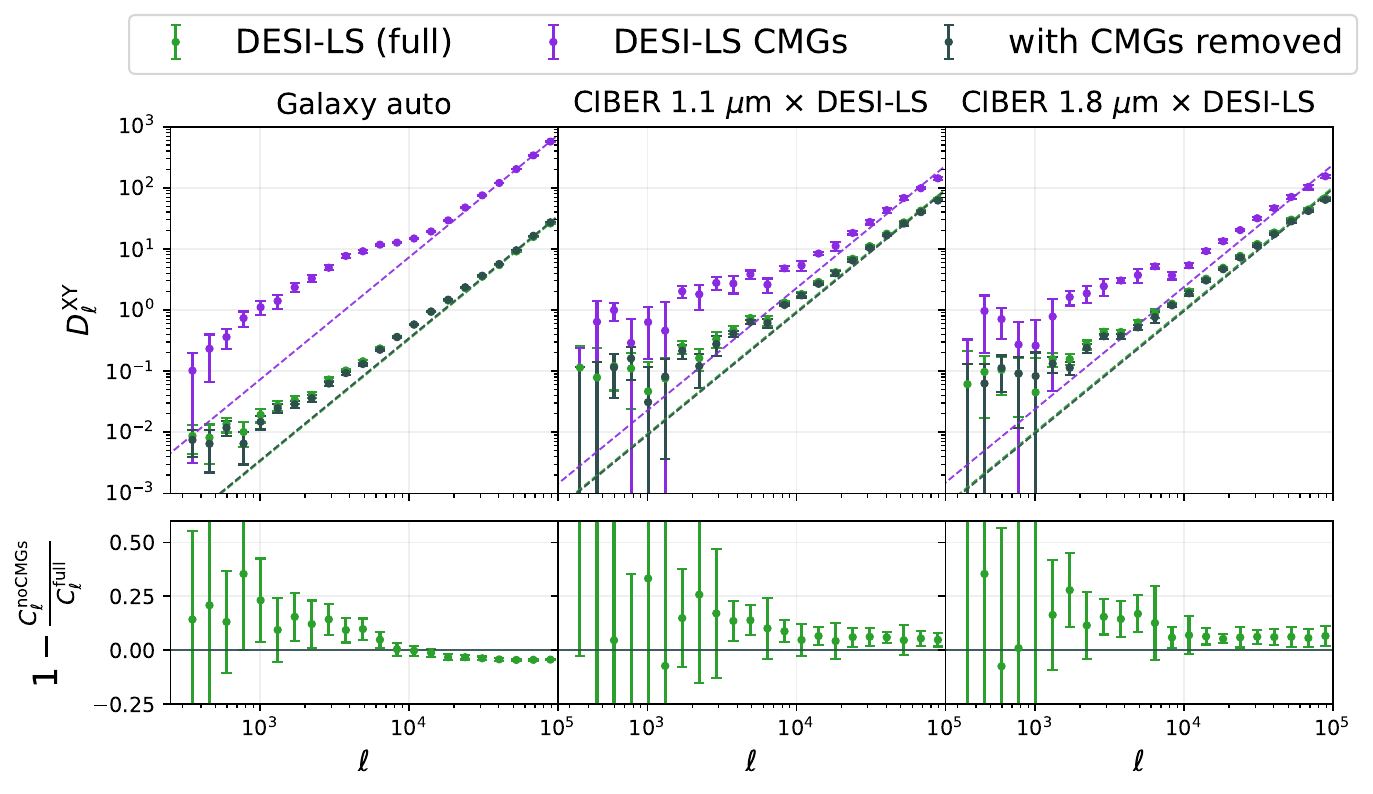}
    \caption{\emph{CIBER} cross-correlation with cluster member galaxies (CMGs) and impact on DESI-LS auto- (left) and cross-power spectra (middle, right). The cluster sample spans a redshift range that peaks at $z=0.6$ and resides largely below $z=1$. The CMG-only measurements are shown in purple, while those for the full DESI-LS sample and with CMGs  are shown in black and green, respectively. Dashed lines indicate the best-fit Poisson level for each power spectrum. As discussed in the text, the CMG-based selection leads to a stronger satellite $\times$ satellite one-halo contribution than that of our primary \emph{CIBER} $\times$ galaxy cross-spectra (c.f. Fig. \ref{fig:pop_terms_censat}). In the bottom row, we plot the fractional contribution of CMGs to the \emph{CIBER} $\times$ DESI-LS cross-spectra.}
    \label{fig:ciber_x_cmgs}
\end{figure}

It is well known that galaxy clusters are the largest gravitationally bound structures in the Universe, comprising the upper tail of the halo mass function. Given the high linear bias of cluster populations, it is natural to ask whether cluster member galaxies (CMGs) and associated intra-cluster light (ICL) contribute significantly to our cross-correlations. In Figure \ref{fig:ciber_x_cmgs} we present cross-spectra between \emph{CIBER} and the CMG sample described in \S \ref{sec:phot_samples}, along with the CMG auto spectrum. Both the auto- and cross-spectra show clear deviations from Poisson fluctuations for $\ell<10,\!000$. Because we select on cluster member galaxies rather than clusters, the CMG auto- and cross-spectra should be dominated by satellite $\times$ satellite one-halo contributions rather than the central $\times$ satellite terms that dominate our primary cross-spectra. Indeed, while we do not attempt to model the CMG-based measurements, they are qualitatively similar to the ``high-mass quiescent" one-halo predictions from Fig. \ref{fig:pop_terms_censat}. Due to this extended one-halo profile, separating the two-halo contribution in intensity is non-trivial from the scales probed in this analysis. 

Given that the CMG sample is a subset of the full DESI-LS $z<22$ sample\footnote{with the caveat that the CMG sample is derived from DESI-LS DR9 products while our primary sample comes from DR8.}, we then remove CMGs from the primary DESI-LS sample and mask them in the \textit{CIBER} maps in order to assess their contribution to clustering from the full sample. We estimate the fractional cluster contribution through the quantity $1-(C_{\ell}^{\rm w/o CMGs}/C_{\ell}^{\rm full})$. For multipoles $1000<\ell<10000$, our results indicate CMGs and associated structure are $15-20\%$ of the total \emph{CIBER} $\times$ DESI-LS cross-power, with no significant differences between 1.1 $\mu$m and 1.8 $\mu$m. 

%% file: modeling.tex
\section{Modeling and Interpretation}
\label{sec:model_interp}
\subsection{Overview}
Thus far, we have examined the cross-power spectra between \emph{CIBER} and DESI-LS and HSC, finding significant discrepancies with respect to the ``standard IGL" model detailed in \S \ref{sec:cib_predictions}. 
A key question arises: are the large-angle cross-spectrum discrepancies caused by a higher-than-expected intensity redshift kernel (i.e., $W_I(z) = b_I \times dI/dz$ from linear clustering), an incomplete EBL description on one-halo scales, or both? To probe this question, we pursue more flexible halo model fits to the data that we can compare against our fiducial predictions. We use redshift bins with $\Delta z = 0.2$; this binning is sufficient to capture the redshift evolution of the correlated EBL signal while remaining robust to redshift errors associated with the tomographic samples\footnote{For symmetric errors and slowly varying $dN/dz$, the impact of redshift errors on cross-spectra is minimal; however, the galaxy autos inherit a negative bias due to de-correlation of the projected clustering from outliers.}.

\begin{figure*}
    \centering 
    \includegraphics[width=\linewidth]{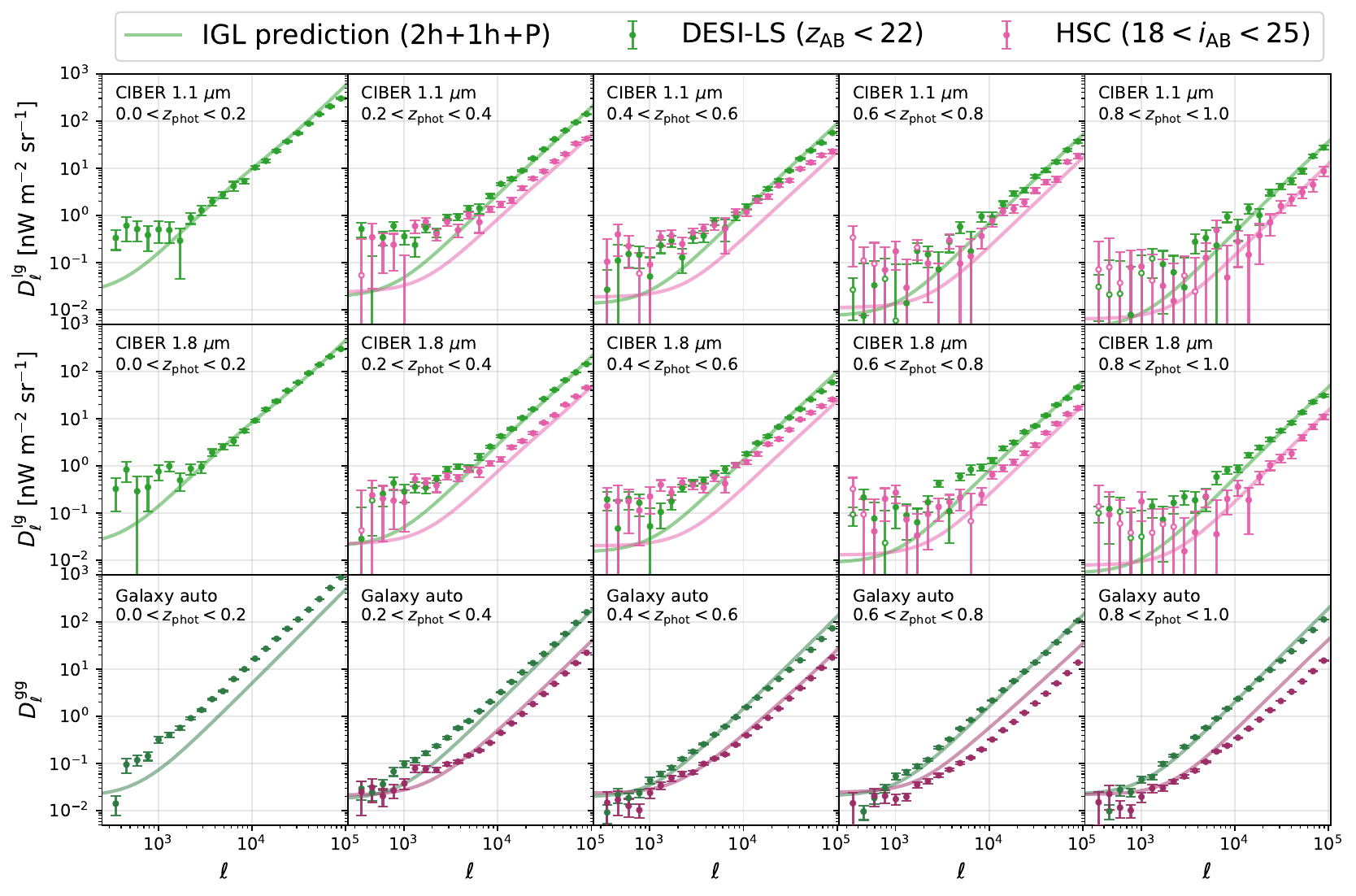}
    \caption{\emph{CIBER} $\times$ DESI-LS (green) and HSC (pink) cross-power spectra, in five $\Delta z= 0.2$ bins spanning $0<z<1$. The top and middle rows show the cross-spectra for \emph{CIBER} 1.1 $\mu$m and 1.8 $\mu$m maps, respectively, while in the third row we show the corresponding galaxy auto spectra. The solid curves show the standard IGL predictions described in \S \ref{sec:cib_predictions}. Open circles indicate negatively-valued bandpowers. We exclude the HSC $0.0<z<0.2$ cross-spectrum measurements from our halo model analysis in \S \ref{sec:model_interp} due to bright-end catalog selection effects.}
    \label{fig:ciber_ls_hsc_byz}
\end{figure*}

We show the cross-power spectra used for modeling in Fig. \ref{fig:ciber_ls_hsc_byz}, along with the corresponding galaxy auto spectra. These measurements reinforce the findings from Fig. \ref{fig:omnibus_cross} -- the same model that reproduces the low-$\ell$ galaxy clustering significantly underestimates the galaxy $\times$ intensity cross-power on the same scales\footnote{One exception to this is the DESI-LS $z\in [0.2, 0.4)$ galaxy auto, which exceeds our predictions. However, for the same redshift bin, the HSC galaxy auto results are in closer agreement while the deviation in the cross persists.}. As the HSC catalog is limited to $i_{\rm AB}>18$ due to saturation issues, brighter galaxies that dominate the $0.0<z<0.2$ signal are omitted, leading to incompleteness in the HSC sample. As a result, we omit the HSC $0.0<z<0.2$ cross-spectrum measurements from our primary results.

\subsection{Parametric fits and comparison to mock predictions}
\label{sec:parametric_fits}
\subsubsection{Model and fitting procedure}
\label{sec:fitting_procedure}
We use a multi-component parametric model for our cross-spectrum fits. For a given redshift bin with center $z_g$, 
\begin{equation}
    C_{\ell}^{\rm Ig}(z_g) = \left[A_{\rm 2h}^{\rm Ig}\mathcal{C}_{\ell}^{\rm 2h}(z_g) + A_{\rm 1h}^{\rm Ig}\left(f_{\rm pop}\mathcal{C}_{\ell}^{\rm 1h, Q} + (1-f_{\rm pop})\mathcal{C}_{\ell}^{\rm 1h, SF}\right) + A_{\rm Poisson}\right] \times \mathcal{P}_{\ell}.
    \label{eq:fit_model}
\end{equation}
Here, $\mathcal{C}_\ell^{\rm 2h}(z_g)$ is our two-halo template, which we compute using the linear galaxy power spectrum following Eq. \eqref{eq:limber}, while $\mathcal{C}_{\ell}^{\rm 1h, SF}$, $\mathcal{C}_{\ell}^{\rm 1h, Q}$ denote one-halo templates corresponding to halos with star-forming and quiescent centrals, respectively (see \S \ref{sec:onehalo_pred}). Using a linear weighting of templates (controlled by parameter $f_{\rm pop}$) for $C_{\ell}^{\rm Ig}$ provides a better fit to the data than a fixed template, and can be interpreted as a proxy for more general scale-dependence in the one-halo signal, as high-/low-mass halos have differing multipole dependence. We rescale the one- and two-halo templates using amplitude parameters $A_{\rm 2h}^{\rm Ig}$ and $A_{\rm 1h}^{\rm Ig}$, and include parameter $A_{\rm Poisson}$ to model the Poisson level. We then account for source alignment errors with $\mathcal{P}_{\ell}$ using a Gaussian damping term:
\begin{equation}
    \mathcal{P}_{\ell} = \exp(-(\ell/\ell_0)^2/(2\sigma_{\rm damp}^2)).
\end{equation}
We determine $\sigma_{\rm damp}=2.3''$ and $2.1''$ for 1.1 $\mu$m and 1.8 $\mu$m, respectively, by fitting the Poisson-dominated \emph{CIBER} $\times$ \emph{Gaia} star crosses (App. \ref{sec:gaia_gal_cross}).

For each cross-spectrum, we perform MCMC sampling with the affine-invariant \texttt{emcee} sampler \citep{2013PASP..125..306F} to obtain full posteriors over $\lbrace A_{\rm 2h}, A_{\rm 1h}, f_{\rm pop}, A_{\rm Poisson}\rbrace$. We assume uniform priors over parameters $\lbrace A_{\rm 2h}, A_{\rm 1h}, A_{\rm Poisson}\rbrace$ in the range $A_{\rm 2h} \in [0, 10]$, $A_{\rm 1h}\in[0, 100]$, $A_{\rm Poisson} \in [0, 10^{-5}]$, respectively. For each configuration, we run thirty-two chains with $N_{\rm step}=4{,}000$, of which we discard the first 1000 as burn-in samples. All convergence statistics indicate sufficient chain mixing ($\hat{R}-1<0.01$) and effective sample size. 

In Appendix \ref{sec:app_parametric_fits}, we test the stability of the fitted results as a function of scale cut, with $\ell_{\rm max}\in \lbrace 30{,}000, 50{,}000, 70{,}000, 90{,}000\rbrace$. While the reduced $\chi^2$ values are largely consistent across variations, we find more significant degradation for $\ell_{\rm max}=90{,}000$. This approaches the Nyquist limit of the \emph{CIBER} observations, where the statistical precision is extremely high and the fits are sensitive to residual errors in the beam correction ($B_{\ell=90000}\sim 0.3$). Furthermore, on such small angular scales ($\theta \sim 10''$), the light profiles of low-redshift, resolved galaxies can lead to further small-scale suppression that may bias our fits. Accounting for such effects requires more detailed modeling that goes beyond the scope of this work, and so we choose a conservative scale cut of $\ell_{\rm max}=30{,}000$ for our fiducial analysis. The two-halo constraints are effectively unchanged across $\ell_{\rm max}$ variations, while the one-halo constraints vary by $1{-}2\sigma$ between $\ell_{\rm max}=30{,}000{-}90{,}000$, which we discuss further in \S \ref{sec:one_halo_vs_redshift}.

\subsubsection{Fit results}
\label{sec:fits_and_degeneracies}

We summarize our halo model fit results for \emph{CIBER} $\times$ DESI-LS and \emph{CIBER} $\times$ HSC in Table \ref{tab:a1h_a2h_constraints}. Our model provides acceptable fits to the cross-power spectra across redshift bins, \emph{CIBER} bands and tracer catalog combinations, with caveats noted in Appendix \ref{sec:app_parametric_fits}. We find evidence for significant one- and two-halo clustering across several redshift bins. We derive detection significances by comparing against ablated model fits (summarized in Table \ref{tab:chi2_comparison}), finding that both components are needed to describe the cross-spectra without considerable degradation in $\chi^2$. In particular, our fits prefer the inclusion of the one-halo component at $>3\sigma$ significance for nearly all DESI-LS combinations, peaking at 6.4$\sigma$ for $0.2<z<0.4$. These measurements build on the stacking analysis of \cite{chengihl}, where the authors detected $A_{\rm 1h}$ at 1.8 $\mu$m using the same \emph{CIBER} maps and low-redshift SDSS galaxies. For the first time, we detect the two-halo component in our first three redshift bins, reaching $>5\sigma$ for bins $0.2<z<0.4$ and $0.4<z<0.6$. The phenomenological parameter $f_{\rm pop}$ shows a mild decreasing trend with redshift, however the uncertainties are large and partially degenerate with $A_{\rm 1h}^{\rm Ig}$, which we discuss further in \S \ref{sec:one_halo_vs_redshift}.

\begin{table*}
\caption{Halo model parameter constraints for our fiducial cross-spectrum fits ($\ell_{\mathrm{max}}=30{,}000$). Note that the one-/two-halo significances are defined using the $\Delta \chi^2$ metric (see \S \ref{sec:app_parametric_fits}), which deviates from component amplitude-based estimates at low-SNR. The phenomenological parameter $f_{\rm pop}$ encodes the linear weighting of one-halo templates from star-forming (low-mass) and quiescent (high-mass) centrals.}\vspace{4pt}
\label{tab:a1h_a2h_constraints}
\centering
\small
\renewcommand{\arraystretch}{1.3}
\begin{tabular}{llccc}
\toprule
Redshift & $\lambda_{\rm CIBER}$ & $D_{\rm 1h}^{\rm Ig} (\ell=5000)$ [nW m$^{-2}$ sr$^{-1}$] & $A_{\rm 2h}^{\rm Ig}$ & $f_{\rm pop}$ \\
\midrule
\multicolumn{5}{c}{\textbf{CIBER $\times$ DESI-LS} ($z_{\rm AB} < 22$, $A_{\times}=20$ deg$^2$)} \\
\midrule
\multirow{2}{*}{$0.0$--$0.2$}
& $1.1\,\mu\mathrm{m}$ & $1.238^{+0.277}_{-0.275}\,(4.5\sigma)$ & $0.413^{+0.108}_{-0.108}\,(3.7\sigma)$ & $0.509^{+0.318}_{-0.311}$ \\
& $1.8\,\mu\mathrm{m}$ & $0.742^{+0.228}_{-0.226}\,(3.3\sigma)$ & $0.570^{+0.132}_{-0.132}\,(4.1\sigma)$ & $0.577^{+0.288}_{-0.350}$ \\
\multirow{2}{*}{$0.2$--$0.4$}
& $1.1\,\mu\mathrm{m}$ & $0.614^{+0.121}_{-0.124}\,(5.2\sigma)$ & $0.321^{+0.051}_{-0.051}\,(6.0\sigma)$ & $0.613^{+0.248}_{-0.256}$ \\
& $1.8\,\mu\mathrm{m}$ & $0.540^{+0.085}_{-0.085}\,(6.4\sigma)$ & $0.228^{+0.045}_{-0.045}\,(5.0\sigma)$ & $0.567^{+0.281}_{-0.247}$ \\
\multirow{2}{*}{$0.4$--$0.6$}
& $1.1\,\mu\mathrm{m}$ & $0.326^{+0.079}_{-0.084}\,(4.0\sigma)$ & $0.090^{+0.024}_{-0.024}\,(3.7\sigma)$ & $0.431^{+0.324}_{-0.193}$ \\
& $1.8\,\mu\mathrm{m}$ & $0.321^{+0.058}_{-0.060}\,(5.5\sigma)$ & $0.136^{+0.025}_{-0.025}\,(5.4\sigma)$ & $0.476^{+0.316}_{-0.210}$ \\
\multirow{2}{*}{$0.6$--$0.8$}
& $1.1\,\mu\mathrm{m}$ & $0.267^{+0.078}_{-0.080}\,(3.4\sigma)$ & $<0.024\,(\mathrm{95\%\ U.L.})$ & $0.522^{+0.296}_{-0.218}$ \\
& $1.8\,\mu\mathrm{m}$ & $0.265^{+0.062}_{-0.061}\,(4.3\sigma)$ & $<0.056\,(\mathrm{95\%\ U.L.})$ & $0.516^{+0.293}_{-0.200}$ \\
\multirow{2}{*}{$0.8$--$1.0$}
& $1.1\,\mu\mathrm{m}$ & $0.161^{+0.085}_{-0.080}\,(2.1\sigma)$ & $<0.018\,(\mathrm{95\%\ U.L.})$ & $0.412^{+0.357}_{-0.228}$ \\
& $1.8\,\mu\mathrm{m}$ & $0.191^{+0.059}_{-0.060}\,(3.3\sigma)$ & $<0.034\,(\mathrm{95\%\ U.L.})$ & $0.421^{+0.328}_{-0.196}$ \\
\midrule
\multicolumn{5}{c}{\textbf{CIBER $\times$ HSC} ($18<i_{\rm AB}<25$, $A_{\times}=4$ deg$^2$)} \\
\midrule
\multirow{2}{*}{$0.2$--$0.4$}
& $1.1\,\mu\mathrm{m}$ & $0.451^{+0.101}_{-0.101}\,(4.5\sigma)$ & $0.328^{+0.074}_{-0.074}\,(4.3\sigma)$ & $0.600^{+0.265}_{-0.241}$ \\
& $1.8\,\mu\mathrm{m}$ & $0.396^{+0.066}_{-0.066}\,(5.9\sigma)$ & $0.337^{+0.061}_{-0.061}\,(5.4\sigma)$ & $0.607^{+0.251}_{-0.209}$ \\
\multirow{2}{*}{$0.4$--$0.6$}
& $1.1\,\mu\mathrm{m}$ & $0.385^{+0.091}_{-0.089}\,(4.4\sigma)$ & $0.173^{+0.050}_{-0.050}\,(3.3\sigma)$ & $0.664^{+0.216}_{-0.197}$ \\
& $1.8\,\mu\mathrm{m}$ & $0.269^{+0.057}_{-0.058}\,(4.7\sigma)$ & $0.247^{+0.043}_{-0.043}\,(5.8\sigma)$ & $0.597^{+0.236}_{-0.180}$ \\
\multirow{2}{*}{$0.6$--$0.8$}
& $1.1\,\mu\mathrm{m}$ & $0.107^{+0.065}_{-0.060}\,(2.1\sigma)$ & $<0.066\,(\mathrm{95\%\ U.L.})$ & $0.486^{+0.312}_{-0.250}$ \\
& $1.8\,\mu\mathrm{m}$ & $<0.137\,(\mathrm{95\%\ U.L.})$ & $0.070^{+0.031}_{-0.031}\,(2.0\sigma)$ & $0.374^{+0.368}_{-0.233}$ \\
\multirow{2}{*}{$0.8$--$1.0$}
& $1.1\,\mu\mathrm{m}$ & $<0.091\,(\mathrm{95\%\ U.L.})$ & $<0.071\,(\mathrm{95\%\ U.L.})$ & $0.241^{+0.402}_{-0.178}$ \\
& $1.8\,\mu\mathrm{m}$ & $<0.096\,(\mathrm{95\%\ U.L.})$ & $<0.050\,(\mathrm{95\%\ U.L.})$ & $0.286^{+0.387}_{-0.196}$ \\
\midrule
\bottomrule
\end{tabular}
\end{table*}

\begin{figure}
    \centering
    \includegraphics[width=0.95\linewidth]{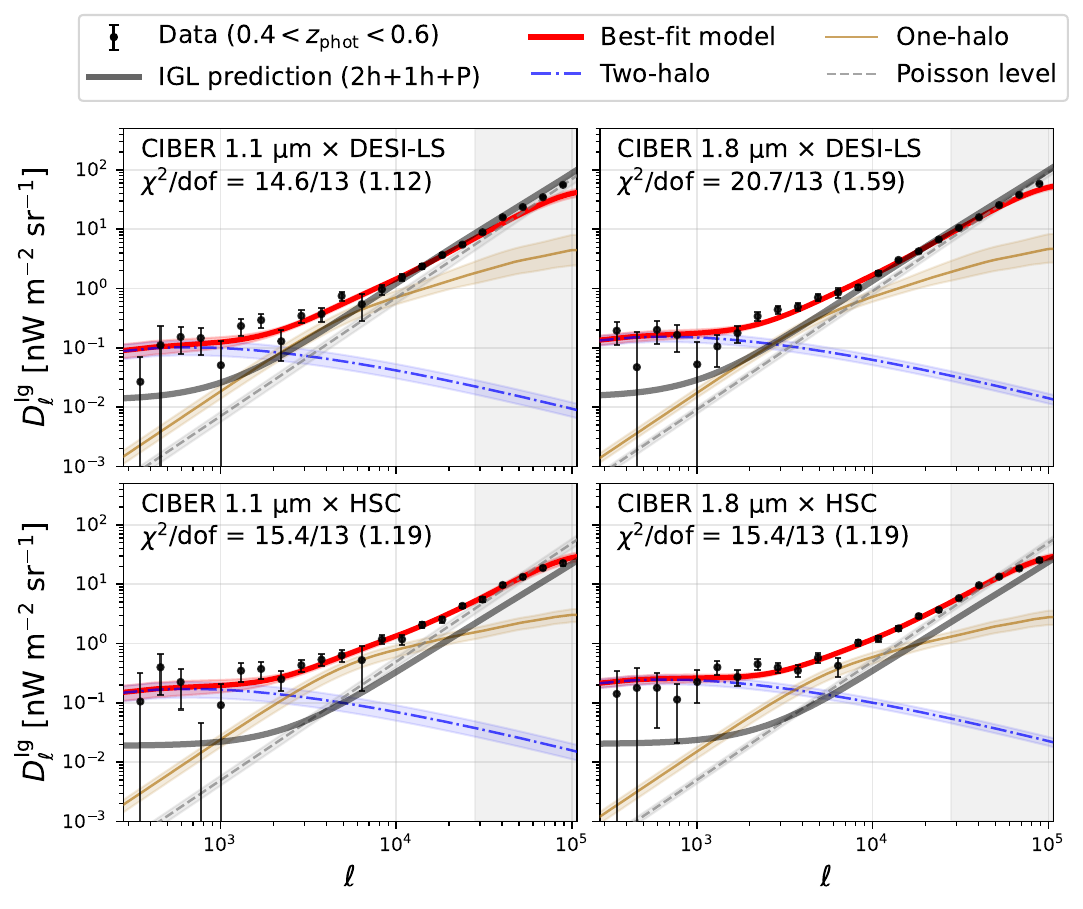}
    \caption{Parametric fits to \emph{CIBER} $\times$ DESI-LS (top) and HSC (bottom) cross-spectra, for tomographic bin $z_{\rm phot} \in [0.4, 0.6)$. We compare these fits with the baseline IGL predictions described in \S \ref{sec:cib_predictions} (dark grey solid curves). The grey shaded region in each panel highlights bandpowers excluded in our power spectrum fits ($\ell>30{,}000$).}
    \label{fig:ps_fit_0p4_z_0p6}
\end{figure}

While we present the full set of parametric fits in Appendix \ref{sec:app_parametric_fits}, we highlight the $0.4<z<0.6$ results for both \emph{CIBER} bands and galaxy tracers in Fig. \ref{fig:ps_fit_0p4_z_0p6}. For this redshift bin, $\chi^2_{\rm red} = 1.1-1.2$ for three of the four combinations, while the reduced $\chi^2$ is slightly higher for the \emph{CIBER} 1.8 $\mu$m $\times$ DESI-LS combination ($\chi^2_{\rm red} = 1.6$). The parametric fits shown in Fig. \ref{fig:ps_fit_0p4_z_0p6} deviate from our standard IGL predictions toward low-$\ell$, exceeding the predicted power at $\ell=1000$ by a factor of five for the \emph{CIBER} $\times$ DESI-LS crosses and nearly an order of magnitude for \emph{CIBER} $\times$ HSC. Our fits suggest that two-halo clustering dominates the fluctuation power on these angular scales, while matching the contribution from one-halo clustering at $\ell=2{,}000{-}3{,}000$. 

In Appendix \ref{sec:app_parametric_fits} we show corner plots for the 1.1 $\mu$m crosses in this redshift bin. These illustrate that, while the two-halo amplitude posteriors are largely uncorrelated with the other model parameters, the parameters $\lbrace A_{\rm 1h}^{\rm Ig}, f_{\rm pop}, A_{\rm Poisson} \rbrace$ have non-trivial correlations given their shared support at high-$\ell$. For this reason, we quote the amplitude of the one-halo power at $\ell=5{,}000$, rather than the $A_{\rm 1h}^{\rm Ig}$ parameter directly. In the absence of systematic effects, the one-halo and Poisson components may be further disentangled by pushing to higher $\ell_{\rm max}$. 

\subsubsection{Redshift evolution of one-halo clustering}
\label{sec:one_halo_vs_redshift}

\begin{figure}
    \centering
    \includegraphics[width=0.38\linewidth]{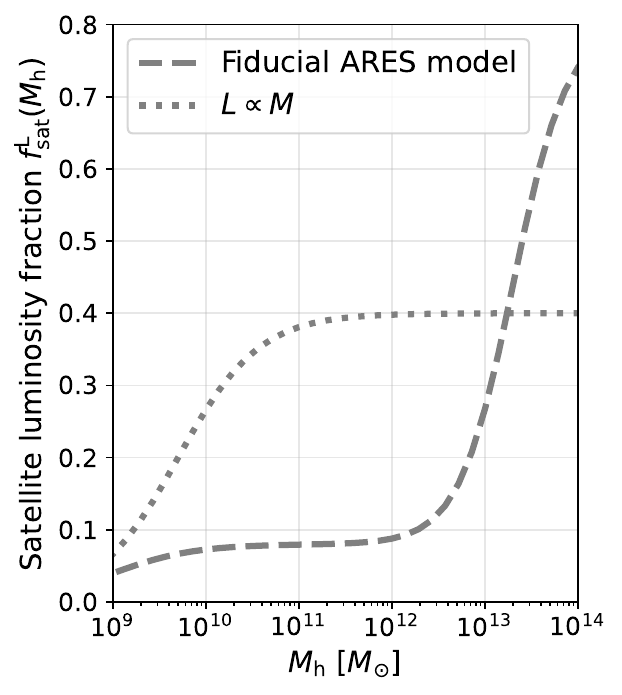}\includegraphics[width=0.61\linewidth]{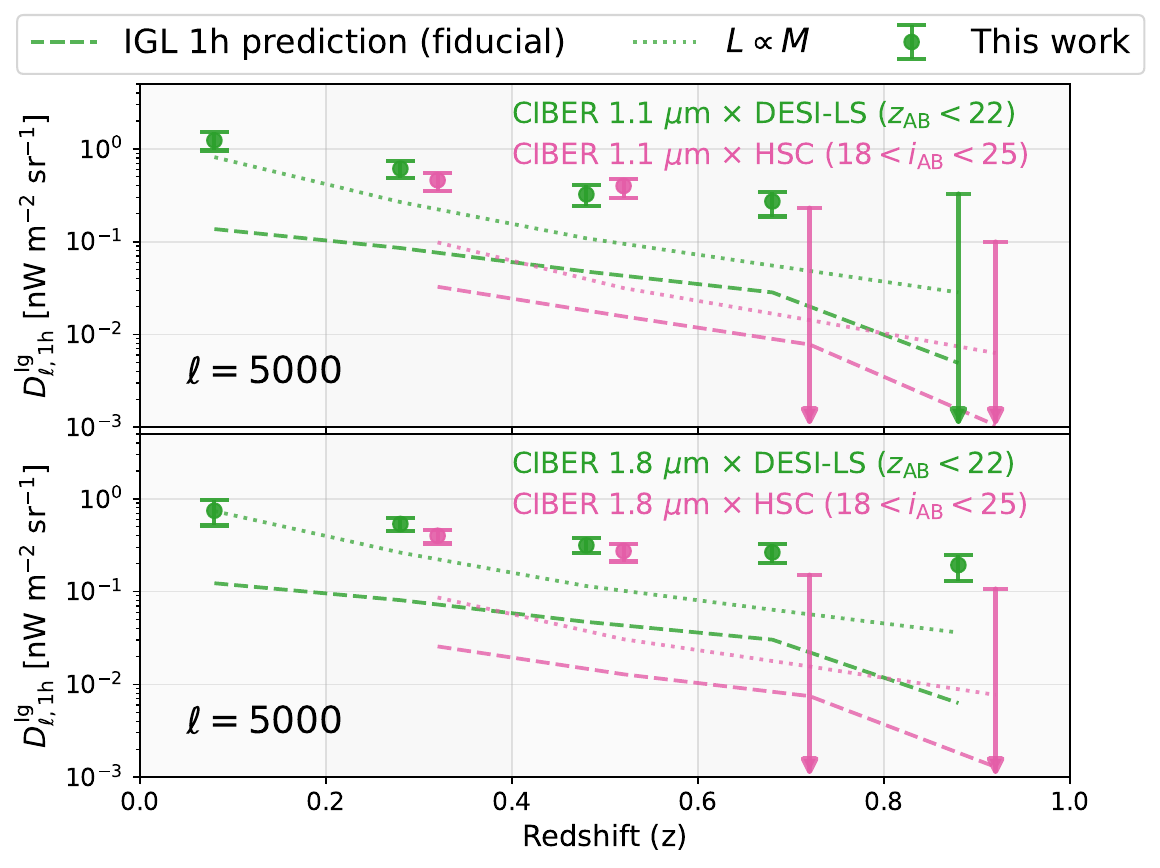}
    \caption{One-halo constraints from our fiducial cross-spectrum fits. On the left, we compare two models for the one-halo component $C_{\ell}^{\rm Ig}$, which depend on $f_{\rm sat}^{\rm L}(M_h)$, the fractional contribution of satellites to the total halo luminosity (see \S \ref{sec:onehalo_pred}). On the right, the inferred one-halo power at $\ell=5000$ is plotted alongside the two sets of model predictions. Downward pointing arrows indicate 95\% upper limits.}
    \label{fig:A1h_vs_redshift}
\end{figure}

In Figure \ref{fig:A1h_vs_redshift}, we show the dependence of the best-fit one-halo power as a function of redshift. In addition to our fiducial one-halo predictions from the \texttt{Ares} model, where $f_{\rm sat}^{\rm L}$ rises sharply for large $M_h$, we show an alternate model in which the per-satellite luminosity scales proportionally with mass, i.e., $L_{\rm sat}/L_{\rm cen} = M_{h}^{\rm sub}/M_{h}$ (which we dub ``$L\propto M$"). Combined with the sub-halo mass function, this results in a roughly constant $f_{\rm sat}^{\rm L}(M) \approx 0.4$ for $M_h \geq 10^{11}$ $M_{\odot}$. For all band+tracer combinations, we observe a clear trend in $D_{\ell, \rm 1h}^{\rm Ig}(z)$, peaking at $z\sim 0$ and declining toward $z=1$. While this trend is consistent with that of our fiducial \texttt{Ares} model predictions, the amplitude of the measured one-halo power is markedly higher, with the HSC measurements and predictions showing the largest discrepancies. In Appendix \ref{sec:param_vs_lmax} where we test fits with higher $\ell_{\rm max}$, $D_{\rm \ell, 1h}^{\rm Ig}$ has a mild decreasing trend within some redshift bins, in particular for \emph{CIBER} $\times$ DESI-LS. The higher $\ell_{\rm max}$ fits may be affected by systematic effects as discussed earlier in the text; however, even these variations taken at face value are not enough to explain the observed differences.

The fact that the one-halo amplitude is similar between DESI-LS and HSC crosses is noteworthy -- if the one-halo signal were solely driven by the most massive halos, one would expect that \emph{CIBER} $\times$ HSC one-halo power would be much lower than from \emph{CIBER} $\times$ DESI-LS, due to the fact that the deeper HSC catalog places more weight on lower-mass halos. Our $L\propto M$ model elevates the predicted one-halo power, but does not fully resolve discrepancies with respect to the measurements. Moreover, while we compare the amplitude of the inferred one-halo power against predictions, these physical scenarios are partially degenerate with the assumed intensity profile $u_I(k|M)$, which is also probed by the shape of the one-halo clustering signal.

Across the two \emph{CIBER} bands, the amplitude of one-halo fluctuation power is roughly equal, consistent with a relatively smooth NIR continuum. Future measurements with higher sensitivity and/or finer spectral resolution may reveal clearer color dependence of the one-halo term, though care is needed to disentangle rest-frame color variations (e.g., due to different stellar populations, metallicity, dust, etc.) from purely geometric k-correction effects (i.e., redshifting). 


\subsubsection{Constraints on bias-weighted intensity kernel}
\label{sec:twohalo_constraints}

\begin{figure}
    \centering
    \includegraphics[width=0.85\linewidth]{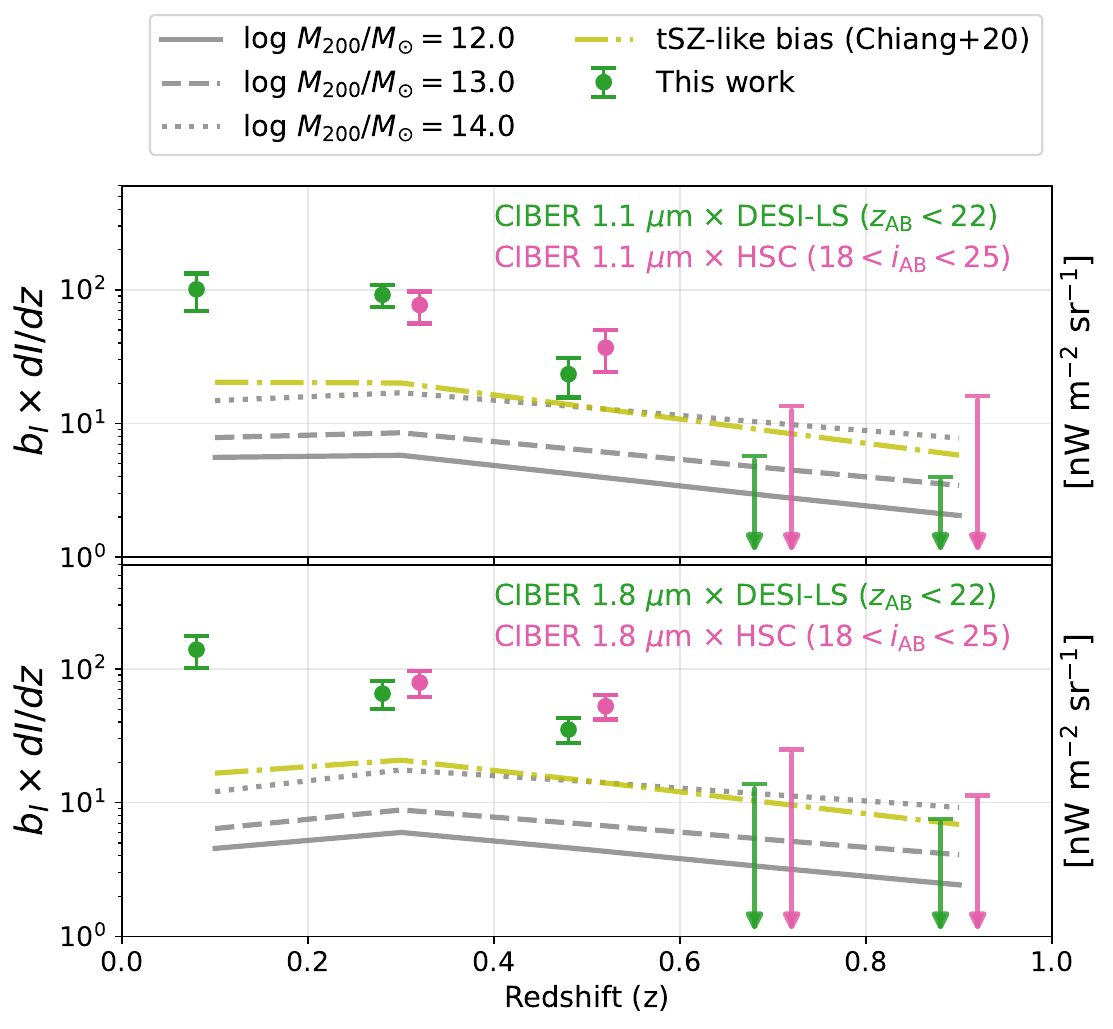}
    \caption{Bias-weighted intensity redshift kernel, $b_I \times dI/dz$, for five tomographic redshift bins spanning $0<z<1$. We derive consistent estimates from DESI-LS (green) and HSC (pink) within uncertainties. The curves show our corresponding IGL predictions for 1.1 $\mu$m and 1.8 $\mu$m, which assume fixed $dI/dz$ from the \texttt{Ares} model (matching existing constraints from galaxy counts, e.g., \citep{2016ApJ...827..108D}) but bracket a range of plausible scenarios for the luminosity-weighted bias $b_I(z)$, corresponding to characteristic halo masses $M_{200}=10^{12}$ $M_{\odot}$ ($b_I=0.8{-}1.2$), $10^{13}$ $M_{\odot}$ ($b_I=1.2{-}2.0$), $10^{14}$ $M_{\odot}$ ($b_I=2.2-4.6$), and a tSZ-like bias ($b_I=2.9{-}3.5$).}
    \label{fig:A2h_vs_z}
\end{figure}

We now turn to the two-halo component, which is well-detected across several band/tracer combinations and redshifts. Using our knowledge of $b_g\times dN/dz$, we convert cross-spectrum amplitudes $A_{\rm 2h}^{\rm Ig}$ to estimates of $b_I \times dI/dz$, which we show in Fig. \ref{fig:A2h_vs_z}. The estimates derived from both tracer catalogs are consistent within uncertainties, following our expectation that the intensity redshift kernel from different tracers should be the same given a fixed source mask. 

As discussed in \S \ref{sec:cib_modl}, it is not possible to directly extract an estimate of the EBL intensity from fluctuation measurements, given the degeneracy between $b_I$ and $dI/dz$. Predicting $b_I$ is challenging as it is determined by the luminosity-weighted halo bias from all galaxies. As such, we consider a range of bias models that bracket our uncertainty on the intensity bias; the first three models assume $b_I$ corresponds to characteristic mass scales $\log M_h \in \lbrace 12.0, 13.0, 14.0\rbrace$, while in the last case we consider a bias matching that of the thermal Sunyaev-Zel'dovich effect, whose signal has a $M^{5/3}$ dependence \citep{ykc_tsz}. Across these configurations, the predicted $b_I \times dI/dz$ varies by a factor of four. Yet, even for the maximal ``tSZ-like" bias, our predictions are unable to reproduce our measurements of $b_I \times dI/dz$ at $z<0.6$, by a large factor. In the last two redshift bins, the 95\% upper limits from \emph{CIBER} $\times$ DESI-LS are in moderate tension with the $\log M_h=14.0$ and tSZ-like bias scenarios, in particular for 1.1 $\mu$m. 

\subsubsection{Impact of nonlinear clustering}

While our halo model provides a good fit to the data, we note that nonlinear clustering modifies the linear theory two-halo term, in particular near the transition between one- and two-halo dominated scales. For $z<1$, the minimum angular multipole of our measurements ($\ell_{\rm min}=300$) probes physical scales in the quasi-linear to nonlinear regime, where the two-halo component is enhanced. Because our two-halo template utilizes the linear theory matter power spectrum, our fits will be biased to some extent, with NL contributions absorbed into the one- and two-halo templates. Assuming our linear theory template absorbs all of the two-halo power, we expect in the worst-case scenario that our estimates of $W_I(z)$ are biased high by a factor of two, which is significant but not enough to explain the observed differences with respect to model predictions.

\subsection{Contribution of \texorpdfstring{$z<1$}{} LSS to CIBER auto power}
\label{sec:recon_auto}

We now turn to the following question: how much do low-redshift galaxies and associated LSS contribute to the \emph{CIBER} auto power? For an intensity field $I$ and galaxy tracer $g$, we can rearrange our earlier expression for the cross-correlation coefficient:
\begin{equation}
    r_{\ell}^{\rm Ig} = \frac{C_{\ell}^{\rm Ig}}{\sqrt{C_{\ell}^{\rm gg}C_{\ell}^{\rm II}}} \rightarrow C_{\ell}^{\rm II} = \frac{(C_{\ell}^{\rm Ig})^2}{C_{\ell}^{\rm gg}}\frac{1}{(r_{\ell}^{\rm Ig})^2}.
\end{equation}
We consider the following estimator, $\hat{C}_{\ell}^{\rm II} = (C_{\ell}^{\rm Ig})^2/C_{\ell}^{\rm gg}$: this is equal to $C_{\ell}^{\rm II}$ for $r_{\ell}^{\rm Ig} = 1$ and represents a lower bound for $r_{\ell}^{\rm Ig}<1$.

In the two-halo clustering regime, for a fixed redshift range, the intensity and galaxy overdensity fields are modeled as linearly biased tracers of the matter field, such that $r_{\ell} \to 1$\footnote{In reality, $r_{\ell}^{\rm Ig} \neq 1$ on two-halo scales because of decorrelation between the matter field and galaxy field.} and our estimator recovers the true two-halo intensity auto power within the associated redshift bin. However, on one-halo scales, the two fields weight halos differently (by luminosity $L_I(M)$ for intensity and by $N_g(M)$ for galaxies). Halos with high $L_I/N_g$ (e.g., massive halos with bright centrals or diffuse intra-halo light) contribute to $C_{\ell}^{\rm II, 1h}$ more strongly than to $C_{\ell}^{\rm Ig, 1h}$, while halos with low $L_I/N_g$ inflate $C_{\ell}^{\rm gg, 1h}$ relative to $C_{\ell}^{\rm Ig, 1h}$. Both cases reduce $r_{\ell}$, and so we interpret $\hat{C}_{\ell}^{\rm II}$ derived from the total clustering $A_{\rm 1h}\mathcal{C}_{\ell}^{\rm 1h}+A_{\rm 2h}\mathcal{C}_{\ell}^{\rm 2h}$ as a lower bound on the contribution of correlated LSS to the \emph{CIBER} auto power. The degree to which $\hat{C}_{\ell}^{\rm II}$ underestimates the true auto power depends on how the number- and luminosity-weighted contributions from galaxies and any unmodeled diffuse components scale with halo mass.

We consider an intensity-weighted galaxy tracer template $\hat{I}_g$:
\begin{equation}
    \hat{I}_g(x, y) = \sum_{j=0}^{N_{\rm gal}} I_j \delta(x-x_j, y-y_j).
\end{equation}
While $\delta_g$ weights each galaxy equally, $\hat{I}_g$ is a better approximation for the luminosity weighting $L_I(M)$ of the intensity field. We construct $\hat{I}_g$ for each tracer, using $z$-band fluxes from DESI-LS and $i$-band fluxes for HSC. We assume catalog flux uncertainties are negligible. We confirm that, while $r_{\ell}^{\rm \hat{I}_g \times I_{CIBER}} \approx r_{\ell}^{\rm \delta_g I_{CIBER}}$ on large angular scales ($\ell<1{,}000$), $r_{\ell}^{\rm \hat{I}_g \times I_{CIBER}}$ is a factor of $1.5-2\times$ higher on scales $1{,}000<\ell<10{,}000$ and even higher for $\ell>10{,}000$. While our flux-weighted templates may be improved further (e.g., using color corrections to match the galaxy fluxes at 1.1 and 1.8 $\mu$m), they provide a substantially stronger bound on $C_{\ell}^{\rm II}$ and so we leave further refinement to future work.

\begin{figure*}
    \centering
    \includegraphics[width=\linewidth]{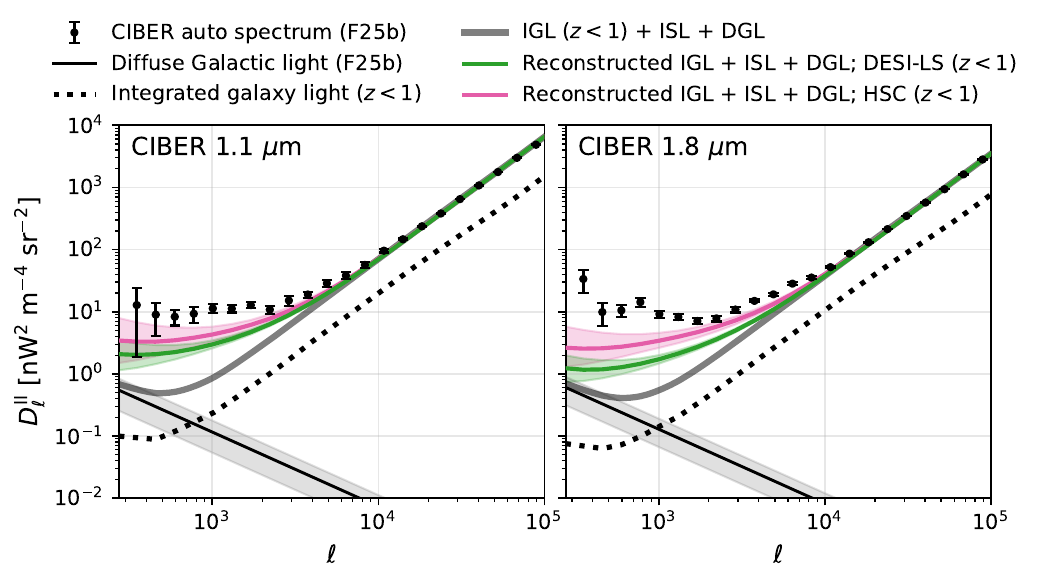}
    \caption{\emph{CIBER} auto power spectrum measurements from \cite{feder25b}, compared against the reconstructed estimate from $z<1$ galaxies and correlated LSS. Also shown are constraints on diffuse Galactic light (DGL) from \cite{feder25b} and IGL mock predictions for $z<1$ (black dotted curves). The auto spectra may have non-negligible Poisson fluctuations from unmasked stars and $z>1$ galaxies, explaining the difference between the IGL predictions and data at high $\ell$. Note that the reconstructed IGL estimates are lower bounds on the true auto power from low-redshift correlated LSS (\S \ref{sec:recon_auto}).}

    \label{fig:recon_auto}
\end{figure*}

In Figure \ref{fig:recon_auto} we compare $\hat{C}_{\ell}^{\rm II}$ with the \emph{CIBER} auto spectrum measurements from \citep{feder25b}, alongside modeled contributions from Poisson fluctuations and diffuse Galactic light (DGL). For all combinations, we derive estimates for $\hat{C}_{\ell}^{\rm II}$ by propagating posterior samples of $A_{\rm 1h}\mathcal{C}_{\ell}^{\rm 1h} + A_{\rm 2h}\mathcal{C}_{\ell}^{\rm 2h}$ from the cross- and auto-spectrum fits. We find that auto-power from correlated LSS at $z<1$ significantly exceeds the same predicted contribution from IGL for $\ell < 2,\!000$. The bounds from HSC are higher than from DESI-LS by roughly a factor of two in both \emph{CIBER} bands. While we measure similar cross-power between both $z<1$ catalogs, the HSC auto-power is lower than from DESI-LS, resulting in a larger ratio $C_{\ell}^{\rm Ig}/C_{\ell}^{\rm gg}$. Once combined with the DGL auto power constraints derived in \cite{feder25b}, the reconstructed power from both catalogs explains a significant portion, though not all, of the measured \emph{CIBER} auto-power.

\subsection{Astrophysical interpretation}
\label{sec:astro_interpretation}
The halo model fits in \S \ref{sec:parametric_fits} establish that both the one- and two-halo cross-power exceed that of our standard IGL predictions by a significant factor, especially for redshifts $z<0.6$, while the Poisson level and galaxy auto power spectrum are well-reproduced by the same model. Any astrophysical explanation must amplify the clustering terms without simultaneously inflating the Poisson fluctuations, as the latter is proportional to the mean NIR intensity per tracer galaxy. This leaves little freedom to enhance the central galaxy NIR emissivity without introducing disagreement at high-$\ell$. 

\subsubsection{Satellite galaxies}
One explanation of the observed cross-power is amplification through a more luminous satellite population than our baseline model assumes.  Detected satellites tend to reside in more massive halos and source one-halo power through satellite-satellite and central-satellite clustering. \citet{cheng22} studied the impact of non-linear galaxy clustering on EBL observables using the MICECAT N-body simulations, finding that satellites source a one-halo term comparable to the two-halo clustering power at $\ell=1,\!000$ at $z=0.5$, however those findings assumed a two-halo contribution much lower than what we measure in this work.

Standard HOD parameterizations \citep{zheng05} model the satellite occupation as a power-law:
$$\langle N_{\rm sat}(M)\rangle \propto 
((M-M_{\rm cut})/M_1)^\alpha,$$ with characteristic mass scales $M_1 
\gtrsim 10^{13}\,M_\odot$ and cutoff mass $M_{\rm cut} = 10^8{-}10^9 M_{\odot}$, assigning satellite 
luminosities via sub-halo abundance matching \citep{behroozi2013} or conditional luminosity 
functions \citep{yang03}. Assuming fixed satellite emissivity, bridging the gap between our measurements and IGL predictions would require satellites to be fainter but more numerous than what standard HODs predict. However, scenarios with elevated star-formation efficiency in low-mass satellite halos at low redshift are not excluded by current data -- in fact, studies combining deep photometric imaging from DECaLS with spectroscopy from DESI find evidence for enhanced activity in such low-mass systems \citep{pacs_desi}. Our NIR measurements directly constrain the luminosity-weighted satellite contribution, complementing stacking and HOD analyses that use discrete tracers. 

\subsubsection{Diffuse intra-halo light}
Another related candidate is diffuse intra-halo light (IHL), which contributes to the NIR intensity field without increasing number counts. The dominant formation channel for IHL is believed to be tidal stripping of satellite galaxies as they orbit within their host halo. \citet{rudick10} predicts that this mechanism accounts for the majority of intra-cluster light (ICL) at $z<0.5$ and that ICL fractions grow to several tens of percent of total cluster light by $z=0$. The existence of extended, low-surface-brightness stellar emission in and around massive halos is established observationally through direct deep imaging \citep{rudick10, merritt16, hsc_huang}, stacking analyses \citep{chengihl}, and fluctuation-based analyses at optical and NIR wavelengths \citep{cooray12, zemcov14, mitchellwynne}. However, most IHL studies are concentrated at group- and cluster-mass scales ($M \gtrsim 10^{13} M_{\odot}$), and models of tidally-stripped IHL predict that diffuse stellar fractions drop substantially toward lower halo mass, with $f_{\rm IHL} \lesssim 5\%$ for Milky Way-mass halos ($M \sim 10^{12} M_{\odot}$) \citep{purcell07}.

Within each tomographic bin, the magnitude-limited galaxy samples used in our cross-correlations span a range of halo masses, complicating direct comparison with IHL models calibrated to halo masses and SMHM relations. Our CMG cross-correlations in \S \ref{sec:cmg_cross} provide the clearest empirical evidence for cluster-mass-scale contributions: both CMG auto and cross-spectra show prominent one-halo power for $\ell<10000$ and CMGs account for 15-20\% of the total \emph{CIBER} $\times$ DESI-LS cross-power on these scales despite comprising $<4\%$ of the full galaxy sample. However, substantial one-halo power remains even after CMG removal from the DESI-LS sample, indicating that group- and galaxy-scale halos also contribute. For IHL to account for the bulk of the unmodeled cross-power across our tomographic bins, either the signal must be dominated by the highly biased upper tail of the halo mass function (which may be possible for a steep luminosity weighting of the intensity field), or the IHL fractions in lower-mass halos must exceed current model predictions. 

We emphasize that the satellite and IHL scenarios are not mutually exclusive and are difficult to separate both observationally and in simulation \citep{sanderson18, proctor23}. A self-consistent model that simultaneously varies the satellite occupation, satellite luminosity function, and IHL fraction as a function of halo mass and redshift is necessary to properly disentangle the contributions of satellites and IHL to our $A_{\rm 1h}(z)$ measurements, and such a model has not yet been formulated that captures the range of scales probed by our analysis. Furthermore, alternative IHL formation channels may contribute differently across halo mass scales, including in-situ star formation in tidal streams and debris, baryonic feedback quenching star formation in satellite galaxies, and the accumulated contribution of disrupted dwarf galaxies over cosmic time \citep{joo2025}.

\subsubsection{A higher NIR background?}

The constraints on $b_I\times dI/dz$ from \S \ref{sec:twohalo_constraints} are in stark disagreement with our IGL predictions, even under maximal tSZ-like bias assumptions. If indeed the intensity bias is high at $1{-}2$ $\mu$m, one might expect the non-linear bias to be non-negligible as well, introducing potential scale-dependent modifications to the linear matter power spectrum, though such corrections are unlikely to explain the factor of several difference observed at $z<0.6$ and our CMG-based results disfavor such a large intensity bias. Absent other effects, our results imply that a higher $dI/dz$ is required to resolve existing discrepancies. As the \texttt{Ares} model is well-calibrated with respect to existing IGL predictions of $\nu I_{\nu}$ \citep{helgason}, this would place our $dI/dz$ estimates in direct tension with constraints from galaxy counts \citep{2016ApJ...827..108D, carter25} and $\gamma$-ray measurements relying on the modeled absorption of blazars \citep{ebl_biteau, banerjee26}, but in closer agreement with absolute photometric measurements from \emph{HST}, \emph{DIRBE} \citep{cosmoglobe_cibmono}, and \emph{CIBER} low-resolution spectrometer \citep{matsuura_ciber}. 

%% file: discussion.tex
\section{Conclusion}
\label{sec:conclusion}
We have presented the first tomographic cross-power spectrum analysis of NIR background fluctuations, cross-correlating \emph{CIBER} 1.1 and 1.8 $\mu$m imager data against photometric galaxy samples from DESI-LS and HSC. These measurements provide a unique window into the cosmic baryon inventory, offering constraints on the how much stellar content exists and its spatial distribution, with minimal impact from Galactic dust or other foregrounds. On small angular scales ($\ell \gtrsim 10,\!000$), the cross-power spectra are in close agreement with standard IGL model predictions, confirming that Poisson levels are reasonably well-described by \texttt{Ares}, our baseline semi-empirical IGL model. On larger scales ($\ell \lesssim 2,\!000$), we find significant excess cross-power relative to these predictions, concentrated primarily at low redshifts $z<0.6$. We find that galaxy clusters and their member galaxies contribute approximately $15{-}20\%$ of the large-angle cross-power, suggesting that group- and galaxy-scale halos contribute the bulk of the EBL signal. Using a phenomenological halo model decomposition, we detect both one- and two-halo clustering contributions at high significance across multiple bands, tracers and redshift bins, with amplitudes that smoothly decline between $z=0$ and $z=1$ but which strongly exceed our standard IGL predictions. 

Using a coherence-based estimator, we then reconstruct the NIR auto-power $C_{\ell}^{\rm II}$ associated with correlated LSS at redshifts $z < 1$. We find that it exceeds our IGL predictions on scales $304< \ell < 2,\!000$ by a factor of $5-10$, accounting for a substantial portion of the \emph{CIBER} auto-power spectrum excess from \cite{feder25b} when combined with estimates of ISL and DGL. Because our reconstruction is a lower bound (for reasons detailed in \S \ref{sec:recon_auto}), the true contribution from $z < 1$ correlated structure may be larger still. This provides direct evidence that EBL clustering from low-redshift LSS is a significant contributor to NIR background fluctuations on several-arcminute scales, with implications for ongoing fluctuation analyses that utilize wavelengths $1< \lambda_{\rm obs} < 2$ $\mu$m. In particular, our results suggest that fluctuation measurements targeting scales near $\ell \sim 1000$, e.g., EoR searches using the Lyman-$\alpha$ line or Lyman break feature \citep{feng19, cheng22_cnib_tomography, euclid_cao, ambrose_spherex} require either substantially larger survey area or a improved foreground mitigation to achieve their stated statistical precision.

A complete physical interpretation of the excess low-redshift NIR power will require more detailed modeling than the phenomenological fits used in this work. Our IGL predictions, derived from the \texttt{Ares} model, may be extended to more detailed central+satellite prescriptions, and to include contributions from diffuse IHL. These extended predictions can then be compared against cosmological hydrodynamical simulations such as FIRE \citep{fire_hopkins}, FLAMINGO \citep{flamingo_joop}, IllustrisTNG  \citep{illustris_tng}, and COLIBRE \citep{colibri_joop}, which self-consistently model the stellar content of halos across a range of masses, providing a check on whether the measured one-halo amplitudes are consistent with modern galaxy formation models. Combining NIR cross-correlations with kinematic Sunyaev-Zel'dovich measurements \citep{schaan21, ksz_lrgs_frank, ksz_elgs_boryana, ksz_guachalla}, which probe the integrated gas content of the same halos, offers a path toward jointly constraining the stellar and gas components of the baryonic budget as a function of angular scale and redshift.

The methods in this work may be directly applied to existing and near-future intensity mapping experiments. \emph{CIBER-2} \citep{zemcov_ciber2} provides six broad bands spanning $0.5{-}2.5$ $\mu$m and collected imager data in the COSMOS field, which contains some of the richest ancillary photometric and spectroscopic data available in the extragalactic sky, offering direct leverage over the optical-to-NIR SED of the EBL fluctuation power. In addition, the 6U-size Cube-Sat VERTECS, which successfully launched on June 12, 2026, will also observe EBL fluctuations in visible light \citep{vertecs_sano, vertecs_takimoto}. 

With over two full-sky surveys completed by the \textit{SPHEREx} mission to date \citep{bock25, korngut26}, NIR intensity mapping has rapidly transitioned from a field limited by statistics and data quality to one with photon noise-limited, multi-wavelength observations covering the entire sky \citep{cukierman_map}. The fine spectral resolution of \textit{SPHEREx} may help exploit the wavelength dependence of one- and two-halo fluctuations and potentially discriminate between known galaxy populations and other astrophysical components. In this work, we have relied on dense photometric samples, which provide the statistical sensitivity needed to detect clustering over the 20 deg$^2$ of \emph{CIBER}-1 imaging and, for some cases, probe samples with lower halo masses than those in current spectroscopic samples. However, with full-sky coverage, cross-correlations with spectroscopic samples (e.g., from DESI and PFS; \citep{desi_dr1, pfs_survey}) can be performed with comparable or better precision on the scales probed by this work, while simultaneously probing larger angular scales dominated by linear clustering. Spectroscopic samples offer well-characterized redshift distributions free of photometric scatter, and the ability to split samples by stellar mass, environment, and other physical properties that are difficult to access through photometry alone. Our work motivates the development of improved models that enable the robust interpretation of what will be an incredibly rich dataset for cross-correlations.

While a mere 220 seconds of imager data from \emph{CIBER}-1 has revealed important information about NIR EBL fluctuations through auto- and cross-correlations, an expanded EBL analysis toolkit and a plethora of novel cross-correlation opportunities will transform our understanding of the NIR in the coming years, providing a complementary view of existing LSS tracers and shedding light on unmodeled sky components.    

%% file: app_phot_dIdz.tex
\section{Validation on mocks}
\label{sec:mock_validation}
\subsection{Auto- and cross-power spectrum recovery}
We test our power spectrum pipeline using an ensemble of synthetic \emph{CIBER} observations generated in a similar fashion to those used in \cite{feder25a}. To generate the full mock astrophysical image, we combine the IGL signal maps with ISL realizations synthesized from the TRILEGAL model \citep{trilegal}, Zodiacal light and an additional fluctuation component designed to match the \emph{CIBER} auto-power spectrum measurements. For simplicity, we assume that the additional fluctuation component is uncorrelated with the galaxy density, such that in our mock recovery tests the primary effect is to add noise on large angular scales. We then add read noise and photon noise, and apply bright source masks with depth matching our fiducial analysis. For simplicity, we do not apply simulated redshift errors to the mocks, though in practice such errors can lead to de-correlation of the true signal if using tomographic bins with redshift uncertainties that are comparable to the bin width.

These mocks are then passed through the same power spectrum pipeline as the real flight images. We test our pipeline on 500 Monte Carlo pseudo-independent realizations, i.e., we pair each of the twenty-five IGL+ISL mocks with multiple realizations of instrument noise, flat-field noise, and additional low-$\ell$ fluctuation power. As with the real data, we apply gradient filtering to the intensity and galaxy overdensity fields to mitigate aliasing effects and remove low-$\ell$ noise from ZL in our cross-correlations. We then use our pseudo-$C_{\ell}$ pipeline to correct for the relevant effects and recover the underlying galaxy auto-/cross-power spectra.

\begin{figure}
    \centering
    \includegraphics[width=0.49\linewidth]{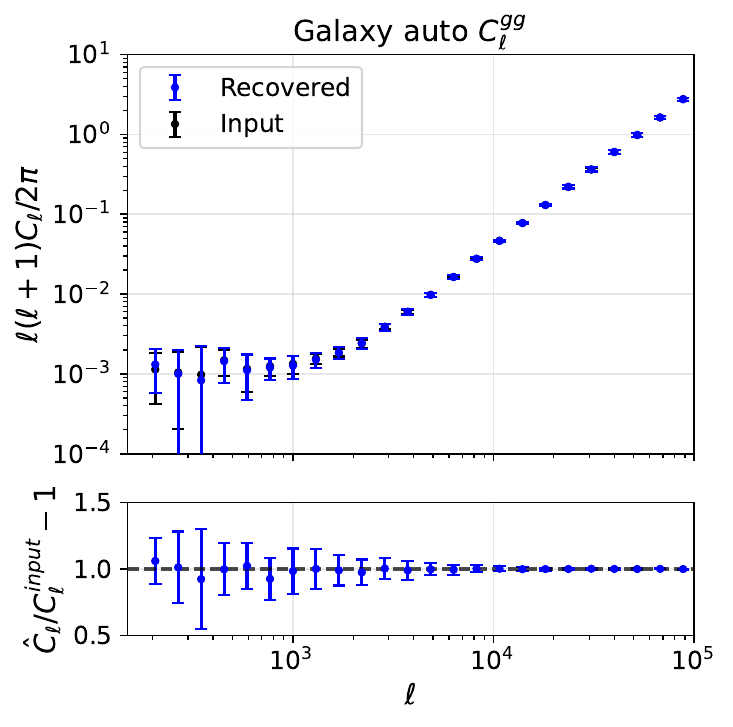}
    \includegraphics[width=0.49\linewidth]{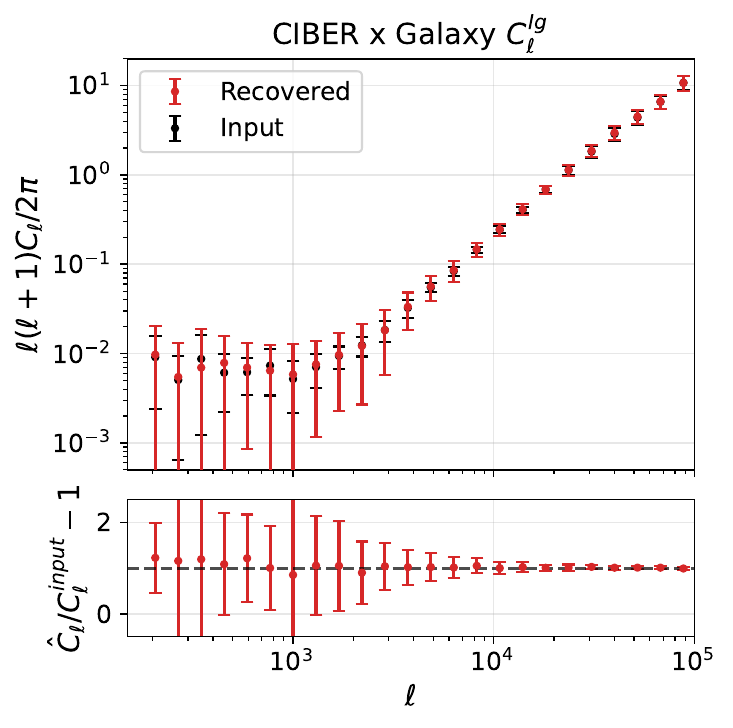}
    \caption{Mock recovery of the galaxy auto spectrum (top) and \emph{CIBER} $\times$ galaxy cross-spectrum (bottom), for a single field with $A_{\rm sur}=4$ deg$^2$ observed at 1.1 $\mu$m, and cross-correlated against a HSC-like tracer with $i_{\rm AB} < 25$. Error bars indicate the dispersion of recovered and true power spectra from 200 lognormal realizations. These results demonstrate our ability to recover unbiased power spectra using our pseudo-$C_{\ell}$ pipeline.}
    \label{fig:mock_recovery_auto_cross}
\end{figure}

In Figure \ref{fig:mock_recovery_auto_cross} we present the results of our mock tests, plotting the input and recovered power spectra along with the fractional errors. For both the galaxy auto- and the \emph{CIBER} $\times$ galaxy cross-spectrum, our pipeline recovers unbiased results across all multipoles. The errors in the cross-spectrum are considerably larger than the galaxy auto due to uncorrelated sample variance, as discussed in \S \ref{sec:sensitivity_forecast}.

\subsection{Impact of source alignment errors}
\label{sec:astr_errs}
One systematic error that uniquely impacts our cross-spectrum measurements arises from mismatch between cataloged sky positions and true source centers in \emph{CIBER} detector coordinates. While position errors have negligible impact on recovered large-angle (i.e., several arcminute) fluctuations, we find that they can drive significant small-scale suppression in the estimated cross-spectra. Such centering biases can be sourced by errors in the initial source identification and fitting, along with errors in the \emph{CIBER} astrometric registration from flight data obtained with \texttt{astrometry.net} \citep{astrometrydotnet}.

\begin{figure}
    \centering
    \includegraphics[width=0.5\linewidth]{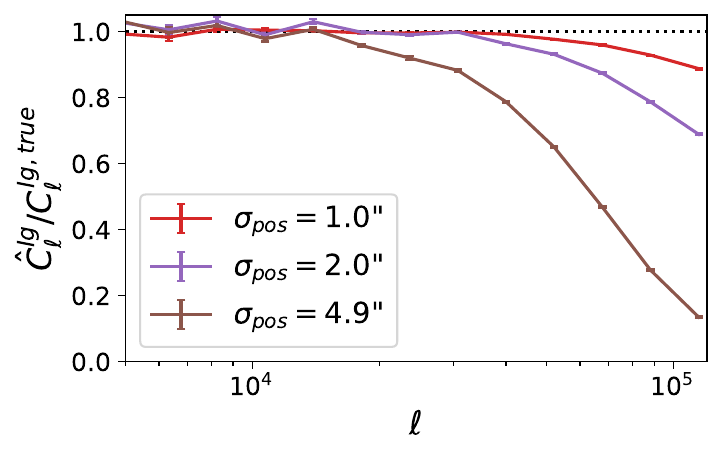}
    \caption{Ratio between recovered and true cross-power spectra as a function of source alignment error, for different effective position uncertainties $\sigma_{\rm pos}$.}
    \label{fig:astr_error}
\end{figure}
In Figure \ref{fig:astr_error}, we show the ratio of recovered to true power spectra as a function of simulated position error $\sigma_{\rm pos}$. We estimate each curve from a set of HSC-like mocks that contain the same elements as our fiducial mock tests. While arcsecond-level position errors lead to at most 10\% errors in the power spectrum for $\ell<10^5$, larger errors approaching the size of the \emph{CIBER} pixel size cause more severe biases.

In general, the positions of galaxies identified in optical surveys from DESI-LS and HSC are known to better than $0.2^{\prime\prime}$. We empirically characterize the effective astrometric registration error from cross-correlations with \emph{Gaia} stars (\S \ref{sec:ciber_gaia}), finding $\sigma_{\rm pos} \approx 2^{\prime\prime}$ in each field ($\lesssim$ one-third of a \emph{CIBER} pixel). While this level of alignment error has a negligible impact on scales $\ell < 30000$, the small-scale suppression reaches $\sim 20\%$ near $\ell=10^5$. We account for this effect by including an effective damping term in our cross-spectrum model (\S \ref{sec:model_interp}).

\subsection{Random catalog construction}

\subsubsection{DESI-LS}
\label{sec:desi_ls_rand}
We use random catalogs derived over the full DESI-LS DR8\footnote{\url{https://www.legacysurvey.org/dr8/}} footprint that are designed to track the survey geometry and minimize large-scale variations in the data. The data-to-random ratio of the full sample is de-trended as a function of Galactic latitude, after which we make a $z_{\rm AB}<22$ selection. We then apply the supplied optical bit masks (\texttt{MASKBITS}) in order to capture the effect of smaller-scale gaps in the survey coverage. Redshifts in the $z_{\rm AB}<22$ sample are resampled in the random catalog, which we use in \S \ref{sec:zdep_clx}. The randoms are resampled sufficiently many times such that $N_{\textrm{rand}}/N_{\textrm{data}}\gg 1$. 

\subsection{HSC-UDS}
\label{sec:hsc_randoms}
For our HSC-UDS analysis, we construct customized randoms that are designed to correct for depth variations across the survey footprint. Our methodology involves deriving a model of the source completeness that depends on both magnitude and sky location. This model is combined with multi-band counts from the \texttt{Ares} semi-empirical model \cite{mirocha26a} (\S \ref{sec:cib_predictions}) to simulate galaxy selection effects across multiple Monte Carlo realizations. This approach not only corrects for fictitious density variations in the galaxy field but also quantifies associated downstream errors due to non-uniformity, informing our choice of magnitude cuts.

The HSC dataset provides 5$\sigma$ depths measured on $\sim 10^{\prime} \times 10^{\prime}$ survey patches across five \textit{grizy} bands. While HSC utilizes all five bands for source detection, the majority of detections are primarily influenced by the HSC $i$-band. Therefore, we specifically focus our completeness modeling and randoms construction on this band. Within this area, the reported depth exhibits variations greater than $0.5$ magnitudes, displaying coherent structures on several arcminute to degree scales.

A key consideration is that the reported HSC depths are derived from a simplified flux sensitivity calculation based on isolated point sources. In reality, effective depth is affected by conditions such as source extension, blending from overlapping sources, sub-threshold source confusion, and possibly other effects. To account for this, we employ both observed and model-derived galaxy counts to achieve independent estimates of galaxy completeness.

\begin{figure}[ht]
    \centering
    \includegraphics[width=0.57\linewidth]{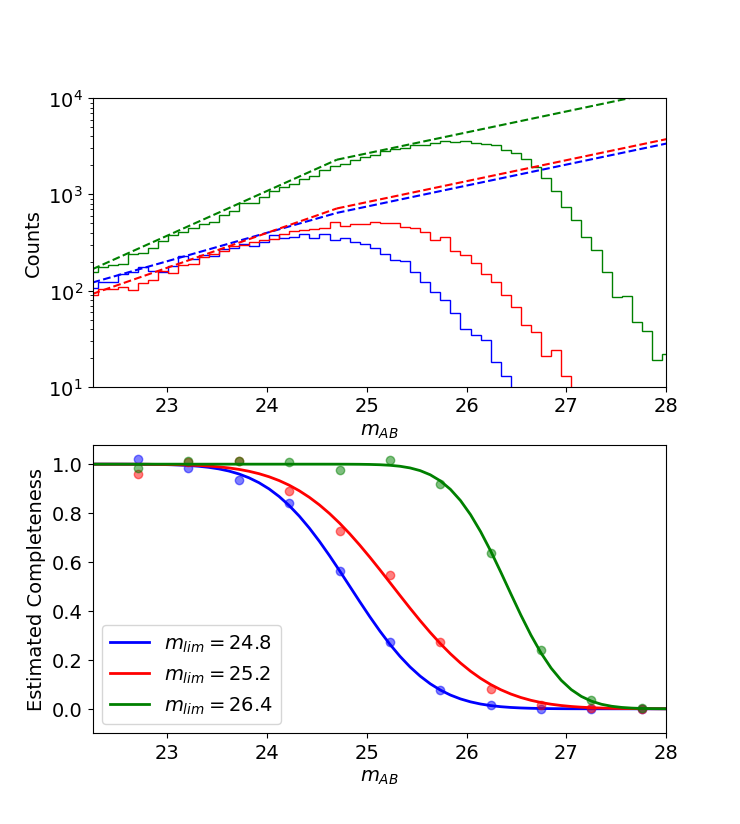}\includegraphics[width=0.4\linewidth]{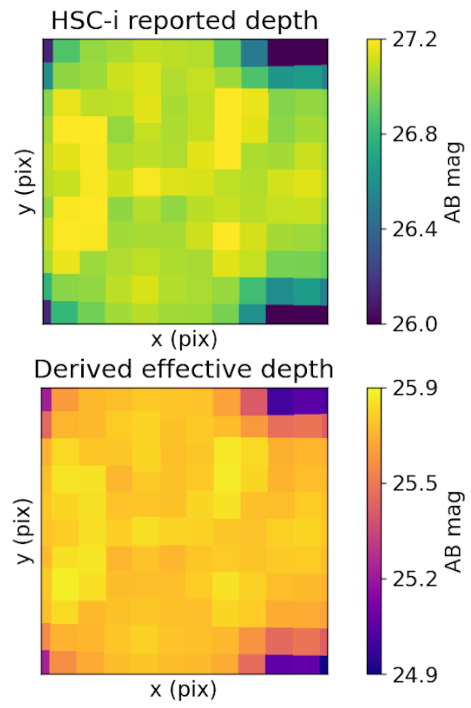}
    \caption{\textbf{Left:} Comparison of HSC $i$-band source counts (top) and completeness (bottom) as a function of $i$-band depth $m_{\rm lim}$, for three $10^{\prime} \times 10^{\prime}$ patches of characteristic depth. The dashed curves show our baseline model of the HSC counts used to estimate completeness, while the solid curves in the bottom panel show the best-fit two-parameter completeness model from \eqref{completeness_eqn}. The colors indicate our results from three characteristic patches with different reported depths. \textbf{Right:} Comparison of HSC i-band 5$\sigma$ point source depth map (top) with our recovered effective depth map using the procedure in \S \ref{sec:hsc_randoms} (bottom), projected onto CIBER 1.1 $\mu$m detector coordinates for the SWIRE ELAIS-N1 field ($2^{\circ} \times 2^{\circ}$). The HSC-reported depth map captures variations on scales $\theta \gtrsim 10^{\prime}$ based on point source flux sensitivity, and correspond approximately to 50\% completeness \citep{hsc_pdr3}.}
    \label{fig:rolloff}
\end{figure}

We first obtain empirical estimates of the completeness as a function of magnitude and position by calculating the (spatially varying) ratio of observed to model-based counts as
\begin{equation}
    \hat{C}(m, \hat{x}) = \frac{N_{\rm obs}(m, \hat{x})}{N_{\textrm{model}}(m)}.
\end{equation}
We employ a double power-law distribution to model $N_{\textrm{model}}(m)$. The bright-end slope is calibrated using the $22.0 < i_{\rm AB} < 23.5$ observed counts (which we assume are complete), while the faint-end slope matches that of \cite{helgason} $i$-band model counts to $i_{\rm AB}=28$. We transition between the two power laws at a pivot magnitude of $i_{\rm piv}=24.7$. We find that floating the bright-end slope as a function of map position captures realistic variations in the observed counts, which may arise from flux boosting effects. While our model is phenomenological, it provides a good match to faint-end counts from the literature. We then fit a smooth function to the estimated completeness at each position $\hat{x}$,
\begin{equation}
C(m, \hat{x}) = \frac{1}{2}\left[1 - \erf\left(\frac{m - m_{\rm lim}(\hat{x})}{\sigma_m}\right)\right].
\label{completeness_eqn}
\end{equation} 
This model smoothly transitions between detected and undetected sources, and accurately captures the roll-off in detection probability due to increasing instrument noise. We find that this two-parameter model, which depends on only $(m_{\rm lim}, \sigma_m)$, provides a good fit to the recovered completeness as a function of magnitude. Given the resolution of the patches that define the HSC depth map and the non-trivial masking that affects some patches more than others, we fit $C(m,\hat{x})$ to sources grouped by patches of similar reported depth, using eight bins in total. In Figure \ref{fig:rolloff} we show the results of this procedure for three bins with varying depths.

Comparing our recovered depths to reported depths, we find depth variations that are well-described by a constant offset and linear trend in $m_{\rm lim}$. In order to further stabilize noisy estimates of the depth from our empirical approach, we perform linear fits to $m_{\rm lim}^{\rm recov} = a\cdot m_{\rm lim} + b$ across our eight bins and use this to interpolate to and correct the per-patch HSC reported depths. We find no significant trends in $\sigma_m$ with respect to $m_{\rm lim}$, and so we fix the best-fit value for all patches in each grouping. Limitations of our model include the fact that finer-scale selection effects from local crowding are not captured or corrected for, and that patch-wise sample variance in counts adds errors to our completeness estimator.

In Figure \ref{fig:rolloff} we compare the reported HSC $i$-band 5$\sigma$ depths along with our derived effective depths over the \emph{CIBER} SWIRE ELAIS-N1 field.
While our completeness model tracks the same large angular scale structure as the reported depth maps by construction, we recover a median 50\% depth of $i=25.8$, over one magnitude shallower than the point source depths. As our semi-empirical estimates are subject to model-based uncertainties related to the assumed counts, we re-derive our completeness model with faint-end power-law slope $\gamma$ varied by $\pm 50\%$ about our fiducial value. We find that such changes lead to shifts in $m_{\rm lim}$ of order $\pm 0.1$ mag, suggesting that faint-end model uncertainties are unlikely to explain the observed difference in reported versus derived depths.

After constructing the completeness model, we use it to generate Monte Carlo realizations of random catalogs. Specifically, we use multi-band mock galaxy catalogs derived from the semi-empirical CIB model of \cite{mirocha26a}. For each CIB mock, we randomize the galaxy positions and evaluate $C(m_i, \vec{x}_i)$ for all sources $i$ to draw per-source detection probabilities $p_{\rm det}\sim C(m, \hat{x})$. We then use $p_{\rm det}$ to probabilistically accept or reject a given source, after which the accepted sources are retained in the random. We repeat this process twenty times in order to generate sufficiently dense catalogs that trace the HSC $i$-band survey depth on large scales. Finally, we retain the redshifts of the shuffled sources in order to construct randoms for our cross-correlations binned by photometric redshift in \S \ref{sec:results}.

\subsection{Impact of randoms corrections on power spectra}
\label{sec:randcorr_app}
\begin{figure}
    \centering
    \includegraphics[width=0.8\linewidth]{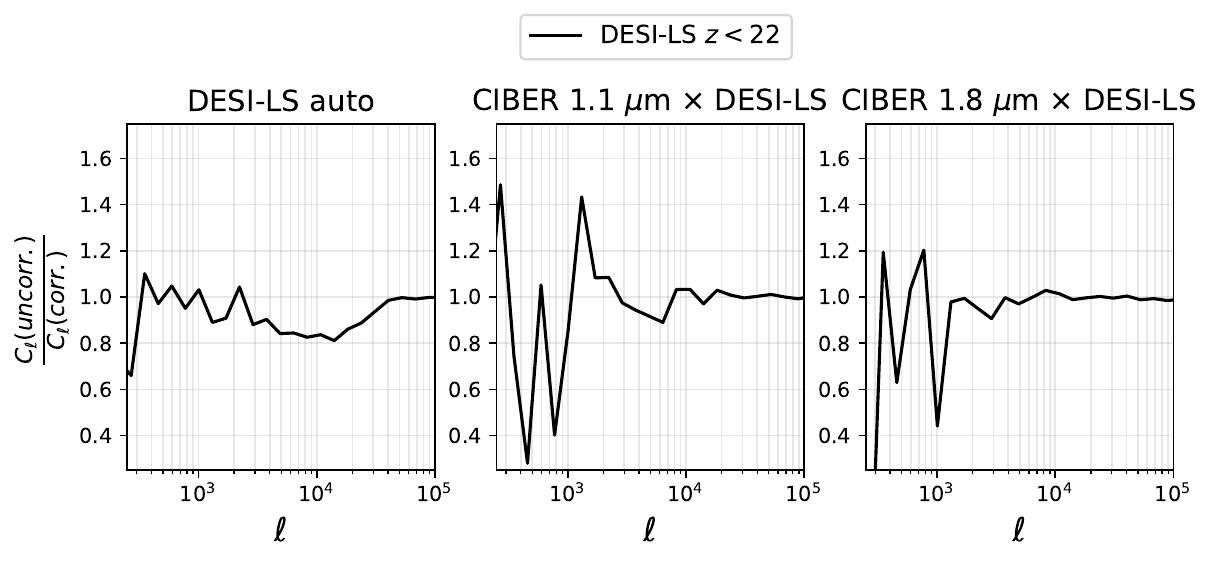}
    \includegraphics[width=0.8\linewidth]{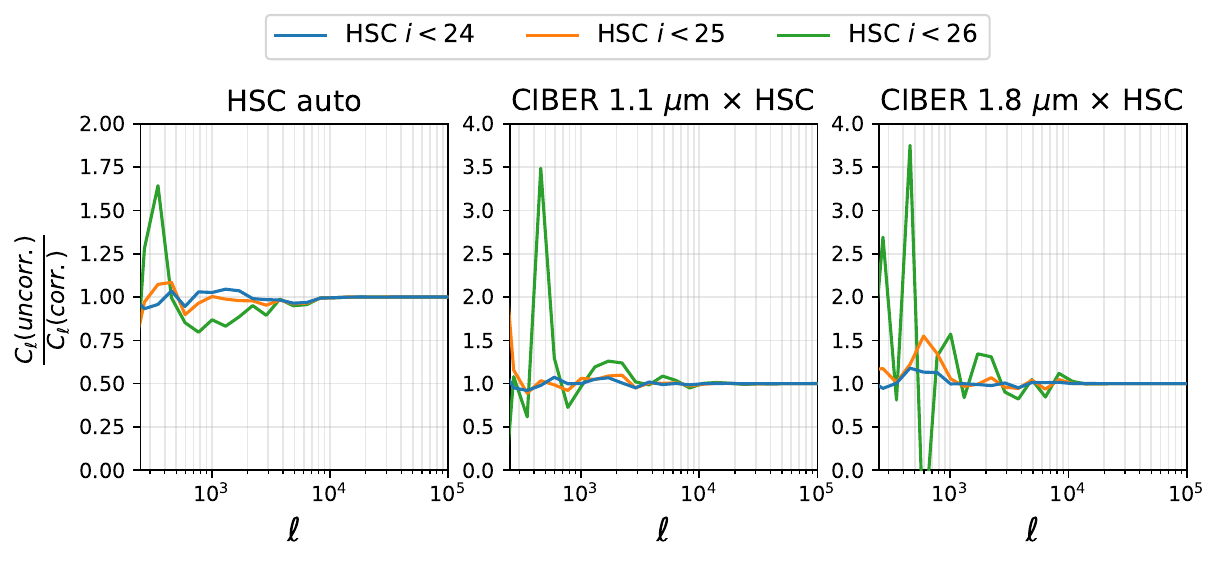}
    \caption{Impact of randoms correction on galaxy auto- and cross-spectra, for the DESI-LS catalog (top row) and for HSC as a function of $i$-band magnitude cut (bottom row). In each panel, we plot the ratio of uncorrected vs. corrected power spectra. In the left column of each row, we compare the auto-spectra, while the right two columns show the same quantities but for cross-spectra.}
    \label{fig:hsc_auto_cross_ratios}
\end{figure}

To assess the impact of the randoms correction on our measurements, in Fig. \ref{fig:hsc_auto_cross_ratios} we plot the ratio of the uncorrected vs. corrected galaxy auto- and cross-spectrum measurements. For each catalog, the galaxy auto typically varies less than the crosses. For DESI-LS, the randoms correction modifies the galaxy auto spectrum by $<20\%$ across the full multipole range. The HSC correction becomes larger with deeper magnitude cuts: while the corrections for $i_{\rm AB}<24$ and $i_{\rm AB}<25$ are small ($<10\%$), the correction for the $i_{\rm AB}<26$ sample has a much larger impact on the auto as well as the crosses, primarily at low-$\ell$. The size of the correction exceeds our formal estimates of the power spectrum uncertainty, highlighting the importance of their accuracy as future cross-correlation measurements push to much higher precision. For HSC, these comparisons and our revised completeness model (see \S \ref{sec:hsc_randoms}) guide our fiducial selections, in which we restrict our galaxy catalog to $i_{\rm AB}<25$.

We note as well that there are potential systematic errors associated with the derivation of the randoms and their associated corrections. While over-/under-correction of depth variations leads to residual biases in the galaxy auto spectrum, the impact on the crosses is primarily to increase the noise, assuming the depth variations that are uncorrelated with the galaxy density field.

\section{Consistency checks for galaxy auto- and cross-spectra}
\label{sec:consistency_checks}
\subsection{Field-field consistency}
\label{sec:field_consistency}

\begin{figure}
    \centering
    \includegraphics[width=0.9\linewidth]{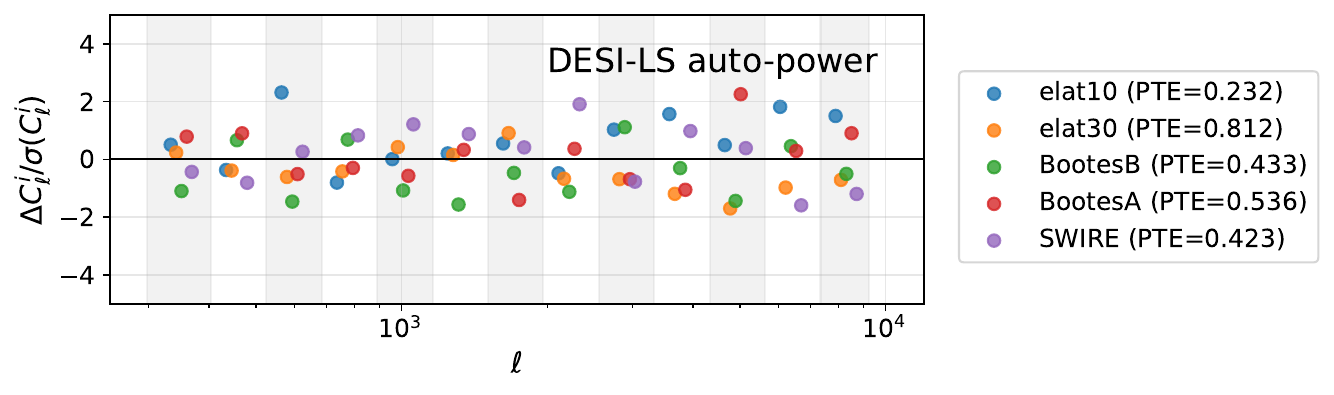}
    \includegraphics[width=0.9\linewidth]{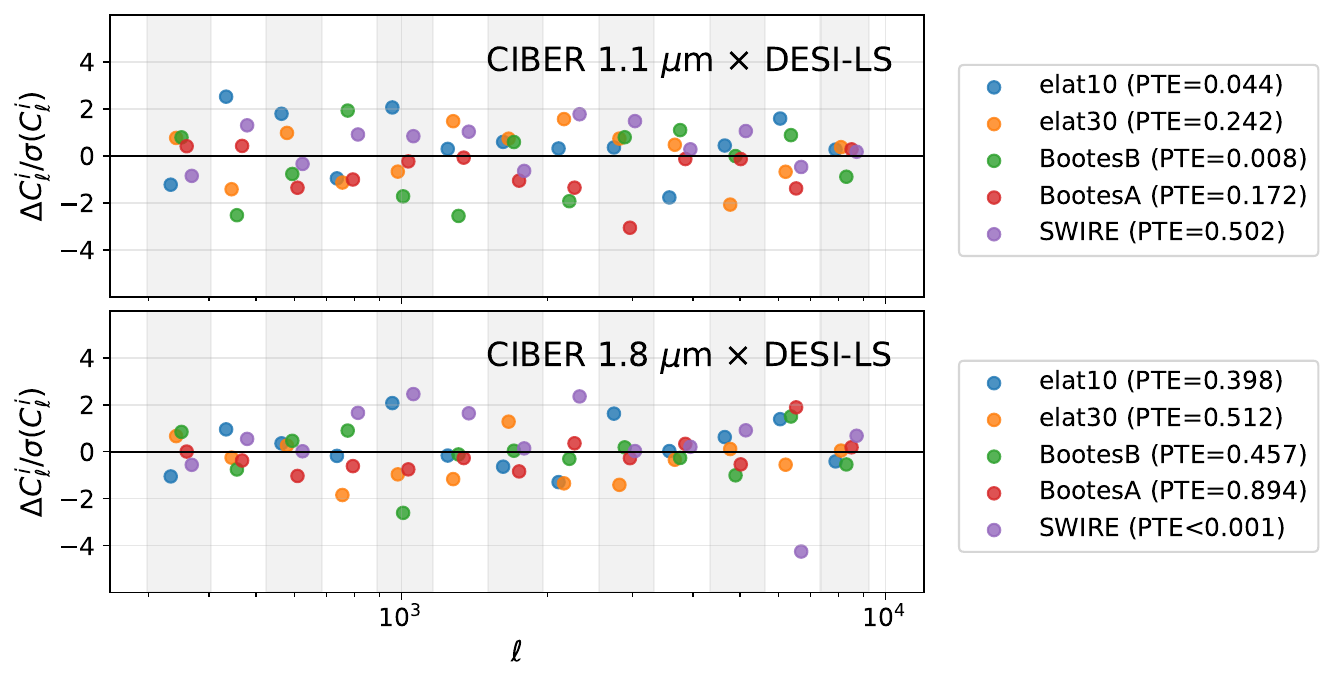}
    \caption{Field consistency of DESI-LS galaxy auto-power spectra (top panel) and \textit{CIBER} $\times$ DESI-LS cross-power spectra (middle/bottom panels). In each panel we plot the $\chi$ statistic, defined as the deviation of the per-field bandpowers from the field-averaged mean, normalized by the uncertainty. We evaluate $\chi$ for all bandpowers with $\ell < 10,\!000$. The Probability-to-Exceed (PTE) value for each field is provided in the legends.}
    \label{fig:cross_field_consistency}
\end{figure}

We assess the internal consistency of our galaxy auto and cross-spectrum measurements across the five \emph{CIBER} fields, following the approach used to quantify field consistency in the \emph{CIBER} auto-power spectra \citep{feder25b}. In particular, we derive per-field uncertainties that follow Eqs. \eqref{eq:auto_ps_uncertainty} and \eqref{eq:cross_ps_uncertainty}, using these to test the null hypothesis that the individual measurements are realizations of a single underlying cross-power spectrum. We show the results of this test in Fig. \ref{fig:cross_field_consistency}, with per-field probability-to-exceed (PTE) values calculated assuming a set of Gaussian-distributed measurements. We restrict to bandpowers $\ell<10,\!000$. 

The galaxy auto-spectrum results indicate strong consistency between fields, with essentially no strong outliers. Our cross-spectrum PTEs also indicate consistency, in particular at low-$\ell$, though there are a handful of outlier bandpowers with $|\chi|>3$. For the \emph{CIBER} 1.8 $\mu$m crosses, the SWIRE field has anomalously low PTE, which are driven by the same set of bandpowers (centered at $\ell=1000$, 2200, and 6500). Despite these caveats, the existing field-field discrepancies are smaller than those between our field-averaged measurements and model expectations at low $\ell$, reinforcing measured tensions with our baseline IGL model. 

\section{Halo model fits to DESI-LS and HSC cross-spectra}
\label{sec:app_parametric_fits}

\begin{figure}
    \centering
\includegraphics[width=0.49\linewidth]{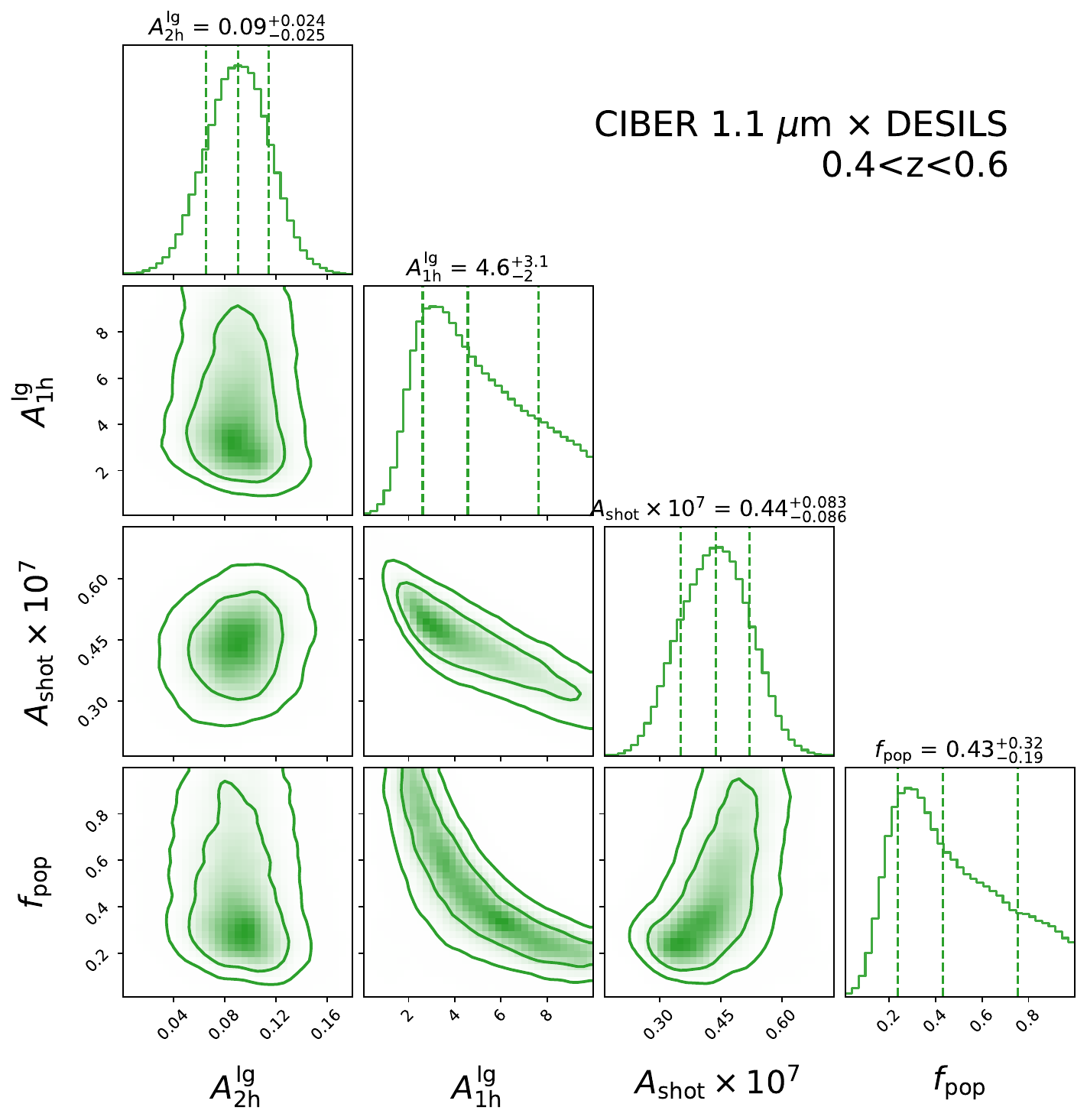}\includegraphics[width=0.49\linewidth]{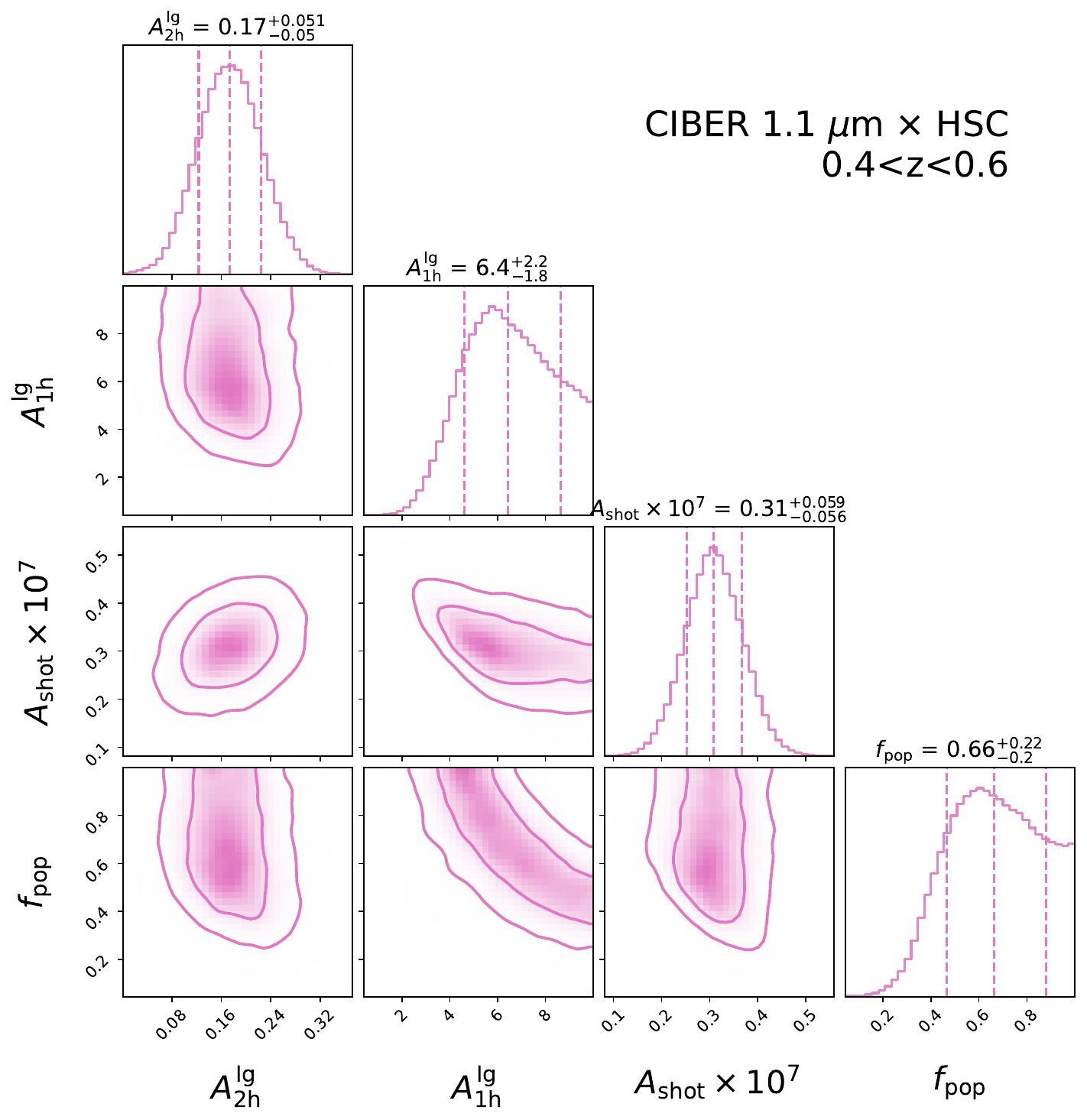}
    \caption{\textbf{Left:} Corner plot for \emph{CIBER} 1.1 $\mu$m $\times$ DESI-LS (left) and HSC (right) galaxies with $0.4 < z_{\rm phot}<0.6$. As the population weighting parameter $f_{\rm pop}$ modifies the shape of the one-halo template, the normalization of $A_{\rm 1h}^{\rm Ig}$ varies, leading us to report the effective one-halo power at fixed multipole in Fig. \ref{fig:A1h_vs_redshift}.}
    \label{fig:corner_plot}
\end{figure}

In Figure \ref{fig:corner_plot} we show the full posterior for our \emph{CIBER} 1.1 $\mu$m $\times$ DESI-LS cross in the same $z\in [0.4, 0.6)$ redshift bin as Fig. \ref{fig:ps_fit_0p4_z_0p6}. 

\begin{figure}
    \centering
    \includegraphics[width=\linewidth]{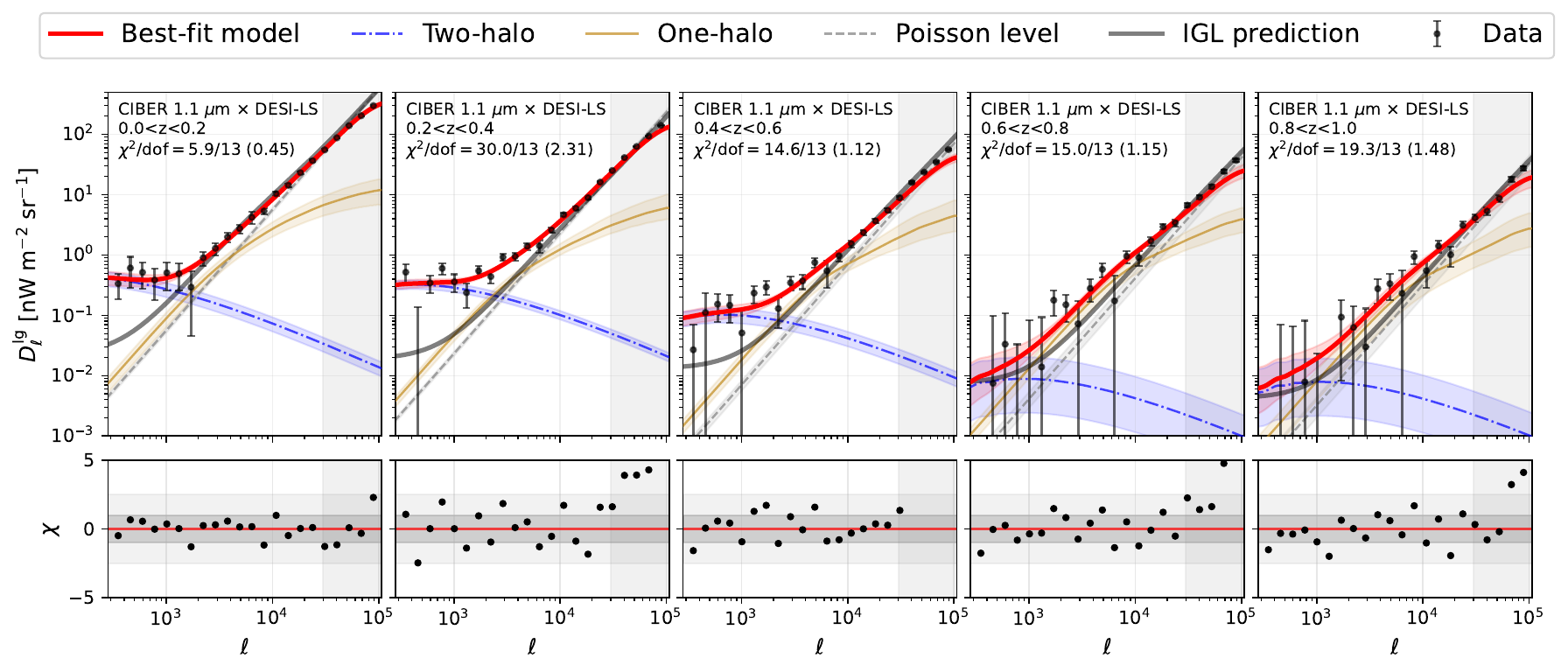}
    \includegraphics[width=\linewidth]{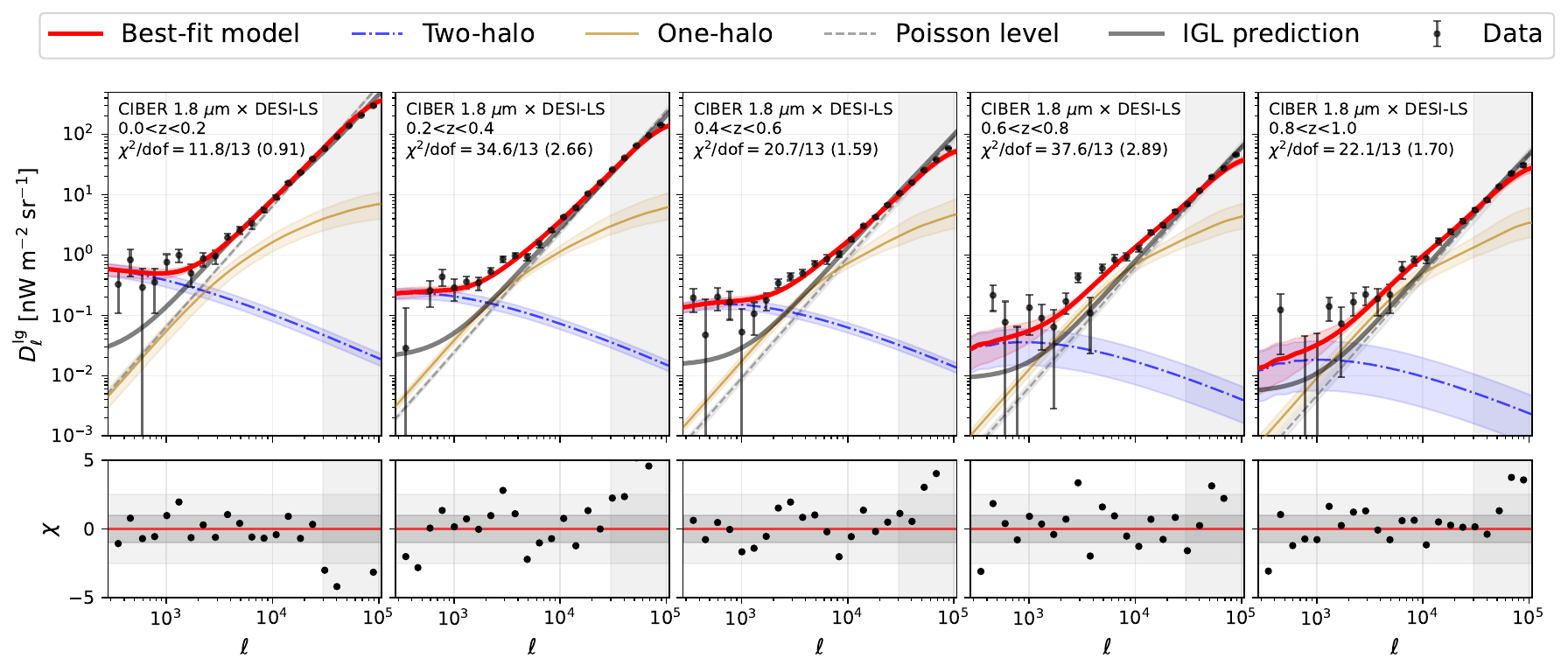}   
    \caption{Parametric fits to \emph{CIBER} $\times$ DESI-LS cross-power spectra, in five redshift bins spanning $0.0<z_{\rm phot}<1.0$. The top row shows the \emph{CIBER} 1.1 $\mu$m-based crosses, while the bottom row shows those for \emph{CIBER} 1.8 $\mu$m. Each model component -- $D_{\ell}^{\rm 2h}$ (2-halo, blue), $D_{\ell}^{\rm 1h}$ (1-halo, green), shot noise (grey) and their total (red) -- is plotted with 1$\sigma$ posterior uncertainties. The IGL predictions from \S \ref{sec:cib_predictions} are shown in magenta. For each combination we show the $\chi$ residuals with respect to $D_{\ell}^{\rm tot}$, with $\pm 1\sigma$ and $\pm2.5\sigma$ regions shaded in dark and light grey, respectively. The vertical shaded region in each panel corresponds to scales not used in our fiducial power spectrum fits ($\ell_{\rm max}>30,\!000$).}
    \label{fig:desils_fits}
\end{figure}

\begin{figure}
    \centering
    \includegraphics[width=\linewidth]{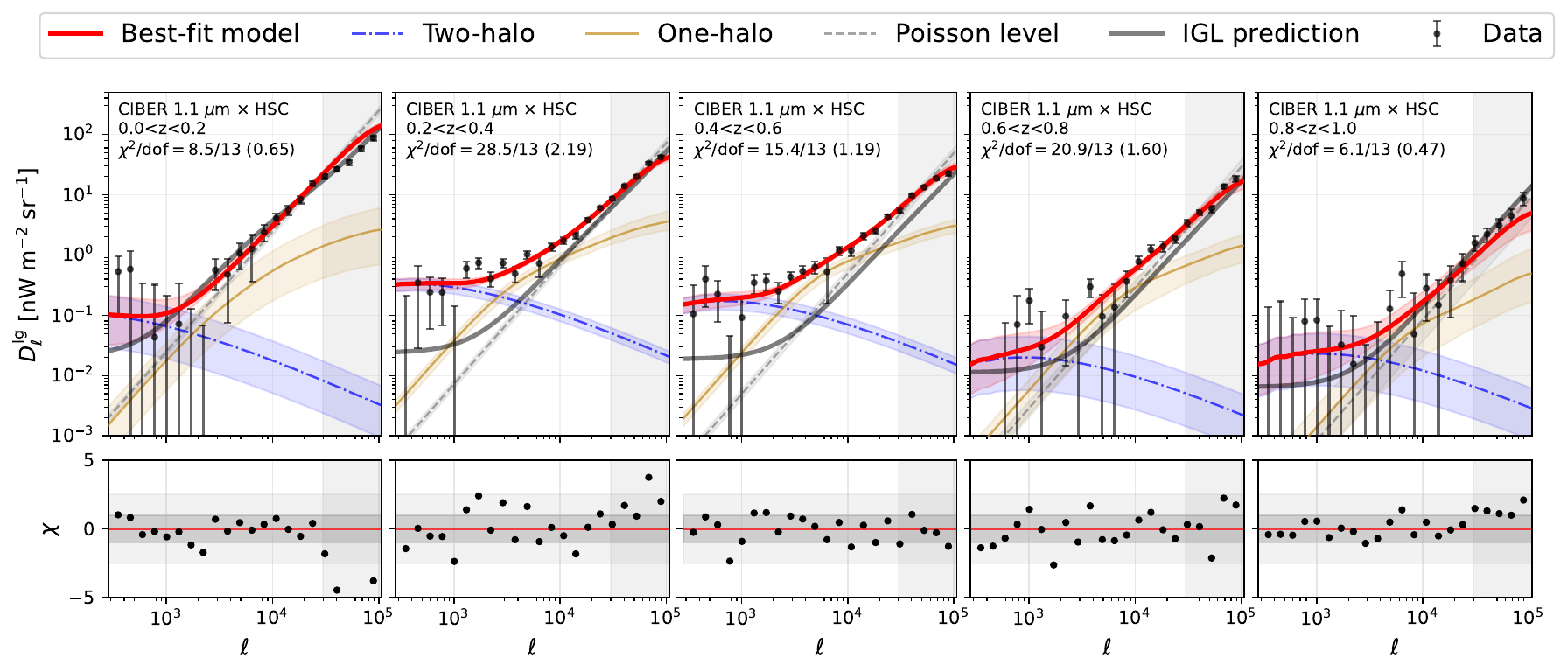}
    \includegraphics[width=\linewidth]{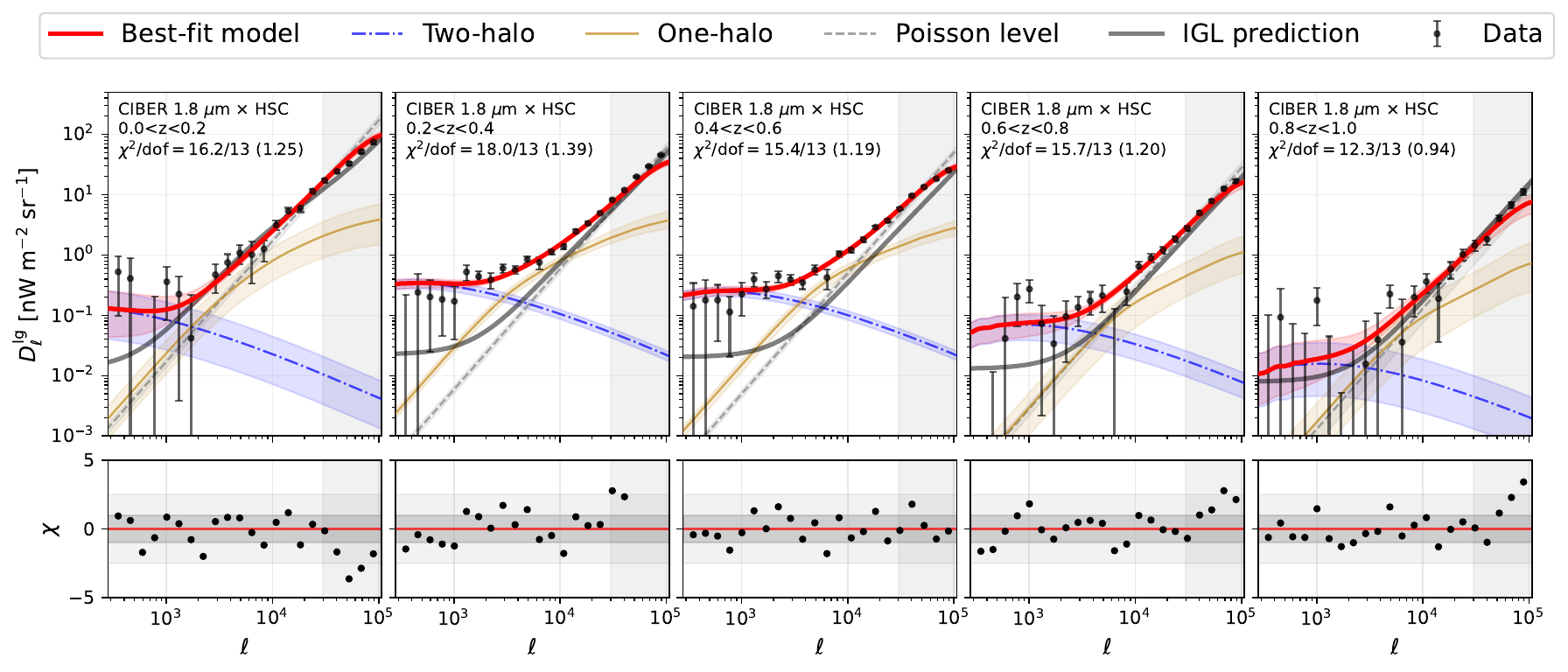}   
    \caption{Same as Fig. \ref{fig:desils_fits} but for HSC. Here we show the fits for $0.0<z<0.2$, however we do not use the HSC results in this bin for our final results due to bright-end selection effects (see \S \ref{sec:model_interp}).}
    \label{fig:hsc_fits}
\end{figure}

In Figures \ref{fig:desils_fits} and \ref{fig:hsc_fits} we show the results of our parametric halo model fits from \S \ref{sec:model_interp} to the \emph{CIBER} $\times$ galaxy cross-spectra, for all combinations and $\Delta z=0.2$ bins spanning $0.0<z<1.0$. We exclude bandpowers above $\ell=30,\!000$ from our fiducial fits, where we lack a detailed model for suppression for resolved galaxy profiles as discussed in \S \ref{sec:results}. Our reduced $\chi^2$ estimates (assuming four model parameters) indicate good fit quality for the majority of cross-spectra, though for DESI-LS we find some exceptions in the $z\in [0.2, 0.4]$ and $z\in [0.6, 0.8]$ bins with $\chi^2_{\rm red}=1.7-2.6$. At the level of our measurement uncertainties, the residuals do not suggest the presence of scale-dependent, coherent deviations between the fitted data and model.

\subsection{Goodness of fit and model preference}
\label{sec:goodness_of_fit_compare}
To assess the model preferences of our data, in Table \ref{tab:chi2_comparison} we present the $\chi^2$ values of our full model, along with the $\chi^2$ from ablated fits in which we remove the one- and two-halo components individually and evaluate the change in $\chi^2$ from each ablated model. The change in $\chi^2$ from removing the components are denoted by $\Delta \chi_{\rm 1h}^2$ and $\Delta \chi_{\rm 2h}^2$.

For our fiducial multipole fitting range ($304<\ell<30,\!000$), we find that both one- and two-halo terms improve the fit quality of cross-spectra for the majority of redshift bins and band/catalog combinations. For \emph{CIBER} $\times$ DESI-LS, we find that $\Delta \chi_{\rm 1h}^2$ ranges from $+4.6$ to $+41.3$, with the largest improvements in the range $0.2<z<0.6$. The improvements in the HSC fits are similar in this intermediate redshift range, however the two higher-redshift bins show less of a statistical preference for a one-halo component. While the HSC catalog is significantly denser than DESI-LS, it covers only one of the five \emph{CIBER} fields, meaning the statistical errors from sample variance are larger. Therefore, we do not interpret the difference between catalogs in this redshift as necessarily arising from astrophysical effects. We find largely similar conclusions for the two-halo ablation test -- both DESI-LS and HSC show strong evidence for a two-halo contribution to the cross-spectra in our first three redshift bins, while the improvement in the last two bins is marginal. 

\begin{table*}
\caption{Goodness-of-fit summary at fiducial $\ell_{\mathrm{max}}=30{,}000$. We report the $\chi^2_{\rm fid}$ values from our fiducial model (2h+1h+Poisson), while $\Delta\chi^2_{\rm 1h}$ and $\Delta\chi^2_{\rm 2h}$ denote the change in $\chi^2$ when the one-halo or two-halo component is removed (positive values indicate that component improves the fit). Entries in bold signify cases where the ablated model component is favored at $>95\%$ confidence ($\Delta\chi^2 \gtrsim 4$ for two-halo, and $\Delta \chi^2 \gtrsim 6$ for one halo).}
\label{tab:chi2_comparison}
\centering
\small
\begin{tabular}{llcccccccccc}
\toprule
& & \multicolumn{5}{c}{\textbf{CIBER $\times$ DESI-LS}} & \multicolumn{5}{c}{\textbf{CIBER $\times$ HSC}} \\
\cmidrule(lr){3-7} \cmidrule(lr){8-12}
Redshift & $\lambda_{\rm obs}$ & $\chi^2_{\rm fid}$ & $\chi^2_{\rm no1h}$ & $\Delta\chi^2_{\rm 1h}$ & $\chi^2_{\rm no2h}$ & $\Delta\chi^2_{\rm 2h}$ & $\chi^2_{\rm fid}$ & $\chi^2_{\rm no1h}$ & $\Delta\chi^2_{\rm 1h}$ & $\chi^2_{\rm no2h}$ & $\Delta\chi^2_{\rm 2h}$ \\
\midrule
\multirow{2}{*}{$0.0$--$0.2$} & $1.1\,\mu$m & 5.9 & 25.9 & $\mathbf{+19.9}$ & 19.5 & $\mathbf{+13.5}$ & --- & --- & --- & --- & --- \\
 & $1.8\,\mu$m & 11.8 & 22.8 & $\mathbf{+10.9}$ & 28.8 & $\mathbf{+17.0}$ & --- & --- & --- & --- & --- \\
\multirow{2}{*}{$0.2$--$0.4$} & $1.1\,\mu$m & 30.1 & 56.8 & $\mathbf{+26.7}$ & 66.6 & $\mathbf{+36.4}$ & 28.5 & 48.7 & $\mathbf{+20.2}$ & 46.9 & $\mathbf{+18.4}$ \\
 & $1.8\,\mu$m & 34.6 & 75.9 & $\mathbf{+41.3}$ & 60.0 & $\mathbf{+25.3}$ & 18.0 & 52.7 & $\mathbf{+34.7}$ & 47.6 & $\mathbf{+29.6}$ \\
\multirow{2}{*}{$0.4$--$0.6$} & $1.1\,\mu$m & 14.6 & 30.2 & $\mathbf{+15.6}$ & 28.2 & $\mathbf{+13.6}$ & 15.4 & 34.9 & $\mathbf{+19.5}$ & 26.6 & $\mathbf{+11.2}$ \\
 & $1.8\,\mu$m & 20.7 & 50.9 & $\mathbf{+30.2}$ & 49.6 & $\mathbf{+29.0}$ & 15.4 & 37.7 & $\mathbf{+22.3}$ & 48.5 & $\mathbf{+33.2}$ \\
\multirow{2}{*}{$0.6$--$0.8$} & $1.1\,\mu$m & 15.0 & 26.5 & $\mathbf{+11.5}$ & 14.0 & $-1.0$ & 20.9 & 25.2 & $+4.3$ & 19.9 & $-1.0$ \\
 & $1.8\,\mu$m & 37.6 & 56.4 & $\mathbf{+18.9}$ & 39.0 & $+1.4$ & 15.7 & 17.4 & $+1.8$ & 19.9 & $\mathbf{+4.2}$ \\
\multirow{2}{*}{$0.8$--$1.0$} & $1.1\,\mu$m & 19.2 & 23.7 & $+4.6$ & 18.1 & $-1.1$ & 6.2 & 6.2 & $-0.0$ & 5.8 & $-0.4$ \\
 & $1.8\,\mu$m & 22.0 & 32.8 & $\mathbf{+10.8}$ & 21.5 & $-0.5$ & 12.2 & 12.9 & $+0.6$ & 11.5 & $-0.8$ \\
\bottomrule
\end{tabular}
\end{table*}

\subsection{Parameter-level consistency checks}
\label{sec:param_vs_lmax}
\begin{figure*}
    \centering
\includegraphics[width=0.495\linewidth]{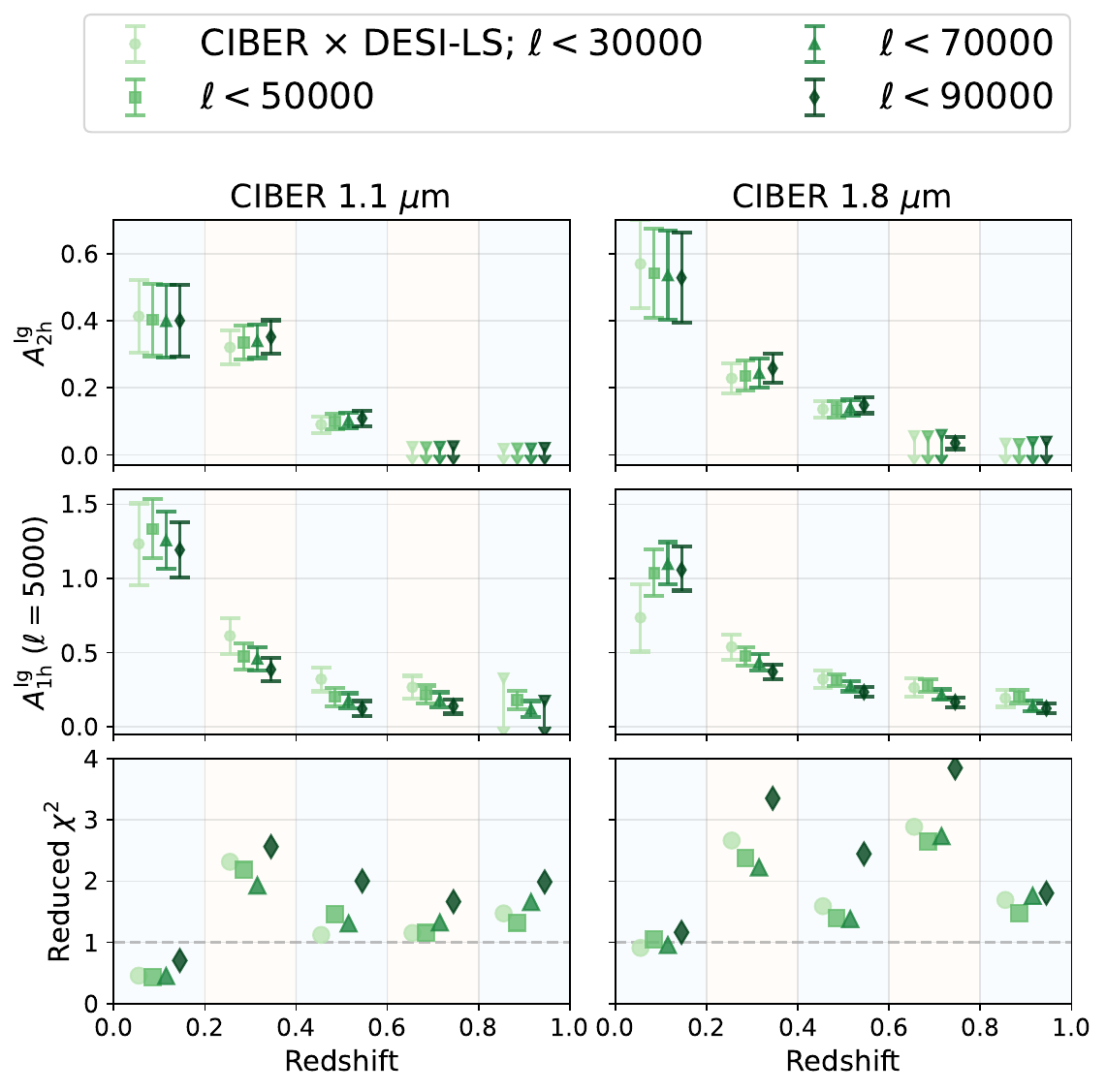}\includegraphics[width=0.495\linewidth]{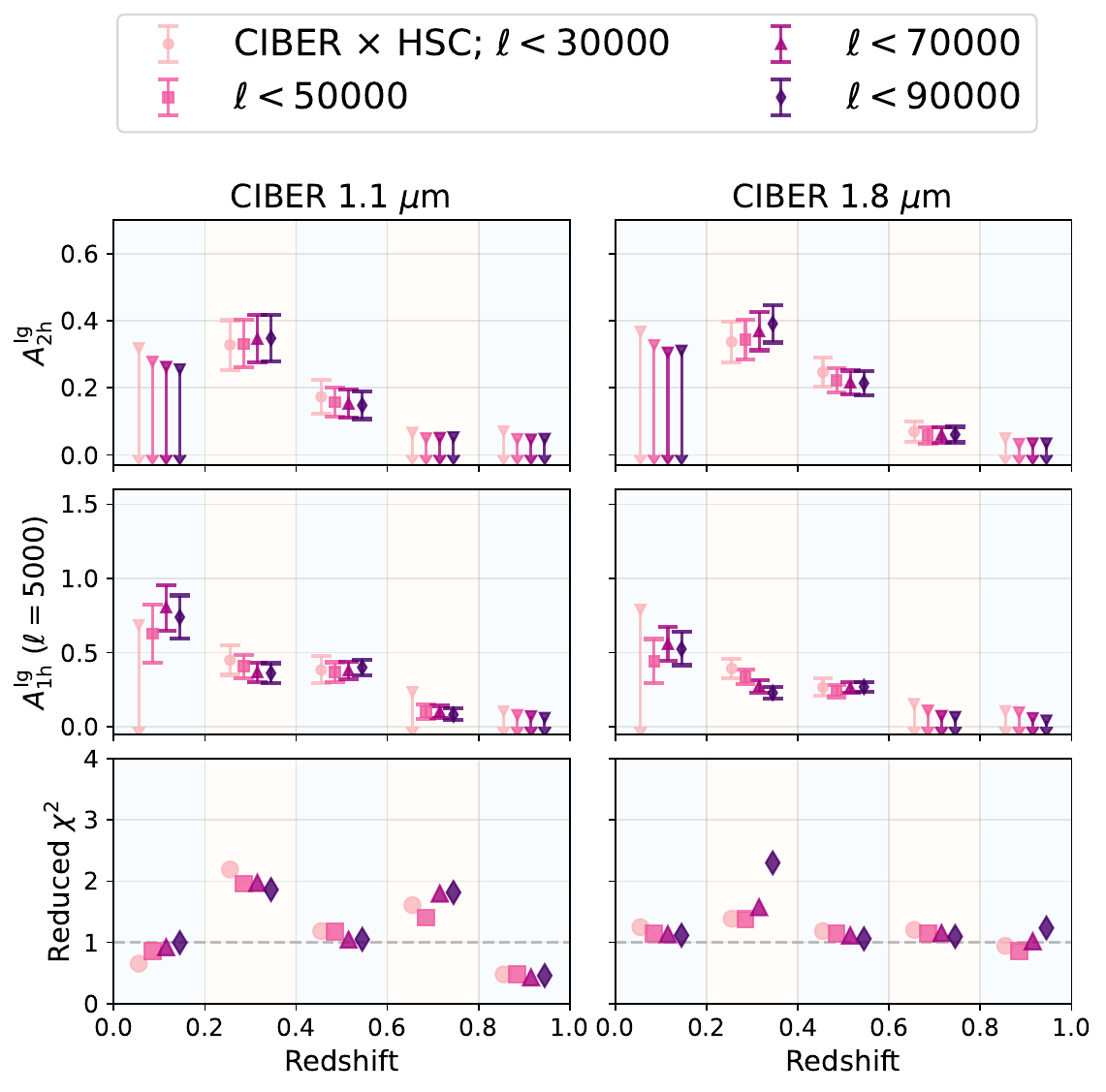}
    \caption{Parameter consistency checks as a function of maximum multipole used in the power spectrum fits presented in \S \ref{sec:model_interp}. The points within each redshift bin are artificially shifted horizontally in order to compare cases. Note that $A_{\rm 2h}^{\rm Ig}$ between catalogs carry different factors of the tracer galaxy bias $b_g$. We show the HSC $z<0.2$ results here, however they are omitted from our main results due to selection effects.}
    \label{fig:param_consistency_vs_lmax}
\end{figure*}

Inadequate modeling of small scales can potentially impact our one- and two-halo fits. To probe the sensitivity of our analysis to choice of maximum multipole, we run four variations with $\ell_{\rm max} \in \lbrace 30,\!000, 50,\!000, 70,\!000, 90,\!000\rbrace$. We show the resulting best-fit parameters across configurations in Fig.~\ref{fig:param_consistency_vs_lmax}. In general, we find that our estimates of $A_{\rm 2h}^{\rm Ig}$ are remarkably consistent across configurations, with shifts for high $\ell_{\rm max}$ at $<0.5\sigma$. On the other hand, the amplitude of the best-fit one-halo component shows a negative trend with increasing $\ell_{\rm max}$ for all but the lowest redshift bin, with shifts between $\ell_{\rm max}=30{,}000$ and $\ell_{\rm max}=90{,}000$ at the $1{-}2$ $\sigma$ level. This variation is more prominent for the DESI-LS crosses than for HSC. For higher $\ell_{\rm max}$, the one-halo fits partially benefits from higher-SNR measurements; however, because the one-halo fitted components are correlated with the Poisson level, any model misspecification on these scales can bias the Poisson level, which can in turn skew the one-halo fits.  

The reduced $\chi^2$ values for all $\ell_{\rm max}$ configurations are largely consistent, with the exception of the $\ell_{\rm max}=90,\!000$ case, where the $\chi^2$ degrades significantly across several redshift bins. The reduced $\chi^2$ is systematically higher for the DESI-LS $0.2<z<0.4$ bin, as well as for combinations with \textit{CIBER} 1.8 $\mu$m for the $0.6<z<0.8$ bin. We find systematically lower $\chi^2_{\rm red}$ for the HSC cross-spectrum fits compared to DESI-LS. With these exceptions in mind, we conclude that our halo model provides acceptable fits to the data. Given that the highest $\ell$ bandpowers primarily constrain the Poisson level and are subject to larger systematic errors, we adopt $\ell_{\rm max}=30,\!000$ as our fiducial choice throughout the analysis.

%% file: app_catalogs.tex
\section{Cross correlation with stars}
\label{sec:ciber_gaia}

\subsection{CIBER \texorpdfstring{$\times$}{x} Gaia cross-spectra}
\begin{figure}
    \centering
    \includegraphics[width=0.48\linewidth]{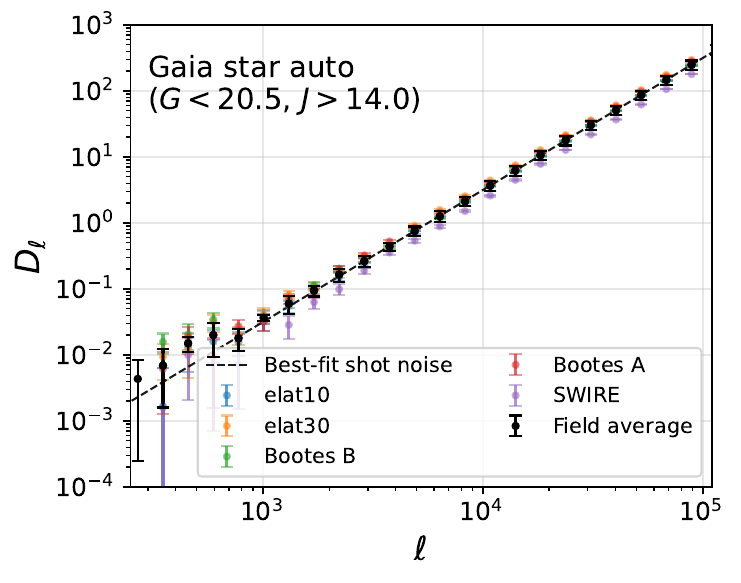}\includegraphics[width=0.51\linewidth]{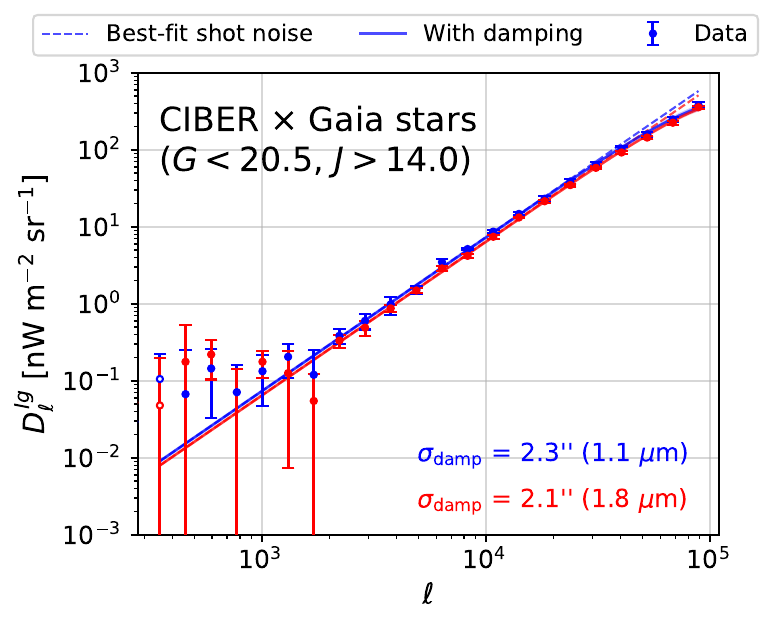}
    \caption{\emph{CIBER} $\times$ \emph{Gaia} stellar counts. The left-hand panel shows the auto-spectrum of \emph{Gaia} $G<20.5$ stars across the five \emph{CIBER} fields, while in the right-hand panel we show the field-averaged cross-spectra against \emph{CIBER} 1.1 $\mu$m (blue) and 1.8 $\mu$m (red).}
    \label{fig:ciber_gaia_star}
\end{figure}

We cross-correlate against Gaia stellar counts in order to assess whether Galactic foregrounds contribute clustering power in cross-correlation. We use the \emph{Gaia} DR3 catalog \citep{gaia_dr3}, and include all sources with $G<20.5$ and which have a point-like probability $p>0.95$. The stellar density across the \emph{CIBER} fields is $1,\!700{-}3,\!200$ deg$^{-2}$, with SWIRE (elat30) having the highest (lowest) density. For these measurements, we use a relaxed source mask, removing sources $J_{\rm AB}<14.9$ and $H_{\rm AB}<15.3$. In Figure \ref{fig:ciber_gaia_star} we show the resulting auto- and cross-spectra with \emph{CIBER} -- for both cases, we recover a power spectrum consistent with Poisson fluctuations, confirming earlier measurements with UKIDSS that found no significant clustering on degree scales and smaller \citep{zemcov14}. 

Following the discussion in \S \ref{sec:astr_errs}, we fit a Poisson + Gaussian damping model to the \emph{CIBER} $\times$ \emph{Gaia} cross-spectra in order to characterize source alignment errors. Using these cross-spectra avoids potential complication from interpreting high-$\ell$ suppression as coming from the resolved profiles of galaxies, which we expect to be present in galaxy cross-correlations. Furthermore, the position errors of these \emph{Gaia} sources are expected to be negligible. Our fits imply that, for both \emph{CIBER} imagers, the mean dispersion is consistent and at the $\sim 2.0-2.5^{\prime\prime}$ level (roughly one-third of a \emph{CIBER} pixel), with the caveat that the cross-Poisson level is flux-weighted and thus may be sensitive to fine astrometry errors around the brightest stars. When we fit to the individual fields, the 1.1 $\mu$m $\times$ \emph{Gaia} results have  $\sigma_{\rm damp} = \lbrace 1.84 \pm 0.04, 2.07 \pm 0.07, 2.00\pm 0.05, 1.59\pm 0.06, 2.24\pm 0.04\rbrace$ arcsec, while for 1.8 $\mu$m $\times$ \emph{Gaia} we obtain $\sigma_{\rm damp} = \lbrace 1.86\pm 0.04, 1.65\pm 0.06, 1.90\pm 0.04, 1.98\pm 0.04, 1.79\pm 0.04\rbrace$ arcsec. Given that both imagers take exposures simultaneously, we attribute the slightly higher estimates from 1.1 $\mu$m as arising due to higher instrument noise, which degrades the fine astrometry solution. We find that the mismatch in $\sigma_{\rm damp}$ between per-field and field-averaged estimates comes from the field weighting -- when we use uniform field weights to average the \emph{CIBER} $\times$ \emph{Gaia} cross-spectra and re-fit our model, we find $\sigma_{\rm damp} = 1.99\pm 0.04^{\prime\prime}$ and $1.89\pm 0.03^{\prime\prime}$ for 1.1 $\mu$m and 1.8 $\mu$m respectively, i.e., consistent with the per-field estimates within uncertainties. 

\subsection{Residual stellar contamination in tracer samples}
\label{sec:gaia_gal_cross}
As an additional point of comparison, we cross-correlate the \emph{Gaia} stellar density maps against our galaxy tracers. While the \emph{Gaia} $\times$ DESI-LS cross-spectra are consistent with zero on all scales, the \emph{Gaia} $\times$ HSC crosses have non-negligible cross-shot noise, with $r_{\ell}^{\rm Gaia, HSC} \approx 0.05$, suggesting our HSC sample has mild stellar contamination. While we do not pursue more detailed contamination modeling at the level of our cross-spectra, these results underscore the importance of stellar contamination for interpreting small-scale fluctuation measurements, such as those in \S \ref{sec:results}.


